\documentclass[10pt]{article} % For LaTeX2e

\usepackage[preprint]{tmlr}

\usepackage{amsmath,amsfonts,bm}

\def\eqref#1{equation~\ref{#1}}
\def\1{\bm{1}}

\DeclareMathAlphabet{\mathsfit}{\encodingdefault}{\sfdefault}{m}{sl}
\SetMathAlphabet{\mathsfit}{bold}{\encodingdefault}{\sfdefault}{bx}{n}

\usepackage{hyperref}
\usepackage{url}
\usepackage{graphicx}
\usepackage{placeins} 
\usepackage{subcaption}
\usepackage{fnpct}
\setcitestyle{numbers,square}
\usepackage{booktabs} 
\usepackage[normalem]{ulem}
\usepackage{multirow}
\usepackage{listings}
\usepackage{xcolor} % Required for coloring the code
\usepackage{rotating}

\title{Sparse Computations in Deep Learning Inference}

\author{
Ioanna Tasou$^{\dagger}$$^{*}$,
Panagiotis Mpakos$^{\dagger}$$^{*}$,
Angelos Vlachos$^{\ddagger}$,
Dionysios Adamopoulos$^{\dagger}$,
Georgios Giannakopoulos$^{\dagger}$,
Konstantinos Katsikopoulos$^{\dagger}$,
Ioannis Karaparisis$^{\dagger}$,
Maria Lazou$^{\dagger}$,
Spyridon Loukovitis$^{\ddagger}$,
Areti Mei$^{\dagger}$,
Anastasia Poulopoulou$^{\dagger}$,
Angeliki Dimitriou$^{\ddagger}$,
Giorgos Filandrianos$^{\ddagger}$,
Dimitrios Galanopoulos$^{\dagger}$,
Vasileios Karampinis$^{\ddagger}$,
Ilias Mitsouras$^{\ddagger}$,
Nikolaos Spanos$^{\ddagger}$,
Petros Anastasiadis$^{\dagger}$,
Ioannis Doudalis$^{\dagger}$,
Konstantinos Nikas$^{\dagger}$,
George Retsinas$^{\S}$,
Paraskevi Tzouveli$^{\ddagger}$,
Christina Giannoula$^{\P}$,
Nectarios Koziris$^{\dagger}$,
Nikela Papadopoulou$^{\#}$,
Giorgos Stamou$^{\ddagger}$,
Athanasios Voulodimos$^{\ddagger}$,
Georgios Goumas$^{\dagger}$
\thanks{Corresponding Authors: Ioanna Tasou, Panagiotis Mpakos, Georgios Goumas \{\href{mailto:itasou@cslab.ece.ntua.gr}{itasou},\href{mailto:pmpakos@cslab.ece.ntua.gr}{pmpakos},\href{mailto:goumas@cslab.ece.ntua.gr}{goumas}\}@cslab.ece.ntua.gr}
\\
\\
\addr{
$^{\dagger}$ Computing Systems Laboratory (CSLab), National Technical University of Athens, Greece \\
$^{\ddagger}$ Artificial Intelligence and Learning Systems Laboratory (AILS), National Technical University of Athens, Greece \\
$^{\S}$ Intelligent Robotics and Automation Laboratory (IRAL), National Technical University of Athens, Greece \\
$^{\P}$ Max Planck Institute for Software Systems (MPI-SWS), Germany \\
$^{\#}$ School of Computing Science, University of Glasgow, United Kingdom
}
}

\usepackage[dvipsnames]{xcolor}

\begin{document}

\thispagestyle{empty}

\maketitle

\begin{abstract}
The computational demands of modern Deep Neural Networks (DNNs) are immense and constantly growing.
While training costs usually capture public attention, inference demands are also contributing in significant computational, energy and environmental footprints. Sparsity stands out as a critical mechanism for drastically reducing these resource demands. However, its potential remains largely untapped and is not yet fully incorporated in production AI systems.
To bridge this gap, this work provides the necessary knowledge and insights for performance engineers keen to get involved in deep learning inference optimization. In particular, in this work we:
a) discuss the various forms of sparsity that can be utilized in DNN inference, 
b) explain how the original dense computations translate to sparse kernels,
c) provide an extensive bibliographic review of the state-of-the-art in the implementation of these kernels for CPUs and GPUs,
d) discuss the availability of sparse datasets in support of sparsity-related research and development,
e) explore the current software tools and frameworks that provide robust sparsity support, and 
f) present evaluation results of different implementations of the key SpMM and SDDMM kernels on CPU and GPU platforms.
Ultimately, this paper aims to serve as a resource for performance engineers seeking to develop and deploy highly efficient sparse deep learning models in productions.
\end{abstract}

%%%%%%%%%%%%%%%%%%%%%%%%%%%%%%%%%%%%%%%%%%%%%%%%%%%%%%%%%%%%%%%%%%%%%%%%%%%%%%
\section{Introduction} \label{sec:introduction}
%%%%%%%%%%%%%%%%%%%%%%%%%%%%%%%%%%%%%%%%%%%%%%%%%%%%%%%%%%%%%%%%%%%%%%%%%%%%%%

Artificial Intelligence (AI) is revolutionizing modern life, with advances in computer vision and Natural Language Processing (NLP) that enable transformative applications such as autonomous driving, intelligent virtual assistants, automated translation systems, precision medicine, financial anomaly detection, and generative models that produce text, images, and other creative content. These developments demonstrate that AI can considerably augment human capabilities, enhance decision-making, and reshape socio-technical systems across diverse domains. All these technological breakthroughs are based on the rapid evolution of Deep Neural Networks (DNNs), which in turn rely on the architectural enhancement of the networks themselves, the availability of huge amounts of data, and the support of large-scale computational infrastructures to host the unprecedented amounts of computations involved in the training and operation of these models. As of Q4 2025, the largest large language models (LLMs) publicly known (e.g., Qwen3-Max, Ling-1t, Samba-1, etc), appear to be approaching or exceeding one trillion parameters, marking a transition beyond the hundreds-of-billions scale. The computational needs to train such models are immense and exponentially increasing \cite{cottier2024rising}. Meta reported that Llama 4 models are trained on a cluster of 100,000 H100 GPUs \cite{wired_article}.
Although the computational needs for training capture more publicity, the operation of these models over time in inference tasks is expected to require significantly higher computational capabilities due to its repetitive nature \cite{katz2010sustainable}. In terms of environmental footprint, recent studies report that the operation of AI models already consumes electricity equal to that of 17,000 US homes \cite{jegham2025hungry}. 

Sparsification comes as a natural mechanism to tame this colossal demand for computing power. Sparse data structures and computational kernels have been resident in computing for several decades, as they precisely represent data and their relations in scientific computing and graph analytics, to name the most popular examples \cite{asanovic2006landscape}. In the context of DNNs, several forms of sparsity can be utilized to avoid a large percentage of the operations, therefore speeding up execution. Researchers and practitioners enforce sparsity by pruning layers or neurons from the networks, exploit the outcome of activation functions that lie at the end of each DNN layer and frequently output zeros, or just implement graph neural networks that are inherently sparse. Finally, other highly popular and effective mechanisms such as the Mixture of Experts (MoE) \cite{jacobs1991adaptive,shazeer2017outrageously,fedus2022switch} or even quantization \cite{gholami2022survey} can be considered sparsification methods in a wider sense. 

In this work, we argue that while sparsification is already a known optimization concept, its potential to further reduce computational requirements in production systems remains largely untapped. We base this claim on three reasons:
First, the industry has entered a competition - the `AI race' -  to deploy accurate AI systems, leaving aside at this stage the computational concern. This, however, will come forward soon to make the operational model sustainable \cite{varoquaux2025hype} and profitable.
Second, the technological field around sparsity is quite complex, since there exist several interacting potential sources of sparsity. Consequently, the performance gains from sparsity are not straightforward to quantify or guarantee, as they involve complex trade-offs between computational efficiency and problem characteristics.
Third, the complexity of the field has, for the moment, limited the involvement of performance engineers in the game, with current solutions in several cases being provided mainly by domain experts. We strongly believe that deeper involvement of performance engineers in sparse DNN optimization will provide much more efficient solutions.  

To that end, this survey aims to explore and enable the incorporation of sparsity mechanisms in production AI systems. It is written with performance engineers in mind with limited or no knowledge in DNNs. We find an interesting analogy from the past: sparse computations and in particular, sparse matrix-vector multiplication are ubiquitous in scientific computing, e.g., in the solution of large, sparse linear systems with iterative methods \cite{saad2003iterative}. The involvement of performance engineers and their collaboration with domain experts has increased the understanding of the interaction of the overarching application, the computational kernels involved, and the target execution platform, ultimately leading to highly effective, production-quality systems that rely on sparse data and computations \cite {balay2019petsc, jasak2009openfoam, heroux2005overview}.  

Since the field is quite complex and fast-evolving, in our initial attempt to capture it, we deliberately limit our focus: we exclude from our discussion the role of sparsity in accelerating training, and discuss only the aspects of training that influence sparse inference. We also restrict our focus on mainstream, commodity hardware platforms like CPUs and GPUs, and leave for future work a large family of research efforts that optimize code for specialized accelerators (e.g., \cite{gerogiannis2024hottiles}) or even co-design accelerators for this purpose (e.g., \cite{song2022sextans, gerogiannis2023spade, zhang2023dynasparse, qin2020sigma}).

We organize the survey around the following key questions:

\begin{enumerate}
    \item In which forms does sparsity arise in DNN inference? (Section~\ref{sec:sparse})
    \item How do the original dense computations transform to sparse kernels? (Sections~\ref{sec:background:dnn_basics} and~\ref{sec:sparse})
    \item What is the current state-of-the-art in the implementation of these kernels? (Section~\ref{sec:sparse_kernels})
    \item What sparse datasets can be used to assist research and development in sparse DNNs? (Section~\ref{sec:data})
    \item What is the current software support  to deploy production systems that rely on sparse DNNs? (Section~\ref{sec:sop})
    \item What is the performance behavior of sparse kernels in modern CPUs and GPUs? (Section~\ref{sec:eval})
\end{enumerate}

We would also like to make a note on the process that led to the production of this survey. We assembled a rather large and diverse team that included last-year undergraduate students, PhD students, postdoctoral researchers, and academics originating from the two disciplines of systems engineering and artificial intelligence. The process was interesting and constructive, revealing the existing gaps between scientists stemming from the two fields and revealing pathways that can bridge these gaps. At the end of the day, the process also served as a learning process for the team itself, and we hope that the community will find the survey useful and enjoyable, as we did with the process that led to it. We envision keeping this as a live document that will constantly capture the evolution in the field. In that sense, if you have comments, have spotted errors, or find that we have not included significant and relevant work, feel free to contact us.

%%%%%%%%%%%%%%%%%%%%%%%%%%%%%%%%%%%%%%%%%%%%%%%%%%%%%%%%%%%%%%%%%%%%%%%%%%%%%%
\section{Background} \label{sec:background}
%%%%%%%%%%%%%%%%%%%%%%%%%%%%%%%%%%%%%%%%%%%%%%%%%%%%%%%%%%%%%%%%%%%%%%%%%%%%%%

%%%%%%%%%%%%%%%%%%%%%%%%%%%%%%%%%%%%%%%%%%%%%%%%%%%%%%%%%%%%%%%%%%%%%%%%%%%%%%
\subsection{The basics of Deep Neural Networks (DNNs)} \label{sec:background:dnn_basics}
%%%%%%%%%%%%%%%%%%%%%%%%%%%%%%%%%%%%%%%%%%%%%%%%%%%%%%%%%%%%%%%%%%%%%%%%%%%%%%
This section gives a brief introduction to how inference works within DNNs and to the characteristics of the operations that power them. This material was included to provide the necessary context for the sparsification discussion in Section~\ref{sec:sparse};  readers already familiar with neural networks may skip it.

DNNs are computational models composed of multiple \textbf{layers} that perform operations towards fulfilling a specific task, typically \textbf{classification}  (e.g., classifying handwritten digits or objects in an image) or \textbf{regression} (e.g., to predict market prices). A typical DNN consists of an input layer, multiple hidden layers, and an output layer, stacked in a pipeline fashion. Conceptually, a neural network layer comprises a set of \textbf{neurons}, which perform linear transformations of the layer's input. These operations depend on a set of meta-parameters, namely the \textbf{weights} and \textbf{bias}, which are learned during the network training process. The hidden layers of a DNN serve as intermediate steps that extract and combine features from their input. By stacking hidden layers, the network can build abstract representations, capturing complex patterns. Nevertheless, a neural network as described up to this point would be a chain of linear operations. In order to enable the model to express more complex, non-linear relationships, \textbf{activation functions} are introduced after the output of each layer, which are consequently non-linear transformations. 

\begin{figure}[ht!]
    \centering
    \includegraphics[scale = 0.5]{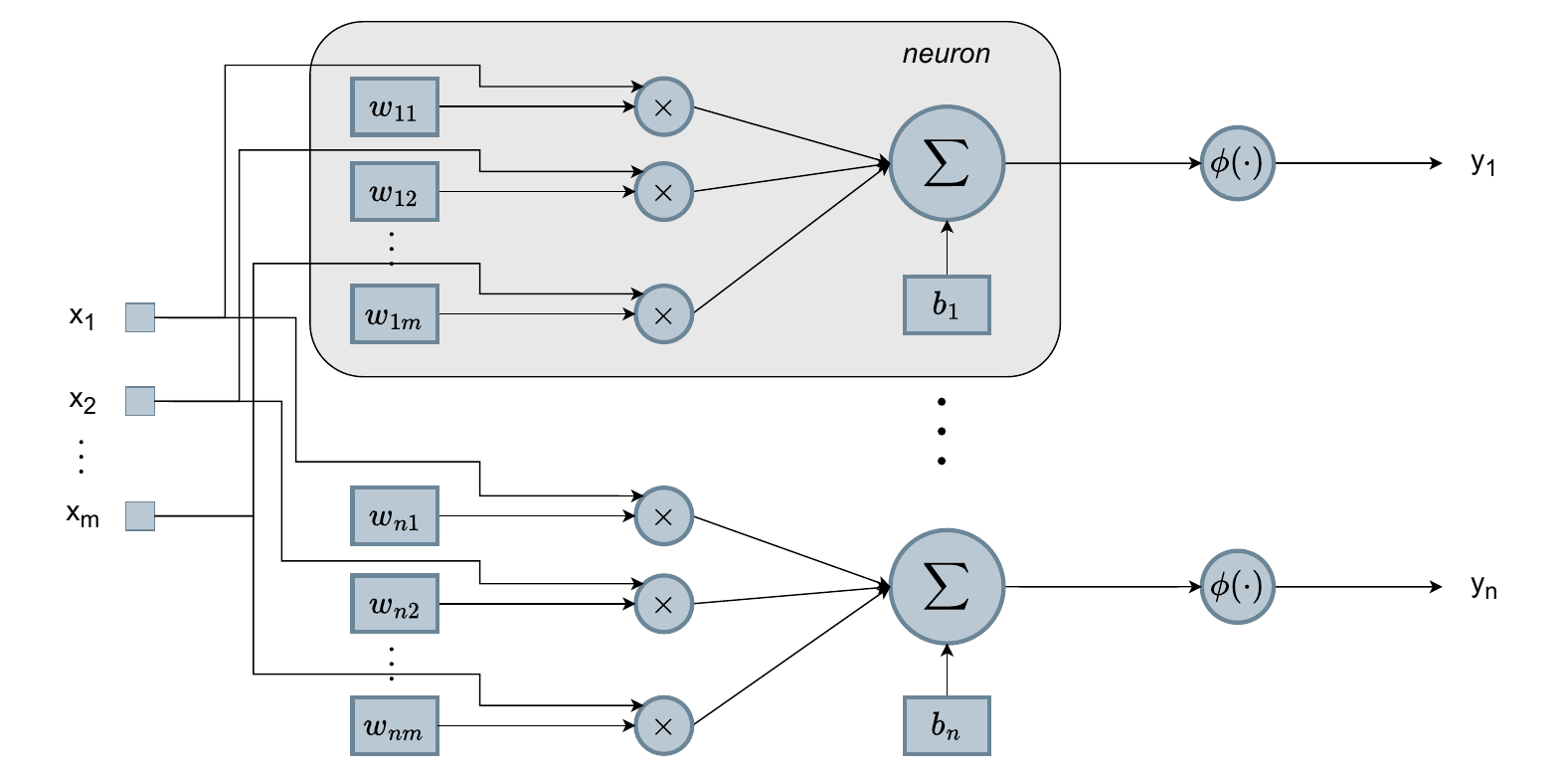}
    \caption{The structure of a basic neural network layer, followed by the activation function ($\phi$).}
    \label{fig:background_nn_layer}
\end{figure}

Figure~\ref{fig:background_nn_layer} presents a typical representation of a simple hidden layer. Input $\mathbf{x} \in \mathbb{R}^{m}$ linearly transformed through $n$ neurons with weights $w_{ij}$ and biases $b_i$, and then passes through a non-linear activation function $\phi$ to produce the output $\mathbf{y} \in \mathbb{R}^{n}$. Note that, in general, inputs and outputs can be multidimensional, representing, e.g., 2-dimensional images with multiple \textbf{features}. In the case of images, features can represent, for example, their RGB content (three features, one per RGB component). In NLP, each \textbf{token} (i.e., word or subword) can have various grammar or syntactical features called \textbf{embeddings}. In these cases, inputs and output are presented with multi-dimensional \textbf{tensors}.  

To accomplish their tasks, DNNs go through a training process during which the DNN calculates (learns) its weights and biases by minimizing a loss function, which measures how far its predictions are from the true targets. Using first-order gradient–based optimizers, such as \textit{Stochastic Gradient Descent (SGD)}, the model updates its parameters iteratively: for each small batch of data, it computes the loss, then applies backpropagation to calculate the gradients of the loss with respect to each weight. These gradients are used to adjust the weights in the direction that reduces the loss, gradually improving the network’s performance. Note that sparsification before or during training \cite{mocanu2018scalable, frankle2018lottery, liu2022unreasonable, evci2020rigging} is beyond the scope of this survey.

\begin{figure}[ht!]
    \centering
    \includegraphics[width=\linewidth]{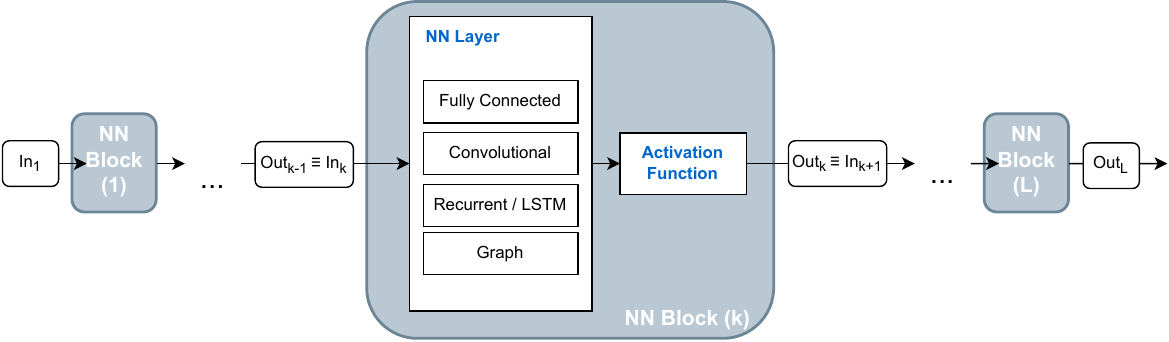}
    \caption{The generic architecture of a DNN.}
    \label{fig:dnn_arch}
\end{figure}

Figure~\ref{fig:dnn_arch}  provides an overview of a simplified but typical DNN architecture. A DNN is composed of multiple hidden building blocks, where each block typically consists of a specific type of hidden layer followed by an activation function. The most common types of hidden layers and activation functions used in modern deep learning are described in the next paragraphs.

%%%%%%%%%%%%%%%%%%%%%%%%%%%%%%%%%%%%%%%%%%%%%%%%%%%%%%%%%%%%%%%%%%%%%%%%%%%%%%
\subsubsection{Activation functions} \label{sec:background:dnn_basics:activation}
%%%%%%%%%%%%%%%%%%%%%%%%%%%%%%%%%%%%%%%%%%%%%%%%%%%%%%%%%%%%%%%%%%%%%%%%%%%%%%
Activation functions are essential components in neural networks, introducing non-linearity into the model that enables the learning of complex, hierarchical patterns beyond simple linear relationships. This nonlinearity is what fundamentally distinguishes dense neural networks from simple linear models. Without an activation function, stacking multiple linear layers would collapse into a single linear transformation, limiting the network's expressive power to that of basic linear regression. To break this limitation, non-linear activation functions are applied after each linear transformation. Historically, early neural networks primarily relied on the sigmoid function:
\begin{equation}
    \sigma(x)=\frac{1}{1+e^{-x}}
\end{equation}
and the hyperbolic tangent:
\begin{equation}
  \tanh (x)=\frac{e^x-e^{-x}}{e^x+e^{-x}}
\end{equation}
These functions are smooth and differentiable, which facilitates optimization using gradient-based methods such as backpropagation. However, they suffer from the vanishing gradient problem, where gradients become too small in saturated regions (e.g., very large positive or negative inputs), impeding learning in DNNs.

To address this issue, the Rectified Linear Unit (ReLU) \cite{nair2010rectified} was introduced as a simple, highly effective alternative. It is defined as:
\begin{equation}
    \operatorname{ReLU}(x)=\max (0, x)
\end{equation}
ReLU is non-linear, avoids saturation in the positive domain and allows gradients to flow efficiently through deep architecture. ReLU also promotes sparse activations by activating only neurons with positive input, reducing computation cost and improving generalization. Due to its simplicity and performance, ReLU has become the most frequent activation function in many modern DNNs.

Despite its advantages, ReLU has a limitation called "the dying ReLU problem", where neurons can become inactive and output zero for all inputs. This can happen when the neuron’s weights are updated in such a way that it only receives negative inputs. If a neuron dies, it stops contributing to the learning process, which can negatively impact model performance, particularly in deeper networks. To address this, ReLU variants have been introduced, such as Leaky ReLU \cite{maas2013rectifier},
\begin{equation}
    \operatorname{LeakyReLU}(x) = \max( ax, x), \quad x>0
\end{equation}
and Parametric ReLU (PReLU), where is a learned parameter \cite{he2015delving}, allowing small gradients for negative inputs. Leaky ReLU allows a small, nonzero gradient for negative inputs, helping to keep more neurons active.

Modern architectures have adopted more sophisticated activation functions. The Gaussian Error Linear Unit (GELU) \cite{hendrycks2016gaussian}, used in Transformer-based models like BERT and GPT, is defined as:
\begin{equation}
    \operatorname{GELU}(x) = x\cdot \Phi (x)
\end{equation}
where $\Phi(x)$ is the cumulative distribution function of the standard normal distribution. Another popular choice is the Swish function \cite{ramachandran2017searching}, defined as:
\begin{equation}
    \operatorname{Swish}(x) = x\cdot \sigma (\beta x)
\end{equation}
with $\beta$ typically a learnable or fixed parameter. Swish and GELU offer smoother transitions and better empirical performance in deep and attention-based models.  A widely used special case of Swish is the SiLU (Sigmoid Linear Unit), given by
    $\mathrm{SiLU}(x) = x \cdot \sigma(x)$, which corresponds exactly to Swish with $\beta = 1$.

Both Swish and GELU provide smooth, non-monotonic behavior and have shown strong empirical performance 
in deep convolutional networks and modern attention-based architectures.

There are also activation functions that operate on vectors. For example, softmax is a function that converts an input vector x into a probability distribution output vector:
\begin{equation}
    \operatorname{softmax}(x_i) = \frac{e^{x_i}}{\sum_{j=1}^{K} e^{x_j}}, \quad \text{for } i = 1, 2, \dots, K
\end{equation}
where $K$ is the number of classes, and is most commonly used at the output layer.

The choice of activation function is closely tied to both the model architecture and the specific learning task. In convolutional neural networks, ReLU and its variants remain dominant largely because they promote activation sparsity, setting a substantial portion of neuron outputs to zero. This sparsity not only reduces computational load but also enhances representational efficiency by encouraging the network to focus on the most salient features. Softmax, by contrast, is applied where normalized probability distributions are required, such as in attention mechanisms or the output layers of multi-class classifiers. Although hidden layers generally use nonlinear activations, output activations are selected according to the prediction objective: softmax for multi-class classification, sigmoid for binary decisions, and linear outputs for regression.

In modern attention-based and large-scale language models, smoother activation functions such as GELU and Swish are preferred because they allow more nuanced control over neuron responsiveness rather than inducing strict sparsity. While they do not produce the hard zeroing effect of ReLU, their soft gating behavior creates a form of graded sparsity, where small inputs are attenuated rather than eliminated. This smooth modulation supports stable gradient flow, improves optimization dynamics, and ultimately enhances performance in deep neural architectures.

%%%%%%%%%%%%%%%%%%%%%%%%%%%%%%%%%%%%%%%%%%%%%%%%%%%%%%%%%%%%%%%%%%%%%%%%%%%%%%
\subsubsection{Fully Connected layers} \label{sec:background:dnn_basics:fully_conn_layers}
%%%%%%%%%%%%%%%%%%%%%%%%%%%%%%%%%%%%%%%%%%%%%%%%%%%%%%%%%%%%%%%%%%%%%%%%%%%%%%
Fully Connected (FC) (dense/linear) layers are a main structural element of DNNs and the central mechanism in \textit{Multi-Layer Perceptrons (MLPs)}. A fully connected layer consists of multiple neurons, where each neuron connects each feature of the input with each feature of the output, and thus performs a linear transformation, which is followed by a non-linear activation function.
Stacking multiple such layers creates an MLP that comprises an input layer, one or more hidden layers, and an output layer. The mathematical formulation of a FC layer for a given input $\mathbf{x} \in \mathbb{R}^{d_{i n}}$, which is typically a vector representing a single input sample with $d_{\text {in }}$ features, which creates an output vector $\mathbf{y} \in \mathbb{R}^{\text {dout }}$ with $d_{\text {out}}$ features, is as follows:
\begin{equation}
    \mathbf{y}=\mathbf{W} \mathbf{x}+\mathbf{b}
    \label{eq:full}
\end{equation}
where, $\mathbf{W} \in \mathbb{R}^{d_{\text {out}} \times d_{\text {in}}}$ is the weight matrix calculated (learned) at training time, $\mathbf{b} \in \mathbb{R}^{d_{i n}}$ is bias vector. 

The formula above implies that the network operates on single input $\mathbf{x}$, and thus the dominant operation is \textit{Matrix-Vector Multiplication}. However, in several cases the network applies inference in batches of input, i.e., $\mathbf{x} \in \mathbb{R}^{d_{i n} \times \mathrm{B}}$ where $\mathrm{B}$ is the batch size, which translates to \textit{Matrix Multiplication}, which in the terminology of software libraries is commonly referenced as \texttt{GEMM} (GEneral Matrix Multiplication). 

FC layers, though theoretically capable of modeling big data, become inefficient and impractical when applied to high-dimensional inputs such as images. For instance, a single RGB image of size $(224 \times 224 \times 3)$ contains over $150,000$ input features; connecting each of these to a dense hidden layer would require millions of parameters, leading to severe risk of overfitting\footnote{Overfitting happens when a neural network learns the training data too specifically leading to poor performance on new data.}, increased memory usage, and computationally expensive training. Furthermore, this architecture requires flattening the image into a one dimensional vector, which dilutes the inherent spatial locality, (i.e., spatial relationships between neighboring pixels, like edges, textures and local patterns), which is crucial for visual content understanding. These challenges highlight the need for architectures that maintain the spatial arrangement of pixels and leverage hierarchical feature extraction, from simple local patterns (e.g., edges, corners) to complex global structures (e.g., entire objects), for efficient visual understanding.

%%%%%%%%%%%%%%%%%%%%%%%%%%%%%%%%%%%%%%%%%%%%%%%%%%%%%%%%%%%%%%%%%%%%%%%%%%%%%%
\subsubsection{Convolutional layers} \label{sec:background:dnn_basics:conv_layers}
%%%%%%%%%%%%%%%%%%%%%%%%%%%%%%%%%%%%%%%%%%%%%%%%%%%%%%%%%%%%%%%%%%%%%%%%%%%%%%
Convolutional layers are the basic building block of Convolutional Neural Networks (CNNs) and offer an efficient alternative that addresses the limitations mentioned above through two key principles: (i) local connectivity, where each neuron is connected only to a local patch of the input, and (ii) weight sharing, where the same filter is applied across different spatial locations. These properties make CNNs both translation-equivariant and highly parameter-efficient, particularly well-suited not only for visual tasks, but also for other structured data domains where spatial or local correlations are important.

\begin{figure}[t!]
    \centering
    \includegraphics[width=0.7\linewidth]{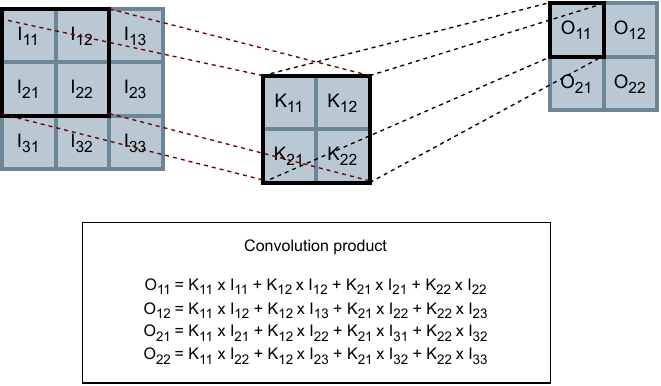}
    \caption{Convolution example: A $3\times 3$ input $I$ is convolved with a $2\times 2$ kernel $K$ using a sliding window to produce the $2\times 2$ output $O$.}
    \label{fig:background_convmatrices}
\end{figure}

The basic operation of a convolutional layer can be described as applying a sliding window over an array, where the window is defined by a learnable \textbf{filter} (or \textbf{kernel}). In image processing terms, kernels operate as `queries' to the specific part of the image as to whether the image contains a specific edge or corner (in early layers of the DNN), or a specific object (e.g., a letter in later layers of the DNN). Layers typically contain multiple kernels (e.g., 64), each one trained to query a specific feature. To illustrate the application of a single kernel, consider a grayscale image $\mathbf{I} \in \mathbb{R}^{3 \times 3}$ and a kernel $\mathbf{K} \in \mathbb{R}^{2 \times 2}$:
\begin{equation}
   \mathbf{I}=\left[\begin{array}{lll}
I_{11} & I_{12} & I_{13} \\
I_{21} & I_{22} & I_{23} \\
I_{31} & I_{32} & I_{33}
\end{array}\right], \quad \mathbf{K}=\left[\begin{array}{ll}
K_{11} & K_{12} \\
K_{21} & K_{22}
\end{array}\right] 
\end{equation}
With stride 1 and no padding, the resulting output $\mathbf{O} \in \mathbb{R}^{2 \times 2}$ is:
\begin{equation}
    \mathbf{O}=\left[\begin{array}{ll}
K_{11} I_{11}+K_{12} I_{12}+K_{21} I_{21}+K_{22} I_{22} & K_{11} I_{12}+K_{12} I_{13}+K_{21} I_{22}+K_{22} I_{23} \\[5pt]
K_{11} I_{21}+K_{12} I_{22}+K_{21} I_{31}+K_{22} I_{32} & K_{11} I_{22}+K_{12} I_{23}+K_{21} I_{32}+K_{22} I_{33}
\end{array}\right]
\end{equation}
Each entry $O_{i j}$ results from an element-wise product between the kernel and the corresponding $2 \times 2$ patch of the input, followed by summation, as shown in Figure~\ref{fig:background_convmatrices}. 

To generalize this operation, consider an input (grayscale) image $I \in \mathbb{R}^{H \times W}$ and a kernel $K \in \mathbb{R}^{m \times n}$ with height $m$ and width $n$. With stride~1 and no padding, the convolution produces an output $O \in \mathbb{R}^{(H-m+1) \times (W-n+1)}$. Each output entry corresponds to applying the kernel to a local $m \times n$ patch of the input, obtained by placing the \textit{top-left corner} of the kernel at position $(i,j)$. Formally,
\begin{equation}
O(i,j) = \sum_{u=0}^{m-1} \sum_{v=0}^{n-1} I(i+u,\; j+v)\, K(u,v),
\label{eq:convolution_grayscale}
\end{equation}
where the valid output indices are:
\[
0 \le i \le H - m,\qquad
0 \le j \le W - n.
\]

In practice, several variations of this basic operation are widely used. Padding is applied to preserve spatial resolution when desired, strided convolutions downsample feature maps by skipping positions, and dilated convolutions expand the receptive field without increasing kernel size. These extensions follow the same underlying principles but are beyond the scope of this work.

The convolution operation, as described in Equation~\ref{eq:convolution_grayscale}, involves nested loops and non-contiguous memory access, which is computationally inefficient on modern hardware. A common strategy followed by many deep learning libraries is to implement convolution as a single, large \texttt{GEMM} operation using the image-to-column (\texttt{im2col}) transformation. This technique converts for example the 2D grayscale input matrix ($I$) into a 2D matrix by taking every local receptive field (patch) that the kernel will operate on, stretching it into a single column vector, and stacking these columns side-by-side. The $K$ kernel (filter) is also flattened into a row vector. The original convolution is then transformed into a matrix multiplication: $\text{Flattened(} \textbf{O} ) = \text{Flattened(} \textbf{K} ) \times \text{im2col}(\mathbf{I})$ as shown in Equation~\ref{eq:im2col_operation}. This transformation maximizes data locality and allows the operation to leverage highly efficient \texttt{GEMM} routines.

\begin{equation} \label{eq:im2col_operation}
    \left[\begin{array}{llll}
    O_{11} & O_{12} & O_{21} & O_{22}
    \end{array}\right]
    =
    \left[\begin{array}{llll}
    K_{11} & K_{12} & K_{21} & K_{22} \\
    \end{array}\right]
    \times
    \left[\begin{array}{llll}
    I_{11} & I_{12} & I_{21} & I_{22} \\
    I_{12} & I_{13} & I_{22} & I_{23} \\
    I_{21} & I_{22} & I_{31} & I_{32} \\
    I_{22} & I_{23} & I_{32} & I_{33}
    \end{array}\right]
\end{equation}

Historically, CNNs gained early recognition with the introduction of LeNet-5 \cite{lecun2002gradient}, which was designed for handwritten digit recognition. Their true breakthrough came with AlexNet \cite{krizhevsky2012imagenet}, which achieved a significant performance leap on the ImageNet Large Scale Visual Recognition Challenge (ILSVRC) \cite{russakovsky2015imagenet}, establishing CNNs as a cornerstone of modern computer vision.  Subsequent architectures such as VGGNet \cite{simonyan2014very}, ResNet \cite{he2016deep}, and DenseNet \cite{huang2017densely} further advanced the field by introducing innovations like increased depth, residual connections, and dense connectivity—enhancing training stability and overall accuracy.
Today, convolutional layers serve as the backbone for most computer vision systems, forming the basis for advanced tasks, such as object detection (e.g., in YOLO \cite{redmon2016you}, Faster R-CNN \cite{ren2015faster}) and image segmentation (e.g., in U-Net \cite{ronneberger2015unet}, DeepLab \cite{chen2017deeplab}). While attention-based models like Vision Transformers (ViTs) \cite{dosovitskiy2020image} are emerging, CNNs remain fundamental due to their efficiency in spatial data processing.

%%%%%%%%%%%%%%%%%%%%%%%%%%%%%%%%%%%%%%%%%%%%%%%%%%%%%%%%%%%%%%%%%%%%%%%%%%%%%%
\subsubsection{Sequence modeling: Recurrent Neural Networks and Long Short-Term Memory networks}
\label{sec:background:dnn_basics:seq_layers}
%%%%%%%%%%%%%%%%%%%%%%%%%%%%%%%%%%%%%%%%%%%%%%%%%%%%%%%%%%%%%%%%%%%%%%%%%%%%%%
Sequence models address tasks where the order of inputs and the temporal relationships, i.e., position and time-based dependencies that exist between elements in a sequence, carry essential meaning.
These requirements occur in areas such as speech recognition, language processing, and time series prediction. Initially, \textit{Hidden Markov Models (HMMs)} \cite{rabiner2002tutorial} provided a primitive statistical approach to sequence modeling, setting the foundation for more advanced architectures \cite{gales2008application} such as the \textit{Recurrent Neural Networks (RNNs)} \cite{elman1990finding} and the \textit{Long Short-Term Memory (LSTM)} \cite{hochreiter1997long} networks which introduced mechanisms to model sequential data and capture temporal dependencies. \textit{Transformers} (see Section~\ref{sec:background:dnn_basics:transformer_attention}) advanced this further by using self attention to process sequences in parallel and enabling more efficient and scalable learning of long-range relationships.   

\begin{figure}[ht!]
    \centering
    \includegraphics[scale = 1]{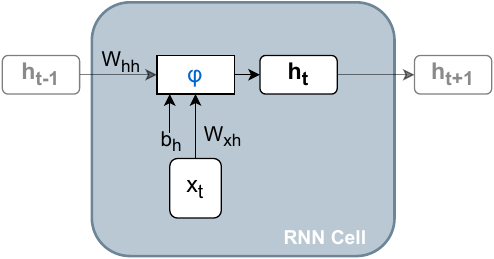}
    \caption{A basic RNN cell, where the hidden state is updated from the previous hidden state and current input, using shared parameters across timesteps.}
    \label{fig:rnn_cell}
\end{figure}

%%%%%%%%%%%%%%%%%%%%%%%%%%%%%%%%%%%%%%%%%%%%%%%%%%%%%%%%%%%%%%%%%%%%%%%%%%%%%%
\paragraph{Recurrent Layers}
%%%%%%%%%%%%%%%%%%%%%%%%%%%%%%%%%%%%%%%%%%%%%%%%%%%%%%%%%%%%%%%%%%%%%%%%%%%%%%
Recurrent layers are the core building block of Recurrent Neural Networks (RNNs) \cite{elman1990finding}, introduced in the late 1980s. Recurrent layers handle sequential data (e.g., time series, language, speech) by using internal loops to retain information from previous inputs, effectively allowing the network to maintain a form of memory \cite{ghojogh2023recurrent}. This mechanism enables the modeling of sequential dependencies by propagating context through the model's hidden states.  
Figure~\ref{fig:rnn_cell} illustrates a recurrent cell, where the hidden state $\mathbf{h}$ receives input $\mathbf{x}$ and is updated over time through recurrent connections to capture temporal dependencies.  
The following equation describes the operation in an RNN cell:
\begin{equation}
\mathbf{h}_t=\phi\left(\mathbf{W}_{h h} \mathbf{h}_{t-1}+\mathbf{W}_{x h} \mathbf{x}_t+\mathbf{b}_h\right)
\end{equation}
At each timestep $t$, the hidden state $\mathbf{h}_t \in \mathbb{R}^{d_h}$ is updated by combining the previous hidden state  $\mathbf{h}_{t-1}$ and the current input $\mathbf{x}_t \in \mathbb{R}^{d_x}$. The learned weight matrices $\mathbf{W}_{hh} \in \mathbb{R}^{d_h \times d_h}$ and $\mathbf{W}_{xh} \in \mathbb{R}^{d_h \times d_x}$, together with the bias $\mathbf{b}_h \in \mathbb{R}^{d_h}$, determine how past and current information are integrated, and the activation function $\phi$ (commonly $\tanh$ or ReLU) introduces nonlinearity.
This mechanism allows the RNN to retain information over time by carrying forward contextual knowledge through the hidden state. 
The hidden state $\mathbf{h}_t$ is then passed to deeper layers (if present) and reused at the next timestep within the same layer. 
Unfolding the recurrent layer across timesteps reveals how the same cell is reused to process input sequences while maintaining memory through the hidden states.

From the above analysis, we can deduce that RNNs are inherently serial networks, since they need to process one input element at a time (e.g., a word within text). Due to this characteristic, the main computational kernel of RNNs is Matrix-Vector Multiplication (commonly realized through \texttt{GEMV}, or \texttt{GEMM} when batched) between the weight matrices and the input and hidden-state vectors.

RNNs found significant use in language modeling and speech recognition tasks; however, they faced considerable challenges, most notably the vanishing and exploding gradient problems, which severely limited their ability to capture long-range dependencies effectively \cite{ghojogh2023recurrent}.

\begin{figure}[ht!]
    \centering
    \includegraphics[width=0.7\linewidth]{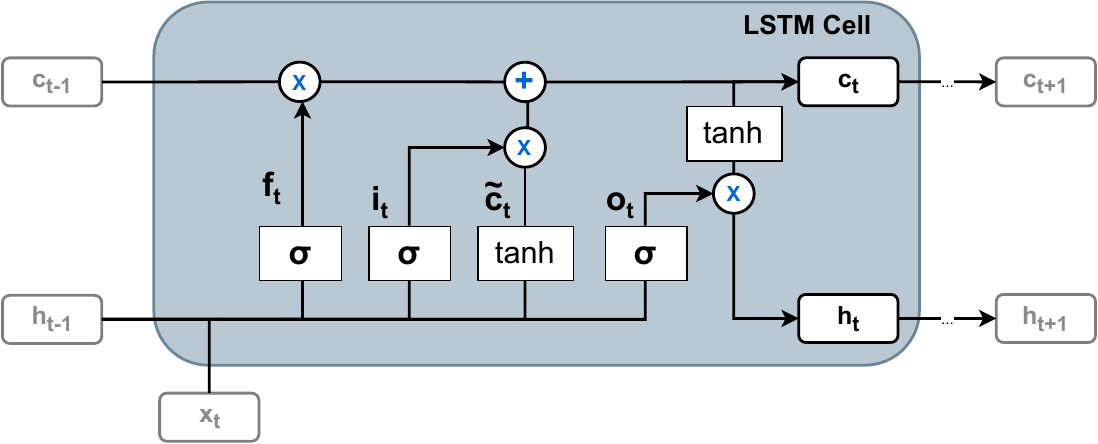}
    \caption{Structure of an LSTM cell with its gating mechanisms and cell state update.} 
    \label{fig:lstm_cell}
\end{figure}

%%%%%%%%%%%%%%%%%%%%%%%%%%%%%%%%%%%%%%%%%%%%%%%%%%%%%%%%%%%%%%%%%%%%%%%%%%%%%%
\paragraph{Long Short-Term Memory (LSTM)}
%%%%%%%%%%%%%%%%%%%%%%%%%%%%%%%%%%%%%%%%%%%%%%%%%%%%%%%%%%%%%%%%%%%%%%%%%
To overcome the limitations of RNNs in capturing long-range dependencies, Long Short-Term Memory (LSTM) networks emerged in the late 1990s, introducing gating mechanisms to better manage information flow and stabilize gradient propagation \cite{ghojogh2023recurrent}. 
LSTMs are also recurrent layers, extending simple RNNs by introducing a dedicated cell state that acts as long-term memory. This cell state flows through the network with minimal modification and is regulated by learned gating mechanisms. The internal structure and operations of an LSTM cell are illustrated in Figure~\ref{fig:lstm_cell}.

Mathematically, let the input at timestep $t$ be $\mathbf{x}_t \in \mathbb{R}^{d_x}$ and let $\mathbf{h}_{t-1}, \mathbf{c}_{t-1} \in \mathbb{R}^{d_h}$ denote the previous 
hidden state and cell state, respectively.
The LSTM architecture defines three gates at each 
timestep: the input gate $\mathbf{i}_t \in \mathbb{R}^{d_h}$, the forget gate $\mathbf{f}_t \in \mathbb{R}^{d_h}$, 
and the output gate $\mathbf{o}_t \in \mathbb{R}^{d_h}$. 
Each gate is associated with:
\begin{itemize}
    \item the forget gate 
    $\mathbf{f}_t=\sigma\!\left(\mathbf{W}_f \mathbf{x}_t + \mathbf{U}_f \mathbf{h}_{t-1} + \mathbf{b}_f\right)$ 
    determines which parts of the previous cell state $\mathbf{c}_{t-1}$ are retained;
    
    \item the input gate 
    $\mathbf{i}_t=\sigma\!\left(\mathbf{W}_i \mathbf{x}_t + \mathbf{U}_i \mathbf{h}_{t-1} + \mathbf{b}_i\right)$ 
    controls how much new information 
    $\tilde{\mathbf{c}}_t=\tanh\!\left(\mathbf{W}_c \mathbf{x}_t + \mathbf{U}_c \mathbf{h}_{t-1} + \mathbf{b}_c\right)$ 
    enters the cell state;
    
    \item the cell state update 
    $\mathbf{c}_t=\mathbf{f}_t \odot \mathbf{c}_{t-1} + \mathbf{i}_t \odot \tilde{\mathbf{c}}_t$ 
    combines retained memory with candidate values;
    
    \item the output gate 
    $\mathbf{o}_t=\sigma\!\left(\mathbf{W}_o \mathbf{x}_t + \mathbf{U}_o \mathbf{h}_{t-1} + \mathbf{b}_o\right)$ 
    filters the updated cell state, producing the hidden state 
    $\mathbf{h}_t=\mathbf{o}_t \odot \tanh(\mathbf{c}_t)$.
\end{itemize}

Here, $\mathbf{W}_f, \mathbf{W}_i, \mathbf{W}_o, \mathbf{W}_c  \in \mathbb{R}^{d_h \times d_x}$ and $\mathbf{U}_f, \mathbf{U}_i, \mathbf{U}_o, \mathbf{U}_c  \in \mathbb{R}^{d_h \times d_h}$ 
are the learned weight matrices, and $\mathbf{b}_f, \mathbf{b}_i, \mathbf{b}_o, \mathbf{b}_c \in \mathbb{R}^{d_h}$ are the bias parameters for their respective gates.
The symbol $\odot$ denotes element-wise (Hadamard) multiplication, and the sigmoid function 
$\sigma(\cdot)$ maps gate values to $(0,1)$.

The gates allow the LSTM to selectively use, write and forget information, making it highly effective for tasks involving long-term dependencies like language modeling, translation and time-series forecasting.

As with RNNs, LSTMs are inherently serial models whose core computational kernels are matrix–vector multiplications (or matrix–matrix when batched). However, each timestep now requires several such operations, one for each gate, making LSTMs computationally heavier than simple RNNs. 
Note that more compact variants, such as Gated Recurrent Units (GRUs), reduce the number of gates and thus require fewer computations while still retaining key gating benefits \cite{chung2014empirical}.

%%%%%%%%%%%%%%%%%%%%%%%%%%%%%%%%%%%%%%%%%%%%%%%%%%%%%%%%%%%%%%%%%%%%%%%%%%%%%%
\subsubsection{Transformer networks and the Attention mechanism} \label{sec:background:dnn_basics:transformer_attention}
%%%%%%%%%%%%%%%%%%%%%%%%%%%%%%%%%%%%%%%%%%%%%%%%%%%%%%%%%%%%%%%%%%%%%%%%%%%%%%
While LSTMs improved long-term dependency modeling, the real breakthrough came with the introduction of Transformers in 2017 \cite{vaswani2017attention}. Transformers replaced recurrence entirely with a \textbf{self-attention} mechanism, which allowed the model to weigh the importance of every token in the input sequence relative to all others simultaneously, significantly enhancing parallel computation and contextual understanding \cite{vaswani2017attention}. 

\begin{figure}[ht!]
\centering
\includegraphics[scale  = 1]{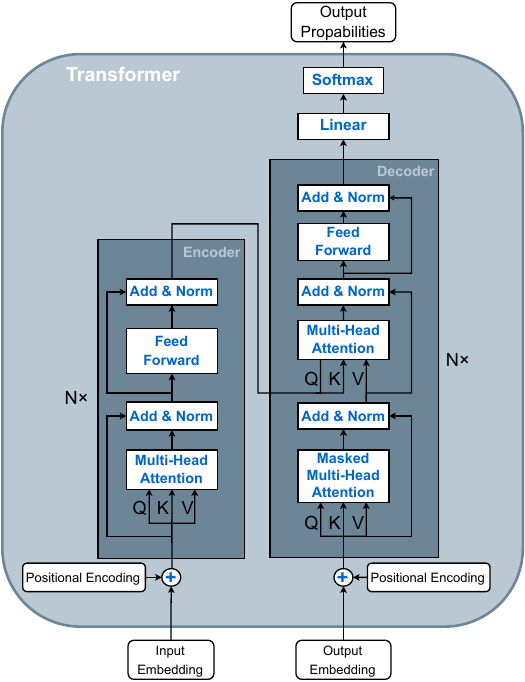}
\caption{Basic building block of the Transformer model architecture \cite{vaswani2017attention}. Dark blue boxes indicate the repeated Transformer blocks in the encoder and decoder. Each block contains the same subcomponents, centered around multi-head attention, and these blocks are stacked to form the full encoder–decoder structure.}
\label{fig:transformer}
\end{figure}

Transformers are deep architectures composed of repeated \textbf{Transformer blocks}, each containing multiple internal sublayers, as shown in Figure~\ref{fig:transformer}. 
The term `Transformer layer' typically refers to one such block, and modern LLMs are built by stacking many of these blocks. 
This structure is considerably richer than the linear, fully connected, and recurrent layers discussed so far. 
A standard Transformer architecture consists of two main components: the \textbf{encoder} and the \textbf{decoder}, each formed by repeated layers with identical subcomponents. 
The encoder processes the entire input sequence and produces a contextualized representation, while the decoder generates the output sequence autoregressively, using both the encoder's representation and the tokens generated so far.

Since Transformers contain no recurrence or convolution, they rely on positional encodings to inject information about token order. These encodings are added to the input embeddings before the first attention layer. A detailed discussion of positional encodings is beyond the scope of this work, but they are included in Figure~\ref{fig:transformer} for completeness.

At the core of the Transformer is the attention mechanism, particularly multi-head self-attention, which applies several attention operations (heads) in parallel. 
This enables each token to attend to every other token in the sequence, dynamically weighting their importance when forming contextualized representations. 
Typically, the encoder applies bidirectional self-attention, allowing every token to see the full sequence simultaneously. 
In contrast, the decoder operates autoregressively: it generates one token at a time, and each prediction depends only on previously generated tokens, enforced through causal masking. 
To avoid recomputing attention over past tokens during generation, modern implementations employ a key--value (KV) cache mechanism \cite{kwon2023efficient}.

Attention can be summarized as follows: An input $\mathbf{X} \in \mathbb{R}^{T \times d_{\text {model}}}$ which is a sequence of $T$ tokens with embedding size $d_\text{model}$ is multiplied with three learned weight matrices, a triplet $\mathbf{W}_Q, \mathbf{W}_K, \mathbf{W}_V\in \mathbb{R}^{d_{model} \times d_k}$ for each one of the $H$ heads $(d_k =d_\text{model} / H)$, to compute the projected $\mathbf{Q}, \mathbf{K} , \mathbf{V}\in \mathbb{R}^{T \times d_k}$ embeddings\footnote{The attention formulation does not require the values to have the same dimensionality as the queries and keys, although standard Transformers typically choose $\mathbf{Q}$, $\mathbf{K}$, and $\mathbf{V}$ to have identical sizes.} as follows:
\begin{equation}
   \begin{aligned}
& \mathbf{Q}=\mathbf{X} \mathbf{W}_Q \\[3pt]
& \mathbf{K}=\mathbf{X} \mathbf{W}_K  \\[3pt]
& \mathbf{V}=\mathbf{X} \mathbf{W}_V
\end{aligned} 
\label{eq:qkv}
\end{equation}
Then attention is computed as:
\begin{equation}
    \operatorname{Attention}(\mathbf{Q}, \mathbf{K}, \mathbf{V})=\operatorname{softmax}\left(\frac{\mathbf{Q} \mathbf{K}^{T}}{\sqrt{d_k}}\right) \mathbf{V} 
    \label{eq:attention}
\end{equation}
The product of $\mathbf{Q}$ and $\mathbf{K}^{T}$ gives raw attention scores, which are then scaled by $\sqrt{d_k}$ (the key dimensionality) to stabilize gradients. A softmax is applied row-wise to convert these scores into attention weights, highlighting the most important parts of the input (with respect to the query). These weights are then used to compute a weighted sum of the value matrix $\mathbf{V}$, allowing the model to focus on relevant information. 
Matrix-matrix multiplication is the core operation of the Transformer's computations. These matrix-matrix multiplications become matrix-vector multiplications in the decoder autoregressive stage when the output is processed sequentially, which is typical for local deployments where the batch size is equal to one.  

\textbf{Masked attention} shown in Figure~\ref{fig:transformer} is a variant of attention where the model is prevented from considering future tokens in the sequence. In practice, this is done by applying a mask to the attention scores ($\mathbf{QK}^T$), setting the values corresponding to `future' positions to $- \infty$ before the softmax, ensuring they receive zero weight after the softmax. As a result, each token can only attend to itself and earlier tokens, never to later ones. 

This architectural innovation has since revolutionized NLP, powering state-of-the-art models such as GPT and BERT, and extending to applications in computer vision and multimodal tasks \cite{devlin2019bert}. Despite their success, Transformers face challenges with scalability due to their quadratic complexity in sequence length, leading to high computational and memory costs. As a result, recent research has focused on developing more efficient variants, including sparse attention mechanisms \cite{child2019generating}, and linearized attention \cite{katharopoulos2020transformers}.

%%%%%%%%%%%%%%%%%%%%%%%%%%%%%%%%%%%%%%%%%%%%%%%%%%%%%%%%%%%%%%%%%%%%%%%%%%%%%%
\subsubsection{Graph Neural Networks (GNNs)} \label{sec:background:dnn_basics:gnn}
%%%%%%%%%%%%%%%%%%%%%%%%%%%%%%%%%%%%%%%%%%%%%%%%%%%%%%%%%%%%%%%%%%%%%%%%%%%%%%
Graph Neural Networks (GNNs) were developed to learn from graph-structured data, where entities (nodes) and their relationships (edges) must be effectively represented. Traditional machine learning methods, designed for Euclidean data such as tables, sequences, or images, struggle with the irregular structure and complex interdependencies in graphs. GNNs extend deep learning to non-Euclidean domains by leveraging node features and graph connectivity to propagate information. Graphs model structured relationships in real-world systems such as social networks (users and friendships), molecules (atoms and bonds), and knowledge graphs (entities and relations). Like traditional DNNs, GNNs use multiple layers to aggregate and transform information from neighboring nodes, building increasingly abstract representations.
By capturing both the topology and node/edge attributes, GNNs can learn meaningful embeddings that reflect both local and global graph properties. 

GNNs take as input graph-structured data with node and edge features.
Each node in the graph can have features that describe the entity it represents (e.g., a user’s profile in a social network, or the atoms in a molecule). Each edge may also have features that describe the relationship between the nodes (e.g., friendship strength between users, or chemical bond types between atoms). 
In a typical use case, the GNN can process either one large graph or multiple smaller graphs. 
For example, in a single graph, the GNN might process a whole social network. In cases where multiple smaller graphs are involved, such as in graph classification tasks where each graph represents a different molecule, the input might consist of several individual graphs, each with its own set of nodes, edges, and associated features.

The output of a GNN depends on the specific task at hand. For node-level tasks (e.g., node classification), the output is a prediction or label for each node. For example, the GNN could predict whether a user is interested in a particular activity based on their connections and profile features. For edge-level tasks (e.g., link prediction), the output is a prediction about the relationship between nodes, such as whether two users in a social network will become friends. Finally, for graph-level tasks (e.g., graph classification), the GNN outputs a prediction for the entire graph, such as determining if a molecule is toxic based on its chemical structure or classifying a network as `healthy' or `risky'. Across all these tasks, the key strength of GNNs is their ability to learn from both the features of the individual nodes and edges as well as their connectivity and structure, allowing them to make predictions based on the overall graph’s configuration and relationships.

A \textbf{GNN layer} is typically represented by a combination of:
\begin{itemize}
    \item \textbf{Graph connectivity} ($\mathbf{A} \in \mathbb{R}^{N \times N}$, typically represented as an adjacency matrix or edge list)
    \item \textbf{Feature representation} ($\mathbf{H}^{(l)} \in \mathbb{R}^{N \times d_l}$, the node feature matrix at layer $l$)
    \item \textbf{Trainable weights} ($\mathbf{W} \in \mathbb{R}^{d_l \times d_{l+1}}$, the learnable weight matrix)
\end{itemize}
Here, $N$ is the number of nodes in the graph, and $d_l$ is the feature dimensionality at layer $l$.

The weight matrix is a key component for transforming the input features of nodes and updating their representations through learned transformations.
GNNs rely on a \textbf{message-passing} (or \textbf{neighborhood aggregation}) mechanism: each node iteratively receives and aggregates information from its neighbors, updating its own representation to reflect both local and global graph context \cite{giannoula2024pygim, scarselli2008graph}. This approach typically involves defining how nodes combine incoming messages (e.g., summing or averaging neighbor features) and how node states are transformed through learnable functions. As a result, GNNs can uncover patterns that arise from the interplay between node features and the structure of the network itself.

Message passing is typically formulated as:
\begin{equation}
    \mathbf{H}^{(l+1)}=\phi \left(\mathbf{A}, \mathbf{H}^{(l)}, \mathbf{W}\right)
\end{equation}

Where $l$ is the layer of the GNN and $\phi$ a typically non-linear function like ReLU.
Standard GNNs like Graph Convolutional Networks (GCNs) \cite{kipf2016semi} use a fixed rule to aggregate neighboring node features. Each node updates its features based on a weighted sum of its own and its neighbors' features in a uniform manner. This actually means that the matrix $\mathbf{A}$ remains constant, precomputed, and captures static information about the graph, like connectivity and node degrees.

GCNs, as introduced by Kipf and Welling, update node features as:
\begin{equation}
    \mathbf{H}^{(l+1)}=\phi\left(\mathbf{\tilde{D}}^{-1 / 2} \tilde{\mathbf{A}} \tilde{\mathbf{D}}^{-1 / 2} \mathbf{H}^{(l)} W\right)
    \label{eq:gcn}
\end{equation}
where:
\begin{itemize}
    \item $\tilde{\mathbf{A}}=\mathbf{A+I}$ is the adjacency matrix with self-loops.
    \item $\tilde{\mathbf{D}}$ is the diagonal degree matrix of $\tilde{\mathbf{A}}$.
    \item $\mathbf{H}^{(l)}$ represents node features at layer $l$.
    \item $\mathbf{W}$ is a learnable weight matrix.
    \item $\phi$ is a non-linearity (e.g., ReLU).
\end{itemize}
The term $\tilde{\mathbf{D}}^{-\frac{1}{2}} \tilde{\mathbf{A}} \tilde{\mathbf{D}}^{-\frac{1}{2}} \mathbf{H}^{(l)}$ computes the messages through matrix multiplication that involves the normalized adjacency matrix $\tilde{\mathbf{D}}^{-\frac{1}{2}} \tilde{\mathbf{A}} \tilde{\mathbf{D}}^{-\frac{1}{2}}$.

Advanced GNN architectures like Graph Attention Networks (GAT) \cite{velivckovic2017graph} introduce mechanisms, such as attention weights, that allow the model to learn the relative importance of neighboring nodes during aggregation. Thus, while the weight matrix helps in feature transformation, the model's architecture determines how neighboring node relationships and their significance are incorporated into the node embeddings. This means that the matrix $\mathbf{A}$ is computed dynamically at each layer of the GNN and each inference pass. GATs introduce an attention mechanism to weigh the importance of each neighbor's contribution. For a given node i and the corresponding feature vector $\mathbf{h}_i$, attention coefficients $\alpha_{i j}$ are computed for each neighbor $j \in \mathcal{N}(i)$ using:
\begin{equation}
e_{i j}=\operatorname{LeakyReLU}\left(\mathbf{a}^T\left[\mathbf{W} \mathbf{h}_i \| \mathbf{W} \mathbf{h}_j\right]\right), \quad \alpha_{i j}=\frac{\exp \left(e_{i j}\right)}{\sum_{k \in \mathcal{N}(i)} \exp \left(e_{i k}\right)}.   
\label{eq:gat}
\end{equation}
These coefficients $\alpha_{i j}$ are then used to compute a weighted sum over the transformed neighbor features. This mechanism enables GATs to focus more on relevant neighbors and adapt to heterogeneity in the graph.

%%%%%%%%%%%%%%%%%%%%%%%%%%%%%%%%%%%%%%%%%%%%%%%%%%%%%%%%%%%%%%%%%%%%%%%%%%%%%%
\subsection{From model architecture to model inference} 
\label{sec:background:model_deployment}
\begin{figure}[tb!]
    \centering
    \includegraphics[scale = 0.7]{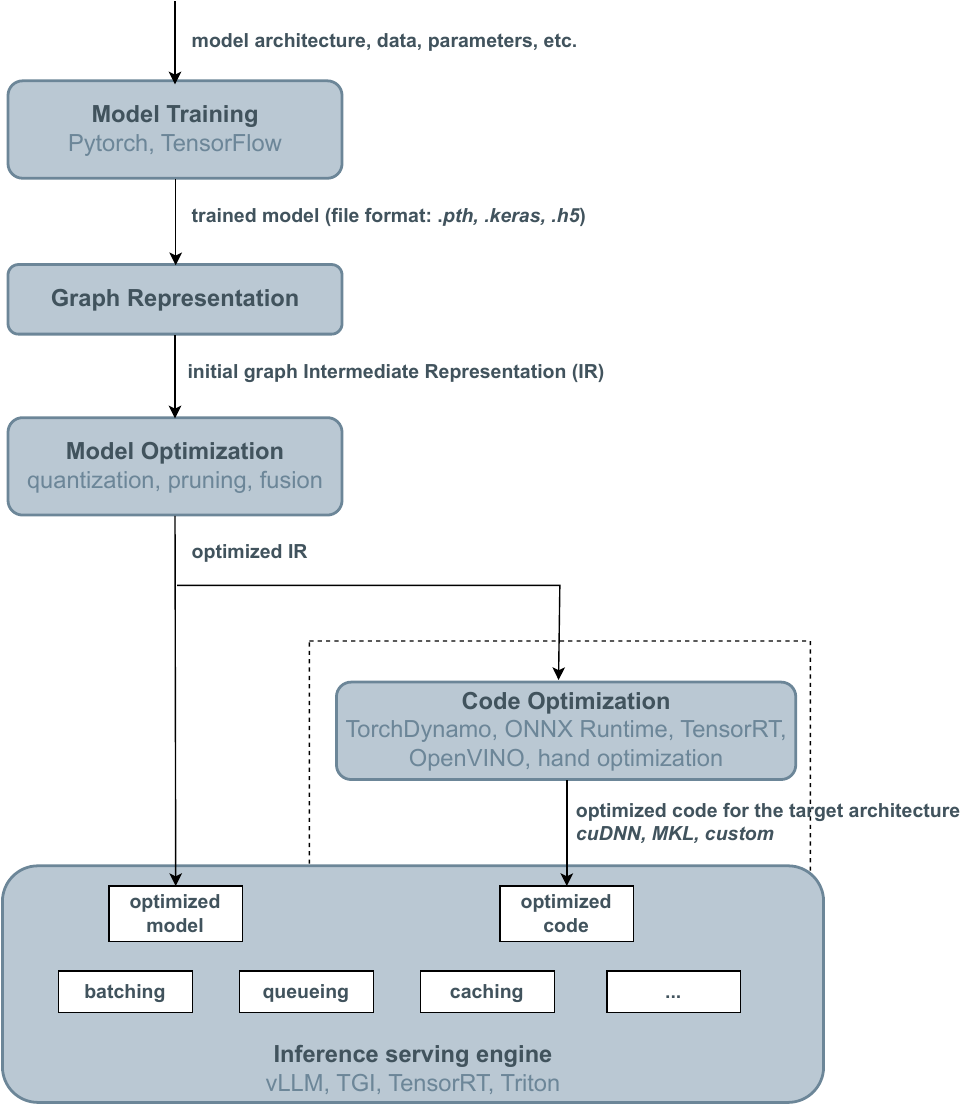}
    \caption{From architecture to model deployment.}
    \label{fig:depl}
\end{figure}

%%%%%%%%%%%%%%%%%%%%%%%%%%%%%%%%%%%%%%%%%%%%%%%%%%%%%%%%%%%%%%%%%%%%%%%%%%%%%%
In this section, we describe how common end-to-end model development and inference pipelines operate. First, we outline a general workflow applicable to most cases and we separate it into three distinct steps: Model training, Model Optimization and Inference. These are depicted in Figure~\ref{fig:depl} and analyzed in the next paragraphs.

%%%%%%%%%%%%%%%%%%%%%%%%%%%%%%%%%%%%%%%%%%%%%%%%%%%%%%%%%%%%%%%%%%%%%%%%%%%%%%
\subsubsection{Model training} \label{sec:background:model_deployment:training}
%%%%%%%%%%%%%%%%%%%%%%%%%%%%%%%%%%%%%%%%%%%%%%%%%%%%%%%%%%%%%%%%%%%%%%%%%%%%%%
The process of Model Training refers to the process that makes a DNN learn to perform a specific task, much like a student learns from practice and correction.

Typically, a model is developed using a popular deep learning framework such as PyTorch \cite{paszke2019pytorch}, TensorFlow \cite{abadi2016tensorflow}, or JAX \cite{jax2018github}, and is trained on a collection of training samples called a dataset. The dataset contains thousands or millions of examples of the type of information we want the model to process (e.g., images for a vision task or text for a language task).
For many common tasks, data are accompanied by specific supplementary information, known as metadata, and, most importantly, ground truth labels. These labels are the `correct answers' (e.g., the label for an image is `cat' or `dog') and are used exclusively in a process called supervised learning.

The core steps of training are as follows:
\begin{enumerate}
    \item \textbf{Architecture Selection:} Based on the task, a suitable network architecture is chosen (like the ones discussed in the previous section).
    \item \textbf{The Loss Function:} To guide this adjustment, a crucial mathematical function called the loss function is used. It measures precisely how far the model’s current prediction is from the correct answer (the ground truth label). A high loss value means the model is performing poorly, and a low loss value means it is performing well.
    \item \textbf{Gradient Computation:} The loss function provides feedback to the network on how the weights should change. This feedback takes the mathematical form of gradients—which indicate the direction and magnitude of change needed for every single weight to reduce the loss.
    \item \textbf{Optimization:} The weights are then updated using an iterative optimization method, such as Gradient Descent, Stochastic Gradient Descent (SGD) \cite{ruder2016overview}, or Adam \cite{kingma2014adam}. 
      These methods use the gradients to modify the weights, aiming to minimize the loss.
\end{enumerate}
In this way, the model gradually, over many iterations, learns to capture the underlying patterns in the data.

%%%%%%%%%%%%%%%%%%%%%%%%%%%%%%%%%%%%%%%%%%%%%%%%%%%%%%%%%%%%%%%%%%%%%%%%%%%%%%
%\subsubsection{Efficiency and Control in Training} \label{sec:background:model_deployment:efficiency}
%%%%%%%%%%%%%%%%%%%%%%%%%%%%%%%%%%%%%%%%%%%%%%%%%%%%%%%%%%%%%%%%%%%%%%%%%%%%%%
To make the training process computationally efficient, especially with massive datasets, several engineering techniques are employed:

\begin{itemize}
    \item \textbf{Batch Training:} Instead of feeding individual data points one by one, models are trained on small groups of samples called batches. The model calculates the loss and gradients for all samples in a batch simultaneously, and then uses the average gradient to update the weights. This method significantly speeds up the training  process by leveraging the parallel processing power of accelerators such as GPUs.
    \item \textbf{Hyperparameters:} Beyond the internal weights, which the model learns, there are external settings called hyperparameters that must be chosen by the engineer. These include the learning rate (how large each weight adjustment step should be) and the batch size itself. Tuning these parameters is crucial for achieving an effective and stable training run.
    \item \textbf{Epochs:}
    A single pass through the entire dataset is called an epoch. Training often requires many epochs to fully learn the data's patterns. However, engineers must monitor the training to prevent overfitting, where the model learns the training examples too well, including their noise, and performs poorly on new, unseen data.
    \item \textbf{Scaling and Distribution:} For the largest models and datasets, training must be scaled across multiple GPUs or even multiple servers, a technique known as distributed training. Frameworks like PyTorch and TensorFlow include tools to efficiently manage this distribution, allowing engineers to train models that would otherwise be impossible on a single machine.
\end{itemize}
Nowadays, the initial training step can often be skipped entirely because many powerful pre-trained models are publicly available online (e.g., via the Hugging Face Hub or PyTorch repositories). These models have already been trained on massive, diverse datasets and can be instantly adapted (fine-tuned) for a specific new task with minimal effort.

%%%%%%%%%%%%%%%%%%%%%%%%%%%%%%%%%%%%%%%%%%%%%%%%%%%%%%%%%%%%%%%%%%%%%%%%%%%%%%
\subsubsection{\texorpdfstring{Model optimization \footnote{Optimization in this case refers to the process of improving the model towards better performance (e.g., latency of throughput) and resource utilization (e.g., smaller memory footprint). This should not be confused with training-time optimization of the model weights.}}{Model optimization}}
%%%%%%%%%%%%%%%%%%%%%%%%%%%%%%%%%%%%%%%%%%%%%%%%%%%%%%%%%%%%%%%%%%%%%%%%%%%%%%

The next goal is to use the trained model to perform inference on new data. There are multiple ways to execute inference, ranging from a simple function call in PyTorch, to a comprehensive pipeline involving exports and hardware-specific optimizations. While the simplest form of inference is common in academic research, our focus here is on identifying how the initial neural network can be enhanced, optimized, and eventually run on the target device. 

The most common formats for representing a neural network are PyTorch (.pth files) or TensorFlow (.keras, .h5). These files may contain either the entire model architecture or just the learned weights. From these formats, models can be exported to other formats to enable optimizations. Model optimization is optional but often essential in order to deploy DNNs efficiently and meet specific latency and throughput targets. Typically, the first step in optimization is converting the model into a computational graph, an intermediate representation (IR) that captures the operations and data flow. Various optimizations can then be applied either to the model’s weights (e.g., quantization, pruning) or to its computations (e.g., operator fusion).

There are many different model representations that can be the starting point for optimizations as such. One common representation is ONNX, an open standard that facilitates interoperability between frameworks. ONNX encapsulates both the architecture and the weights as a computational graph, where nodes represent operations and edges denote data flow. This format can be used directly for inference, but it can also serve as input for further optimization. Other common representations include TorchScript, which allows PyTorch models to be traced or scripted into an IR that can be deployed outside the Python interpreter; TensorFlow SavedModel / GraphDef, which encode the computation graph and variables for TensorFlow pipelines; TensorFlow Lite, tailored for mobile and edge inference with optimizations such as quantization; Core ML, used for deploying on Apple devices; and hardware‐ or vendor-specific formats/engines such as TensorRT or OpenVINO. These formats likewise hold both structure and weights, and each provides its own suite of optimizations (operator fusion, quantization, pruning, static graph transformations, etc.), making them attractive starting points for deployment pipelines. Depending on the use case (edge vs cloud, latency vs throughput, resource constraints), one may convert from one representation to another, or optimize directly on these IRs.

%%%%%%%%%%%%%%%%%%%%%%%%%%%%%%%%%%%%%%%%%%%%%%%%%%%%%%%%%%%%%%%%%%%%%%%%%%%%%%
\subsubsection{Model inference} \label{sec:background:model_deployment:inference}
%%%%%%%%%%%%%%%%%%%%%%%%%%%%%%%%%%%%%%%%%%%%%%%%%%%%%%%%%%%%%%%%%%%%%%%%%%%%%%
At this point, the model is in an optimized state and ready for execution. Each backend selects or generates the kernels required for inference based on the optimized graph and available hardware. These kernels may be dynamically compiled at runtime for flexibility or pre-compiled for efficiency, with modern engines often adopting a hybrid strategy. Beyond this, inference workloads can take different forms: \textbf{batch inference} prioritizes throughput in offline settings, while online inference minimizes latency for interactive tasks. To meet these demands at scale, models are deployed through \textbf{serving systems} or \textbf{engines} (e.g., vLLM \cite{kwon2023efficient}, TGI \cite{HuggingFaceTGI}, TensorRT \cite{TensorRT_LLM}), that manage requests, scaling, caching, versioning, etc. Here, optimization techniques such as quantization or sparsity play a direct role, since they reduce compute and memory overhead, improving both latency and throughput across deployment scenarios.

%%%%%%%%%%%%%%%%%%%%%%%%%%%%%%%%%%%%%%%%%%%%%%%%%%%%%%%%%%%%%%%%%%%%%%%%%%%%%%
\subsection{Sparsity basics} \label{sec:background:sparsity_basics}
%%%%%%%%%%%%%%%%%%%%%%%%%%%%%%%%%%%%%%%%%%%%%%%%%%%%%%%%%%%%%%%%%%%%%%%%%%%%%%
Sparsity in our case refers to working on data, typically represented as matrices or tensors that are sparse, i.e., a large number of their elements are zero. Sparsity can be coarsely categorized into three main types in DNNs:  

\textbf{Structured sparsity} assumes sparse elements in coarse parts of the tensor, as, for example, is the case in weight pruning (see Section~\ref{sec:sparse:weight_pruning}) where one removes components such as neurons, features, filters, or even layers~\cite{li2016pruning, han2015learning}, leading to model simplification that is hardware-friendly and easier to accelerate. Structured sparsity typically leads to the creation of smaller, dense neural networks, thus resorting to dense operations~\cite{wen2016learning}, and for this reason, we will not focus on structured sparsity in this survey. Moreover, the accuracy of such approaches is not satisfactory in several cases, and as a result, alternative sparsification methods are considered.

\textbf{Unstructured sparsity} assumes the most generic distribution of nonzero elements in the tensor, i.e., they can reside anywhere, without any limitation while their number and position is decided by some optimization process (e.g., in the case of pruning by an algorithm that zeros out specific components that do not contribute substantially to the model’s accuracy~\cite{lecun1989optimal}). Unstructured pruning can reach much higher compression ratios without compromising model accuracy, but the performance of the kernels that operate on unstructured sparsity can be disappointingly low~\cite{gale2019state}. 

\textbf{Semi-structured} sparsity refers to a sparsity pattern where zero elements are not randomly distributed but follow a regular, partially predictable structure. It strikes a balance between structured and unstructured sparsity by imposing some regularity while maintaining flexibility. \textbf{N:M sparsity} is a specific and widely adopted subtype of semi-structured sparsity~\cite{mishra2021accelerating}, and it will be the main semi-structured paradigm discussed in this survey. In N:M sparsity, exactly $n$ out of every $m$ elements remain nonzero, enabling both substantial compression (typically lower than unstructured sparsity, nevertheless) and more efficient execution on supported hardware. For example, in 2:4 sparsity, every group of four elements contains exactly two nonzero values.
\textbf{Block sparsity} allows nonzero elements in fixed shaped dense blocks~\cite{gray2017gpu} and utilizes data structures like BSR and specialized kernels that expect the nonzeros in this form.

In the following paragraphs, we elaborate more on unstructured and N:M sparsity that are heavily employed in research and practice. In particular, we discuss data structures that store sparse data in each case, computational kernels involving sparse data and, where relevant, hardware support provided by vendors to improve the performance of these kernels. 

%%%%%%%%%%%%%%%%%%%%%%%%%%%%%%%%%%%%%%%%%%%%%%%%%%%%%%%%%%%%%%%%%%%%%%%%%%%%%%
\subsubsection{Unstructured sparsity} \label{sec:background:sparsity_basics:unstructured}
%%%%%%%%%%%%%%%%%%%%%%%%%%%%%%%%%%%%%%%%%%%%%%%%%%%%%%%%%%%%%%%%%%%%%%%%%%%%%%

%%%%%%%%%%%%%%%%%%%%%%%%%%%%%%%%%%%%%%%%%%%%%%%%%%%%%%%%%%%%%%%%%%%%%%%%%%%%%%
\paragraph{Data structures}
%%%%%%%%%%%%%%%%%%%%%%%%%%%%%%%%%%%%%%%%%%%%%%%%%%%%%%%%%%%%%%%%%%%%%%%%%%%%%%

Two dimensional arrays, which we henceforth call matrices, are the core data structure to store sparse data in algebraic operations heavily used in DNNs. A matrix is considered sparse when a substantial fraction of its elements have zero values, and therefore can be stored in a compressed form by omitting them. This creates the need of somehow expressing and storing the structure of the matrix, i.e., the positions of the remaining nonzero elements.

\begin{figure}[ht!]
    \begin{minipage}{0.45\linewidth}        
            \centering
            \includegraphics[width=\linewidth]{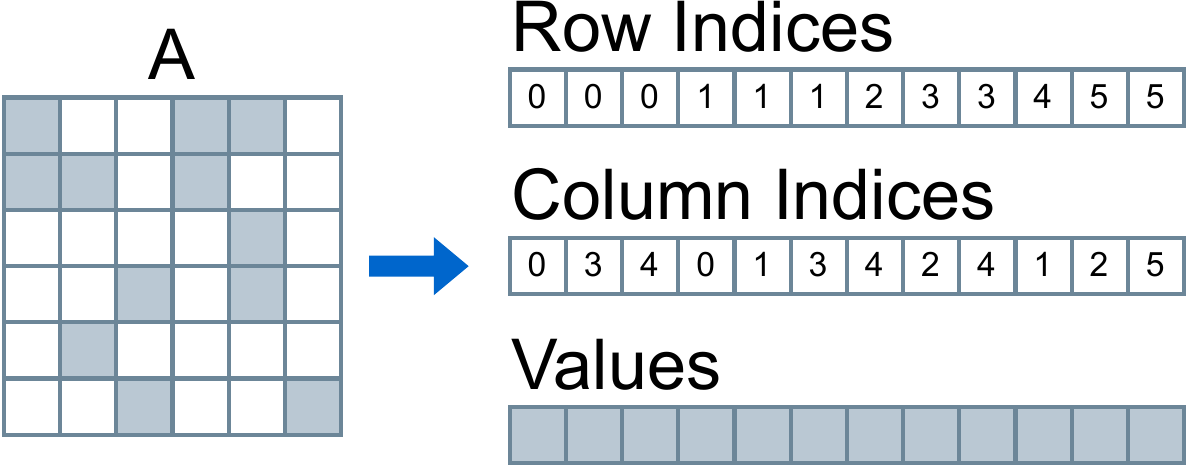}
            \caption{The COO storage format for sparse matrices.}
            \label{fig:background_coo}
    \end{minipage}
    \hfill
    \begin{minipage}{0.45\linewidth}
            \centering
            \includegraphics[width=\linewidth]{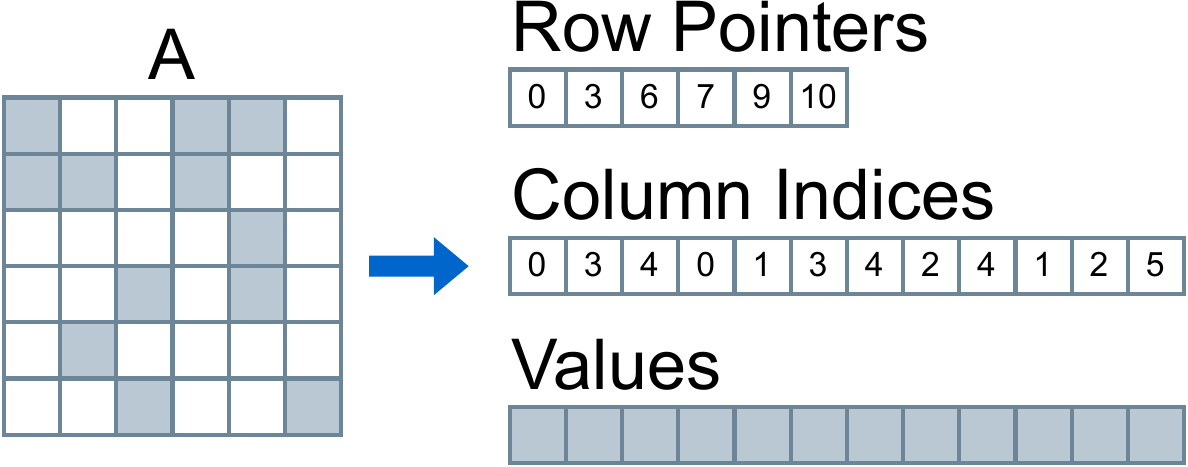}
            \caption{The CSR storage format for sparse matrices.}
            \label{fig:background_csr}
    \end{minipage}
\end{figure}

A straightforward approach to describe a sparse matrix is the so-called \textit{Coordinate (COO)} format (Figure~\ref{fig:background_coo}), the de facto way to store matrices in files. In COO, each nonzero element is stored alongside its indices, i.e., its position inside the matrix (row and column). In several cases, the memory footprint of the sparse matrix is of vital importance, so the state-of-practice approach to store matrices in memory is the \textit{Compressed Sparse Row (CSR)} format (Figure~\ref{fig:background_csr}) which stores as metadata the number of elements per row (in the form of a prefix sum) and the column index of each nonzero element.

Furthermore, the optimization process of several sparse kernels brought forward a variety of format families. The most notable families are the ELL-type formats (a variation of CSR in which we assume a constant row length, enforcing it by padding smaller rows with extra zeros, and therefore does not need the RowPtr array), the DIA-type formats (storing the non-empty diagonals, padding smaller ones to the same length as the number of rows), and Block-type formats (variations where we assume that nonzeros are clustered in 2D blocks of constant size, padding with zeros when needed, and storing the positional data of the blocks instead of each element; most commonly encountered in the form of BSR - Blocked Sparse Row, the blocked variant of CSR). Lastly, more complex formats like SparseX \cite{elafrou2018sparsex}, CSR5 \cite{liu2015csr5}, LCM \cite{cheshmi2022vectorizing} and DIV \cite{galanopoulos2025div} specifically target the optimization of the \texttt{SpMV} kernel (see the next paragraph).

%%%%%%%%%%%%%%%%%%%%%%%%%%%%%%%%%%%%%%%%%%%%%%%%%%%%%%%%%%%%%%%%%%%%%%%%%%%%%%
\paragraph{Computational kernels}
%%%%%%%%%%%%%%%%%%%%%%%%%%%%%%%%%%%%%%%%%%%%%%%%%%%%%%%%%%%%%%%%%%%%%%%%%%%%%%
Below we provide information on the basics of sparse computational kernels that we revisit in this survey.

%%%%%%%%%%%%%%%%%%%%%%%%%%%%%%%%%%%%%%%%%%%%%%%%%%%%%%%%%%%%%%%%%%%%%%%%%%%%%%
\subparagraph{Sparse Matrix - Vector Multiplication (SpMV)}
%%%%%%%%%%%%%%%%%%%%%%%%%%%%%%%%%%%%%%%%%%%%%%%%%%%%%%%%%%%%%%%%%%%%%%%%%%%%%%
\textit{Sparse Matrix-Vector Multiplication} (\texttt{SpMV}) is an operation that takes a sparse matrix and a dense vector as input and produces a dense vector as output. \texttt{SpMV} is a fundamental operation in HPC, and in particular a core building block for the solution of large sparse linear systems \cite{saad2003iterative}. In the case of DNNs, \texttt{SpMV} is employed in RNNs, LSTMs (see Section~\ref{sec:background:dnn_basics:seq_layers}) and in the decoder stage of Transformer networks in cases where input is not processed in batches. 

\begin{figure}[ht!]
    \centering
    \includegraphics[height=3cm]{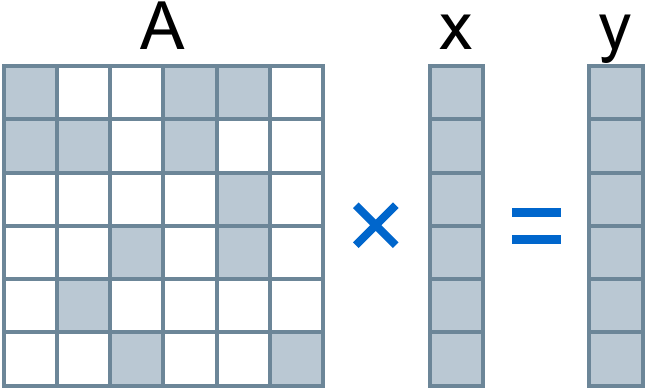}
    \caption{Sparse matrix-vector multiplication (\texttt{SpMV}).}
    \label{fig:background_spmv}
\end{figure}

In the typical implementations of \texttt{SpMV}, only the nonzero elements of the matrix and their positions are stored using CSR or COO. This reduces memory usage and computational effort by skipping unnecessary operations with zero values. We next provide pseudocode for these important cases.

\begin{center}
    \begin{minipage}{0.45\linewidth}
        \begin{lstlisting}[
            caption={\texttt{SpMV} using COO Format}, 
            label={lst:coo_spmv}
        ]
for (j=0; j<nnz; j++)
    y[RowIdx[j]] += Values[j] * x[ColIdx[j]];
//
        \end{lstlisting}
    \end{minipage}
    \hfill
        \begin{minipage}{0.45\linewidth}
        \begin{lstlisting}[
            caption={\texttt{SpMV} using CSR Format}, 
            label={lst:csr_spmv}
        ]
for (i=0; i<rows; i++)
    for (j=RowPtr[i]; j<RowPtr[i+1]; j++)
        y[i] += Values[j] * x[ColIdx[j]];
        \end{lstlisting}
    \end{minipage}
\end{center}

%%%%%%%%%%%%%%%%%%%%%%%%%%%%%%%%%%%%%%%%%%%%%%%%%%%%%%%%%%%%%%%%%%%%%%%%%%%%%%
\subparagraph{Sparse Matrix - Dense Matrix Multiplication (SpMM)}
%%%%%%%%%%%%%%%%%%%%%%%%%%%%%%%%%%%%%%%%%%%%%%%%%%%%%%%%%%%%%%%%%%%%%%%%%%%%%%
\textit{Sparse Matrix - Dense Matrix Multiplication} (\texttt{SpMM}) is an operation that takes a sparse matrix and a dense matrix as input and produces a dense matrix as output. Specifically, each element in the resulting matrix is computed by performing a dot product between a row of the sparse matrix and a column of the dense matrix. During this process, only the nonzero elements of the sparse matrix are involved in the multiplication, while the structure of the dense matrix is fully utilized in each computation. \texttt{SpMM} is used in fields such as scientific computing, machine learning, graph processing, data mining, and computational physics. \texttt{SpMM} is a key computational kernel in sparse inference as discussed in Section~\ref{sec:sparse_kernels}.

\begin{figure}[ht!]
    \centering
    \includegraphics[height=3cm]{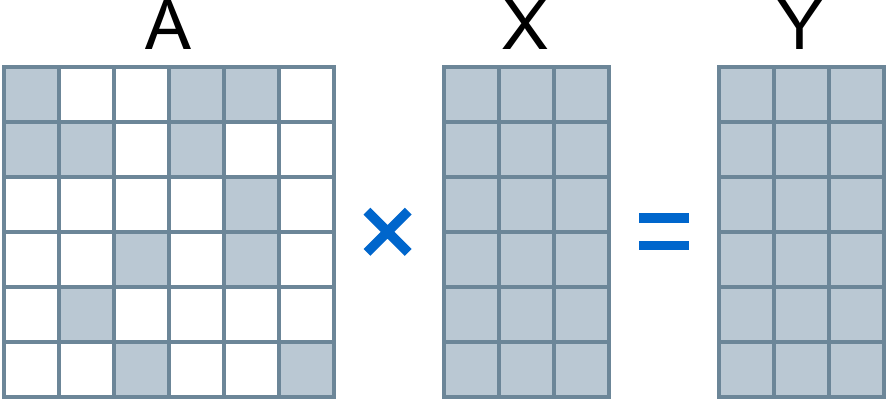}
    \caption{Sparse matrix-matrix multiplication (\texttt{SpMM}).}
    \label{fig:background_spmm}
\end{figure}

Sample code:

\begin{center}
    \begin{minipage}{0.45\linewidth}
        \begin{lstlisting}[
            caption={\texttt{SpMM} using CSR Format}, 
            label={lst:csr_spmm}
        ]
for (i=0; i<rows; i++)
    for (j=RowPtr[i]; j<RowPtr[i+1]; j++)
        for (k=0; k<K; k++)
            Y[i][k] += Values[j] * X[ColIdx[j]][k];
        \end{lstlisting}
    \end{minipage}
\end{center}

%%%%%%%%%%%%%%%%%%%%%%%%%%%%%%%%%%%%%%%%%%%%%%%%%%%%%%%%%%%%%%%%%%%%%%%%%%%%%%
\subparagraph{Sampled Dense-Dense Matrix Multiplication (SDDMM)}
%%%%%%%%%%%%%%%%%%%%%%%%%%%%%%%%%%%%%%%%%%%%%%%%%%%%%%%%%%%%%%%%%%%%%%%%%%%%%%
\textit{Sampled Dense-Dense Matrix Multiplication} (\texttt{SDDMM}) is an operation where two dense matrices are multiplied, but the results are computed only at specific positions defined by a sparse mask. Given dense matrices X1 and X2, and a sparse mask matrix A, indicating the positions of interest, \texttt{SDDMM} computes the dot product of the corresponding row from X1 and column from X2 only at the nonzero locations of A. The output is a sparse matrix with the same sparsity pattern as A, containing the sampled results of the dense matrix multiplication. \texttt{SDDMM} is used in fields like graph neural networks, recommendation systems, natural language processing, and scientific computing.

\begin{figure}[ht!]
    \centering
    \includegraphics[width=.5\linewidth]{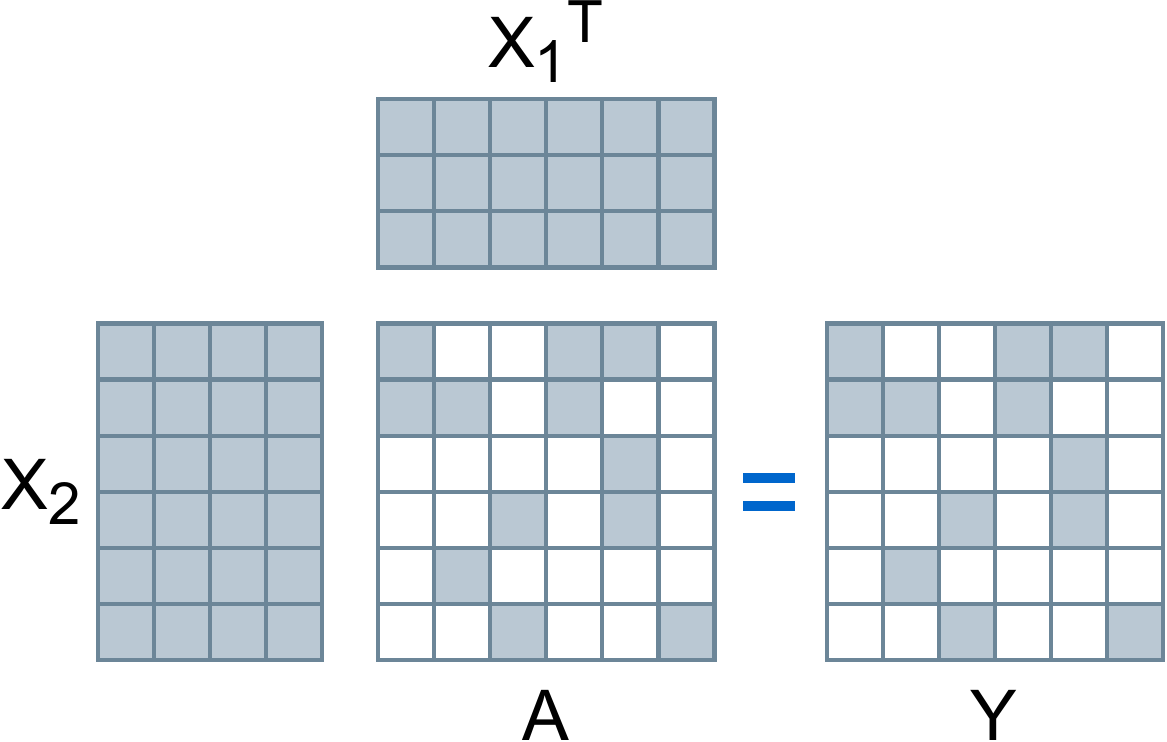}
    \caption{Sampled Dense-Dense Matrix Multiplication (\texttt{SDDMM}).}
    \label{fig:background_sddmm}
\end{figure}

Sample code: 

\begin{center}
    \begin{minipage}{0.45\linewidth}
        \begin{lstlisting}[
            caption={\texttt{SDDMM} using CSR Format}, 
            label={lst:csr_sddmm}
        ]
for (i=0; i<rows; i++)
	for (j=RowPtr[i]; j<RowPtr[i+1]; j++)
		for (k=0; k<K; k++)
			Y_Values[j] += X1[i][k] * X2[k][j];
        \end{lstlisting}
    \end{minipage}
\end{center}

%%%%%%%%%%%%%%%%%%%%%%%%%%%%%%%%%%%%%%%%%%%%%%%%%%%%%%%%%%%%%%%%%%%%%%%%%%%%%%
\subparagraph{Sparse convolution}
%%%%%%%%%%%%%%%%%%%%%%%%%%%%%%%%%%%%%%%%%%%%%%%%%%%%%%%%%%%%%%%%%%%%%%%%%%%%%%
\textit{Sparse convolution} is a variation of standard convolution where the input data or the convolutional filters contain many zero values, and the computation is performed only on the nonzero elements. This reduces the number of operations by skipping computations involving zeros. In practice, sparse convolution applies a convolutional kernel to an input feature map while leveraging the sparsity pattern to optimize performance.

Sample code (Dense input, sparse convolution kernel):

\begin{center}
    \begin{minipage}{0.65\linewidth}
        \begin{lstlisting}[
            caption={Sparse Convolution using COO Format}, 
            label={lst:coo_spconv}
        ]
for (i=0; i<H-K+1; i++)
    for (j=0; j<W-K+1; j++)
        for (n=0; n<nnz; n++)
            output[i][j] += k_val[n] * input[i+k_row[n]][j+k_col[n]];
        \end{lstlisting}
    \end{minipage}
\end{center}

Sparse convolutions can be further categorized depending on where their sparsity originates, leading to three main types with each type requiring specialized optimization strategies \cite{won2023unified}.
\textit{Filter Sparse Convolution} specifically addresses sparsity within the convolution kernels themselves, converting the operation into an \texttt{SpMM} kernel via \texttt{im2col} transformation where applicable.
\textit{Activation Sparse Convolution} focuses on the common scenario where activations after non-linearities (like ReLU) are sparse, designing efficient vectorized routines that only process nonzero activation (input) values.
\textit{Masked Sparse Convolution} is the same as a standard convolution kernel with the addition of an element-wise multiplication with a sparse binary mask that determines the sparsity pattern of the output.
\textit{Sparse Convolution for multidimensional input} or \textit{submanifold convolution} (see Appendix~\ref{app:spconv}) is a specific case of masked convolutions that strictly maintains the input sparsity pattern, ensuring that output nonzeros only appear where input nonzeros existed, which is critical for applications involving intrinsically sparse data like 3D point clouds.

%%%%%%%%%%%%%%%%%%%%%%%%%%%%%%%%%%%%%%%%%%%%%%%%%%%%%%%%%%%%%%%%%%%%%%%%%%%%%%
\subparagraph{Other sparse kernels}
%%%%%%%%%%%%%%%%%%%%%%%%%%%%%%%%%%%%%%%%%%%%%%%%%%%%%%%%%%%%%%%%%%%%%%%%%%%%%%
Other sparse kernels like \textit{Sparse Matrix - Sparse Matrix Multiplication} (\texttt{SpGEMM}) \cite{bulucc2012parallel} or \textit{Sparse Matrix - Sparse Vector Multiplication} (\texttt{SpMSpV}) \cite{azad2017work} have been discussed and implemented for applications in other domains. To our knowledge, they are not heavily used in DNN inference and thus not further analyzed in this survey.

%%%%%%%%%%%%%%%%%%%%%%%%%%%%%%%%%%%%%%%%%%%%%%%%%%%%%%%%%%%%%%%%%%%%%%%%%%%%%%
\subsubsection{Semi-structured N:M sparsity} \label{sec:background:sparsity_basics:semi_structured}
%%%%%%%%%%%%%%%%%%%%%%%%%%%%%%%%%%%%%%%%%%%%%%%%%%%%%%%%%%%%%%%%%%%%%%%%%%%%%%

%%%%%%%%%%%%%%%%%%%%%%%%%%%%%%%%%%%%%%%%%%%%%%%%%%%%%%%%%%%%%%%%%%%%%%%%%%%%%%
\paragraph{Data structures}
%%%%%%%%%%%%%%%%%%%%%%%%%%%%%%%%%%%%%%%%%%%%%%%%%%%%%%%%%%%%%%%%%%%%%%%%%%%%%%
\begin{figure}[ht!]
    \centering
    \includegraphics[width=.65\linewidth]{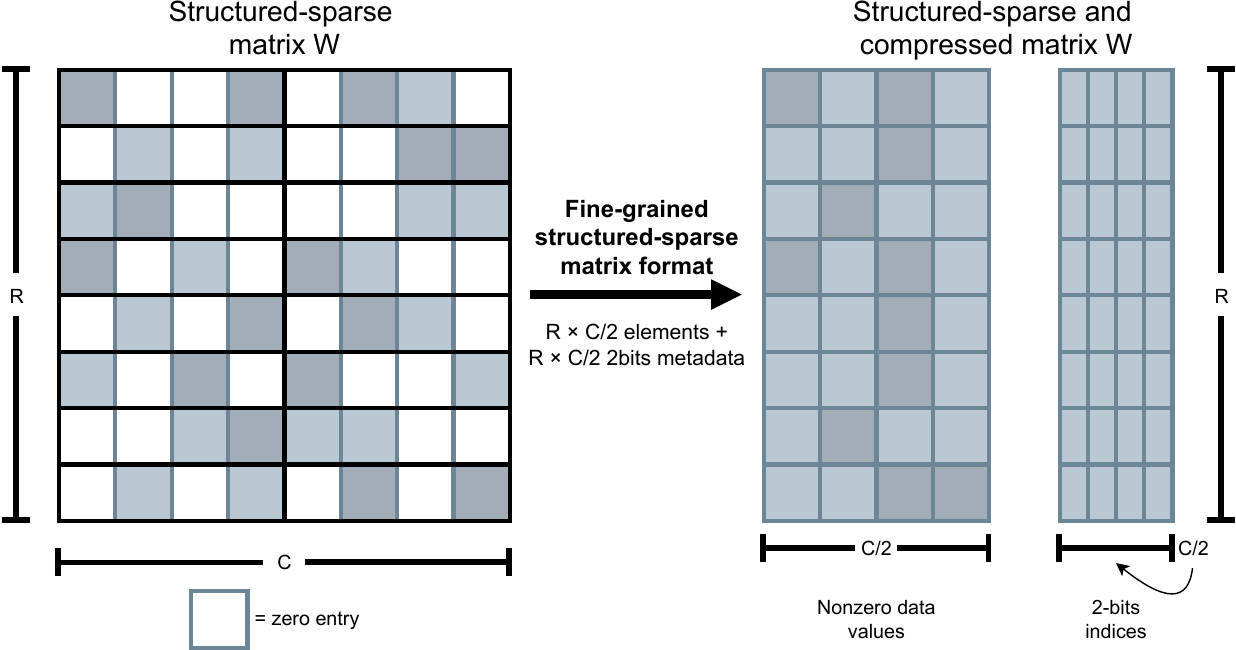}
    \caption{Data structure for semi-structured sparsity \cite{mishra2021accelerating}.}
    \label{fig:semi-structured}
\end{figure}

Existing widely used sparse matrix formats (CSR, COO etc.) are not well-suited for representing matrices with moderate sparsity levels, such as those resulting from n:m structured pruning. To better exploit these regular sparsity patterns, alternative representations have been introduced \cite{mishra2021accelerating,lin2023efficient}, consisting of the data and the index array. The data array stores only the nonzero values, while the index array encodes the positions (ranging from 0 to m-1) of the nonzero elements within each group of m elements. For instance, in 2:4 structured sparsity, which is natively supported by NVIDIA's Sparse Tensor Cores and presented in Figure~\ref{fig:semi-structured}, only 2 out of every 4 weights are kept. Their positions within the group are encoded using 2 bits each, resulting in 4 bits of metadata per group in the index array. Note that, computational kernels for N:M sparsity that operate on such data structures involve state-of-the-art, recent implementations that are discussed in Section~\ref{sec:sparse_kernels}.

%%%%%%%%%%%%%%%%%%%%%%%%%%%%%%%%%%%%%%%%%%%%%%%%%%%%%%%%%%%%%%%%%%%%%%%%%%%%%%
\paragraph{Hardware support}
%%%%%%%%%%%%%%%%%%%%%%%%%%%%%%%%%%%%%%%%%%%%%%%%%%%%%%%%%%%%%%%%%%%%%%%%%%%%%%
NVIDIA’s Sparse Tensor Cores, introduced with the Ampere GPU architecture, are an extension of the Tensor Core architecture, for accelerating matrix operations under structured sparsity. They support the 2:4 sparsity pattern, where exactly two nonzero elements are retained in every group of four consecutive weights. The cores are optimized to perform sparse matrix–dense matrix multiplication (\texttt{SpMM}) efficiently by skipping unnecessary multiplications with zero, effectively halving the number of compute operations and achieving nearly a 2× speedup. The sparse operand is stored in a compressed format, using 2-bit metadata per group to encode the nonzero positions, for fast decoding and minimal storage overhead. Sparse Tensor Cores support a range of input formats, FP16, BF16, INT8, and TF32, and nearly double the arithmetic throughput compared to their dense counterparts.

%%%%%%%%%%%%%%%%%%%%%%%%%%%%%%%%%%%%%%%%%%%%%%%%%%%%%%%%%%%%%%%%%%%%%%%%%%%%%%
\section{Sparsity in DNNs} \label{sec:sparse} 
%%%%%%%%%%%%%%%%%%%%%%%%%%%%%%%%%%%%%%%%%%%%%%%%%%%%%%%%%%%%%%%%%%%%%%%%%%%%%%

In this section we aim to elaborate on all forms of sparsity that are discussed in the context of deep neural networks. Our main focus, however, is to further analyze those forms of sparsity that lead actually to representation of sparse data and invocation of sparse computational kernels.
As we clarify in this section, some of the cases that use terms relevant to sparsity, do not ultimately resort to sparse computations.

In general, there are three sources of sparsity in the inference of deep learning tasks: \textit{enforced sparsification}, \textit{natural sparsity}  and \textit{ephemeral sparsification}. 

\textbf{Enforced Sparsification} refers to the process of deliberately sparsifying the inference process in order to save computation overhead and data representation. Enforced sparsification can take form in two ways: \textbf{model} (\textbf{weight}) \textbf{pruning} and \textbf{attention sparsification}. The main elements that can be sparsified (pruned) are: weights, neurons and filters in convolutional layers~\cite{mao2017exploring}. Attention sparsification, on the other hand, uses properly selected masks to avoid some tokens in the attention computation.

\textbf{Natural sparsity} is inherent to the problem at hand and mainly refers to GNNs where the problem is represented with a graph, a naturally sparse data structure. We also discuss in Appendix~\ref{app:spconv} the special case of sparse input in 3D point clouds \cite{wu2019pointconv, tang2022torchsparse, hong2023exploiting, yang2024minuet}, but as we discuss there, although they carry the term `sparse', the computations involved are dense.

\textbf{Ephemeral Sparsification} is applied during the computation of each example individually in an inference step and it is only relevant for this example. This case refers to the activation functions like ReLU and SoftMax that lead to sparsified outputs of DNN layers. Ephemeral sparsification is dynamically updated for each example and configured during inference and training.

Finally, we also consider the special cases of the \textbf{Mixture of Experts (MoE)} \cite{jacobs1991adaptive,shazeer2017outrageously,fedus2022switch} and \textbf{Quantization} \cite{gholami2022survey}. A MoE is a DNN architecture that divides a large model into multiple smaller specialized components, called experts. For each input, only a subset of these experts is activated, allowing the model to focus on the most relevant knowledge while reducing computation costs. This approach enables high model capacity with improved efficiency compared to using a single large model for all tasks. 
Since in MoE models only a small subset of experts (and thus parameters) is activated per input, this approach is considered to apply `conditional sparsity'.
However, MoE models do not use sparse data structures and computations and as such, we do not consider them further and provide more information on this popular approach in Appendix~\ref{app:moe}. 

Quantization is a technique that reduces the precision of numerical values (such as model weights and activations) from high-precision formats like 32-bit floating point to lower-precision ones like 8-bit integers. This decreases the memory footprint and computational requirements of a model, enabling faster inference and lower energy consumption. While it can introduce small accuracy losses, quantization is often used to make large models more efficient for deployment on limited hardware. 
Clearly, quantization cannot be considered a form of sparsity, as it relies on dense computations with lower precision. However, since it is frequently discussed in par with sparsification methods, we provide some information on quantization in Appendix~\ref{app:quantization}.

Figure~\ref{fig:dnn_sparsity} provides an summary of the various forms of sparsification together with the computations involved. In the following paragraphs we further discuss those forms that involve sparse data and computations. 

\begin{figure}[tb]
    \centering
    \includegraphics[scale =0.55]{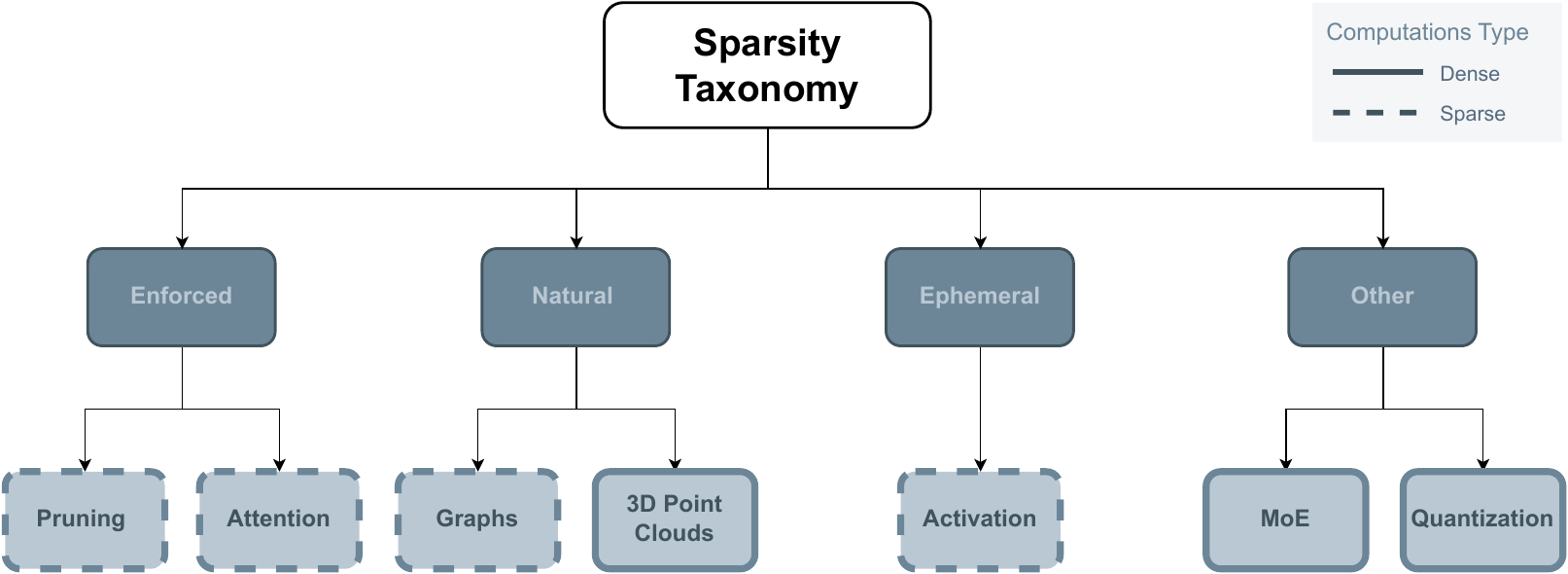}
    \caption{Forms of sparsity in DNNs.}
    \label{fig:dnn_sparsity}
\end{figure}

%%%%%%%%%%%%%%%%%%%%%%%%%%%%%%%%%%%%%%%%%%%%%%%%%%%%%%%%%%%%%%%%%%%%%%%%%%%%%%
\subsection{Weight pruning } \label{sec:sparse:weight_pruning}
%%%%%%%%%%%%%%%%%%%%%%%%%%%%%%%%%%%%%%%%%%%%%%%%%%%%%%%%%%%%%%%%%%%%%%%%%%%%%%
Weight pruning is the process of removing weights from the layers of a DNN with a primary goal to improve model performance, such as inference time, required flops and latency, along with reducing memory footprint. Secondarily, the process can have a positive impact on other goals like generalization, since sparsity can act as a form of regularization~\cite{hoefler2021sparsity}.
Three major types of pruning have been extensively discussed: \textbf{structured pruning}, which removes whole neurons, channels, or filters; \textbf{semi-structured pruning}, which removes weights in regular blocks or patterns; and \textbf{unstructured pruning}, which removes individual weights. 
Model sparsification can be applied at different stages of the training pipeline, including before training (e.g., sparse initialization), during training (gradual pruning), or after training (post-hoc pruning) \cite{hoefler2021sparsity,gale2019state}. 
In practice, different layers and architectures exhibit different sensitivities to pruning, and pruning methods therefore assign sparsity levels per layer using heuristic or empirically driven criteria rather than a universal rule \cite{hu2016network, evci2020rigging, liu2017learning}. 
Some layers are typically pruned more conservatively (for example, downsampling or projection layers in residual networks), while others can sustain higher sparsity without substantial accuracy degradation \cite{molchanov2016pruning, he2017channel}. 
It is also important to note that sparsifying different layers can lead to very different computational savings. 
In CNNs, the FLOP contribution of a layer depends on both its spatial resolution and its channel dimensionality, which vary inversely across the network: early layers operate on large spatial maps but with fewer channels, while deeper layers have many more channels but operate at much smaller spatial sizes. 
As a result, the FLOP reduction achieved by pruning a given layer is not uniform across the network, and choosing how to distribute sparsity—given an overall sparsity or FLOP-reduction target—requires balancing accuracy considerations with practical efficiency.
Therefore the sparsification process and the related design choices drastically affect the level and pattern of sparsity created, with a high subsequent impact on the efficiency of the relevant operations at inference time.
More details about methods for model sparsification like random pruning, magnitude pruning and regularization can be found in Appendix~\ref{app:pruning}.

Sparsity has been applied to a wide range of architectures and layer types, including fully connected layers, recurrent networks, convolutional networks, and transformer-based models. 
Fully connected layers have been extensively explored in pruning research, with both structured (e.g., neuron or channel removal) and unstructured weight pruning investigated in early and recent works \cite{lecun1989optimal,han2015learning}. 
LSTM and RNN layers can be sparsified similarly to fully connected layers, supporting both structured and unstructured pruning, as demonstrated in methods like ESE (block-balanced sparsity) \cite{han2017ese} and MASR (bitmap scheme with moderate sparsity) \cite{gupta2019masr}.
Convolutional layers have also been widely studied, with approaches covering unstructured, structured, and channel/filter-level pruning \cite{molchanov2016pruning,he2018soft}. 
More recently, transformer-based models have become an important target for pruning, with sparsification applied to attention projections, feed-forward layers, and embeddings \cite{kurtic2022optimal,frantar2023sparsegpt}. 
In contrast to earlier deep architectures, which were often trained in a parameter-rich but data-limited regime, modern large language models (LLMs) are typically trained on hundreds of billions to trillions of tokens, resulting in multiple tokens per parameter~\cite{hoffmann2022training}. This higher data-to-parameter ratio alters the model’s redundancy profile, and consequently, simpler sparsification techniques that proved effective in previous settings tend to be less reliable for today’s LLMs.

Following the analysis on the operations involved for each one of the layers in Section~\ref{sec:background:dnn_basics}, the following apply:
\begin{itemize}
    \item When a Fully Connected layer is pruned, the weight matrix $\mathbf{W}$ from Equation~\ref{eq:full} becomes sparse. Thus, if the operation is applied on single input, the computation involved is transformed from matrix-vector to sparse matrix-vector multiplication (\texttt{SpMV}). If the operation is applied on batches of input, the relevant matrix multiplication is transformed to sparse matrix - matrix vector  multiplication (\texttt{SpMM}).
    \item When a Convolutional layer is pruned, then the convolution operation is transformed into \textit{sparse convolution}. In the cases where the \texttt{im2col} and matrix multiplication is employed, the process is transformed again to \texttt{im2col} and sparse matrix - dense matrix multiplication (\texttt{SpMM}).
    \item When a Recurrent or LSTM layer is pruned, the relevant weight matrices (see Section~\ref{sec:background:dnn_basics:seq_layers}) become sparse. The operations involved are thus transformed from matrix-vector to sparse matrix - dense vector multiplication (\texttt{SpMV}). Again, if the input is batched, a sparse matrix - matrix vector  multiplication is used (\texttt{SpMM}).
    \item In Transformer and LLM architectures, most components (including attention projections, feed-forward layers, and embeddings) are implemented as linear layers. When these linear layers are pruned (e.g., in the computation of $\mathbf{Q}$, $\mathbf{K}$, and $\mathbf{V}$, the weight matrices $\mathbf{W}_Q$, $\mathbf{W}_K$, and $\mathbf{W}_V$ in Equation~\ref{eq:qkv} are becoming sparse), their associated dense matrix multiplications are transformed into sparse matrix–dense matrix multiplications (\texttt{SpMM}).
\end{itemize}

%%%%%%%%%%%%%%%%%%%%%%%%%%%%%%%%%%%%%%%%%%%%%%%%%%%%%%%%%%%%%%%%%%%%%%%%%%%%%%
\subsection{GNNs} \label{sec:sparse:gnn}
%%%%%%%%%%%%%%%%%%%%%%%%%%%%%%%%%%%%%%%%%%%%%%%%%%%%%%%%%%%%%%%%%%%%%%%%%%%%%%
Sparsity may arise in GNNs at multiple levels of the model and data pipeline. We can attest that sparsity in GNNs arises in all three forms. In this paragraph, we comment on the most frequently discussed form:  the natural sparsity of GNNs. In Section~\ref{sec:sparse:combined} we make a brief discussion on other potential forms that may be relevant to GNNs.

GNNs inherently operate on graph-structured data, where natural sparsity arises from the topology of real-world graphs. These include social networks, biological systems, knowledge graphs, and citation networks, which typically show low average node degrees (in some cases lower than 10) compared to the total number of nodes. As such, the adjacency matrix representing the graph is highly sparse (in several cases less than 1\% of the elements are nonzeros), and this structural sparsity is central to GNN design. Additionally, in some domains (e.g., natural language processing or biology), node and edge features can also be sparse - for instance, multi-hot encoded attributes or bag-of-words representations. Because of this, graph connectivity is almost always represented using sparse formats like edge lists or sparse matrices.

This form of sparsity is not only widespread but foundational. Nearly all GNN architectures are designed to exploit this input sparsity by performing computations only along nonzero entries in the adjacency matrix. Message passing frameworks and libraries such as Deep Graph Library (DGL) \cite{wang2019deep} and PyTorch \cite{fey2019fast} are structured around this assumption, using sparse matrix operations as their core computational primitives.

As discussed in Section~\ref{sec:background:dnn_basics:gnn}, the typical message-passing operation in GNNs is of the form:
\begin{equation}
    \mathbf{H}^{(l+1)} = \phi \left(\mathbf{A}, \mathbf{H}^{(l)}, \mathbf{W} \right)
\end{equation}
where $\mathbf{A}$ is a sparse matrix following the pattern of the adjacency matrix of the graph, $\mathbf{H}^{(l+1)}, \mathbf{H}^{(l)}$ the node features at layers $l+1$ and $l$ respectively and $\mathbf{W}$ is a learnable weight matrix.

Message-passing in Graph Convolutional Networks (see Equation~\ref{eq:gcn}) utilizes a static matrix $\mathbf{A}$ and the computation involves the computation of $\mathbf{H}^{(l+1)}$ using $\mathbf{A}, \mathbf{D}$ (a degree matrix – also static) and $\mathbf{H}^{(l)}$ $\mathbf{W}$. The computational kernel involved is \texttt{SpMM}. Other, more advanced networks like GraphSAGE
\cite{hamilton2017inductive}, Relational Graph Convolutional Networks (R-GCN) \cite{schlichtkrull2018modeling} and Graph Isomorphism Networks (GIN) \cite{xu2018powerful} use more elaborate functions for message passing, but in terms of sparse computations they fall into the same category as GCNs, relying on the \texttt{SpMM} operation.

On the other hand, Graph Attention Networks (see Equation~\ref{eq:gat}) need to recalculate matrix $\mathbf{A}$ at each layer dynamically introducing essentially an attention mechanism. Matrix $\mathbf{A}$ retains the same structure as the adjacency matrix, but its values need to be recomputed at each step. The computation of this attention matrix $\mathbf{A}$ is based on a Sampled Dense-Dense Matrix Multiplication (\texttt{SDDMM}) using the adjacency matrix of the graph as mask on the product. This operation is then followed by \texttt{SpMM} to compute the final matrix $\mathbf{H}^{(l+1)}$.

%%%%%%%%%%%%%%%%%%%%%%%%%%%%%%%%%%%%%%%%%%%%%%%%%%%%%%%%%%%%%%%%%%%%%%%%%%%%%%
\subsection{Sparse activations} \label{sec:sparse:activations}
%%%%%%%%%%%%%%%%%%%%%%%%%%%%%%%%%%%%%%%%%%%%%%%%%%%%%%%%%%%%%%%%%%%%%%%%%%%%%%
Inference can be accelerated by utilizing ephemeral sparsity, i.e., the type of sparsity that occurs when sparse output is produced by the activation operation of a DNN layer. Modern DNNs use an activation function like ReLU or alternatives (see Section~\ref{sec:background:dnn_basics:activation}), which inherently produce sparse activations by clipping a large number of layer outputs to zero. We frame the sparse activation landscape in the two following groups, \textbf{inherent} and \textbf{explicitly induced sparsity}, which are discussed next. 

In the inherent activation sparsity zeros emerge automatically from the activation itself. To be more elaborate, classic ReLU yields sparsity by deterministically mapping all negative pre-activations to zero. Thresholded variants, in the manner of FATReLU \cite{kurtz2020inducing}, generalize this mechanism by introducing a positive threshold, suppressing small but non-informative responses. In practice, this can be deployed in a ``static'' method that simply replaces the activation in a pre-trained model without retraining, or a ``dynamic'' thresholding approach, more of which will be discussed in the following group.

Although ReLU-based techniques remain the most widely adopted ones, mainly due to their plug-and-play simplicity and already very high zero yield, several significant alternative approaches achieve comparable or even superior sparsity control and efficiency. Ranking-based gating functions such as k-WTA \cite{xiao2019enhancing} apply a global top$-k$ filter to an $N-$dimensional activation vector $\mathbf{y}$: only the $k$ largest entries are preserved and all others are set to zero. To adapt across layers of differing width, $k$ is typically chosen via a fixed sparsity ratio $\gamma \in (0,1)$, so that $k = \lfloor \gamma \times N \rfloor$ in each layer. The runtime cost of selecting the top-k elements is $O(N)$,  equivalent to a standard ReLU pass. 

In all the cases mentioned, the zero pattern is a deterministic function of the current pre-activations alone:
\begin{equation}
    \begin{array}{c}
        S_{\operatorname{ReLU}} = \left\{i \mid z_{i}\leq 0 \right\},\\[5pt]
         S_{\operatorname{FATReLU}} = \left\{i \mid z_{i}\leq \tau \right\},\\[5pt]
          S_{k-\operatorname{WTA}} = \left\{i \mid \operatorname{rank} \left(z_{i}\right) >  k \right\}
    \end{array}
\end{equation}
with $\tau$ and $k$ set as hyperparameters and no auxiliary sparsity-inducing loss, since zero values arise inherently during the forward pass from the activation itself.

Consistent with this view, Mirzadeh et al. \cite{mirzadeh2023relu} demonstrate that swapping GELU with plain ReLU in billion-scale language models produces extremely sparse feed-forward layers, which can be rapidly fine-tuned to recover any lost accuracy. Again here, sparsity is a direct by-product of the activation form.

While inherent activation functions deterministically produce zeros from the current pre-activations and are governed by architectural hyperparameters $(\tau, k)$, their density may not align with the designated task or hardware-level targets. To obtain finer, layer-adaptive and data-adaptive control, one can explicitly induce sparsity during training or adaptation. In explicitly induced activation sparsity the network keeps its usual non-linearity, but auxiliary mechanisms, like density-penalizing loss terms, curriculum schedules or quantization deliberately drive intermediate activations toward zero. A good example of this category is L1 Regularization, which adds a penalty $\lambda \| \alpha \|_{1}$ on the intermediate activations $a$, shrinking small responses toward zero and increasing at exactly zero without changing the activation function itself. A closely related but often stronger alternative is the square Hoyer regularizer, proportional to $\left(\| \alpha \|_{1} \right)^{2}/ \| \alpha \|_{2}^{2}$, which is scale-invariant and concentrates energy into fewer coefficients, typically yielding higher sparsity at similar accuracy \cite{kurtz2020inducing}. Building on these induced distributions, Activation Map Compression \cite{georgiadis2019accelerating} augments training with an L1 term and then quantizes  activations to 8-bit followed by entropy coding, exploiting the elevated fractions of zeros and near-zeros to cut bandwidth and memory traffic for end-to-end acceleration.

Building on penalty-based and compression approaches, DEFT (Density-aware parameter-Efficient Fine Tuning) \cite{runwal2025peft} treats activation density as an explicit training signal during fine-tuning of Transformer MLPs: a lightweight density loss is added to steer intermediate outputs toward zero, typically reducing activation density substantially with minimal impact on accuracy. Extending this explicitly induced principle to generative settings, SQ-DM \cite{fan2025sq} targets diffusion models via temporal sparsity, coupling aggressive 4-bit activation quantization with time-step-aware gating/routing; sparse and dense channels are then directed to specialized compute units, translating induced zeros and near-zeros into end-to-end speed and energy gains.
Collectively, these methods are classified as explicitly induced because the zero pattern is shaped by external objectives, losses, schedules, or quantization policies rather than being a fixed consequence of the activation function itself.

Returning to the previously mentioned FATReLU, while its `static' approach is in the inherent sparsity category, the FATReLU in its `dynamic' form is better understood as a hybrid of the two categories. It consists of an inherent component, which is the thresholded activation that zeroes all responses below a per-layer level $T_{\ell}$ and an explicitly induced component, wherein the threshold set is not fixed but rather scheduled or learned under external objectives and accompanied by brief fine-tuning, along with explicit regularization during adaptation (L1, square-Hoyer) so that weights reorganize around progressively stricter gates. Such external objectives are target activation density, accuracy-loss tolerance or latency/energy budgets. Operationally, thresholds are raised per layer via a simple curriculum or a sensitivity analysis that selects the largest $T_{\ell}$ consistent with a small accuracy-degradation bound; short recovery phases at each step preserve performance, while the activation distribution shifts toward higher sparsity. Because the attained density is governed by this optimization loop rather than by the nonlinearity alone, dynamic FATReLU is a hybrid mechanism: an inherent thresholded gate operated by an induced controller.

Extending this hybrid pattern, ReLUfication pipelines \cite{song2024prosparse} and SAFS exemplify  mixed methods that couple inherent and induced sparsity. In ReLUfication, GELU/Swish are replaced with ReLU to establish an inherently sparse baseline, then a staged L1 schedule and FATReLU–style dynamic thresholds are applied to progressively prune weaker activations—achieving very high sparsity in large language models without performance loss. In parallel, SAFS \cite{loni2023learning} performs AutoML-driven activation search, jointly evolving piecewise-linear activation functions and training hyperparameters to maximize sparsity in already weight-pruned networks. 

This dichotomy distinguishes sparsity arising from the activation rule from sparsity treated as an explicit optimization objective. In practice, the two are frequently combined (e.g., sparse activations followed by density-aware fine-tuning) to achieve higher, better-controlled sparsity that guides methodological choices and hardware co-design.

Interestingly, the sparsification opportunities at the activation stage can be much more significant than those in weight pruning. This contrast is readily quantified by a layer-level analysis. Consider the second $3\times 3$ convolution in the Inception-V3 initial block (stem), the illustrative case used by Georgiadis \cite{georgiadis2019accelerating}.
That layer consumes a $149\times 149 \times 32$ activation tensor and emits a $147\times 147 \times 32$ one, summing up to an amount of $1.401.920$ total input-plus output activations. In reality, its kernel holds only
\begin{equation*}
    32\times 32\times 3\times 3 = 9216
\end{equation*}
parameters. Hence the activation volume exceeds the parameter volume by a factor of
\begin{equation*}
    \frac{1401920}{9216}\approx 1.52 \times 10^{2} (\approx 150\times).
\end{equation*}
Even in single-precision,
\begin{equation*}
        \frac{1401920\times 4\mathrm{B}}{1048576}\approx 5.6\mathrm{MB},  \quad  \frac{9216\times \mathrm{4B}}{1024}\approx 36\mathrm{KB}.
\end{equation*}
This means $\sim 5.6\mathrm{MB}$ of activations versus merely $\sim 36\mathrm{kB}$ of weights. Such a two-orders-of-magnitude disparity confirms that aggressively sparsifying activations can yield far greater memory bandwidth and compute saving than further pruning the already compact weight tensor for this, and many other, CNN layers.

\begin{figure}[tb]
    \centering
    \includegraphics[width = 0.55\linewidth]{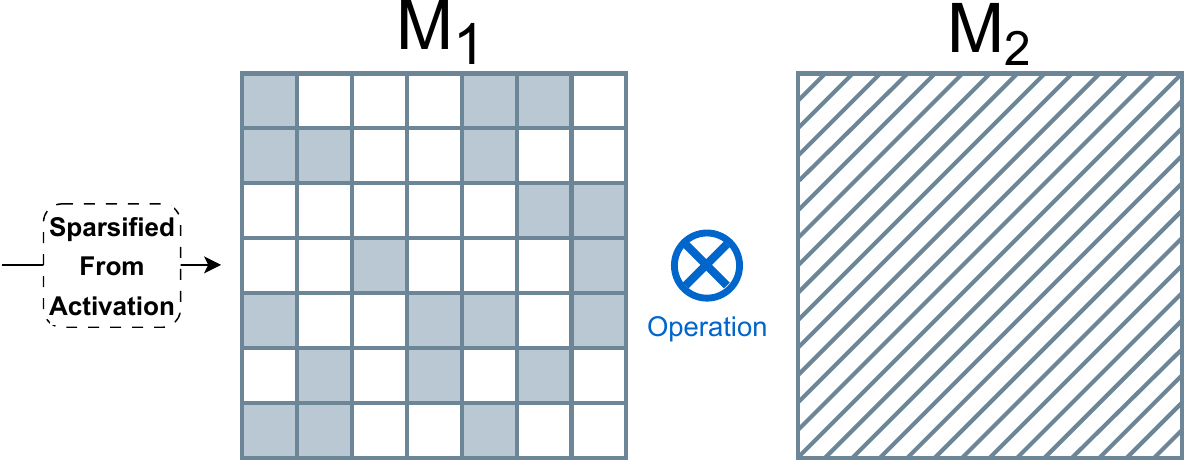}
    \caption{Operations with sparsified activations from the previous layer.}
    \label{fig:sparse_activation_operation}
\end{figure}

In an analogous way as in the case of weight pruning, a sparse tensor is entering the layer of a DNN because of the sparsification that took place at the output of the previous layer. In this case, the operation involved is determined by the following layer. If the layer is Fully Connected, then the operation is again \texttt{SpMM}. If the layer is a convolutional layer, the operation is a sparse convolution with the sparsity resident at the input rather than the weights. Figure~\ref{fig:sparse_activation_operation} depicts the process. The $\mathbf{M}_{1}$ matrix represents a sparse activation tensor resulting from a previous layer’s non-linearity (e.g., ReLU, FATReLU, or $k-$WTA), where zero entries are omitted. In both cases, the sparsity resides in the input $\left(\mathbf{M}_{1}\right)$ rather than the weights $\left(\mathbf{M}_{2}\right)$, which remain dense. Note that the operation for computing the weight gradients for layers with sparse inputs, during the backward pass, can also be considered as a sparse operation, leading to potential benefits during training. 

Due to the ephemeral nature of sparsity in this case, sparse tensors need to be generated in each inference example (as opposed to weight pruning, where sparsified weights are inserted in the inference engine at deployment time). This is a very important feature of this type of sparsity that greatly influences the optimization opportunities and process. Performance engineers need to consider the efficacy of the process to manage this form of sparsity, i.e., there is no luxury in heavy-weight approaches to store the sparse tensor. On the contrary, on the fly, effective and lightweight schemes need to be employed. 

%%%%%%%%%%%%%%%%%%%%%%%%%%%%%%%%%%%%%%%%%%%%%%%%%%%%%%%%%%%%%%%%%%%%%%%%%%%%%%
\subsection{Sparse Attention } \label{sec:sparse:attention}
%%%%%%%%%%%%%%%%%%%%%%%%%%%%%%%%%%%%%%%%%%%%%%%%%%%%%%%%%%%%%%%%%%%%%%%%%%%%%%
Sparse attention modifies the dense attention operation shown in Equation~\ref{eq:attention}, by introducing a binary mask $\mathbf{M}\in \{0, 1\}^{N\times N}$, which selectively zeroes out certain entries in the attention matrix.

The sparse attention operation is then implemented as:
\begin{equation}
    \operatorname{Attention}_{\operatorname{sparse}}(\mathbf{Q,K,V}) = \operatorname{softmax_{nz}}\left(\frac{\mathbf{M}\odot \mathbf{QK}^{T}}{\sqrt{d}} \right)\mathbf{V}
\end{equation}
where $\odot$ denotes element-wise multiplication that is implemented by the Sampled Dense-Dense Matrix Multiplication (\texttt{SDDMM}). $softmax_{nz}$ is the softmax operation defined in Section~\ref{sec:background:dnn_basics:activation} applied only to the nonzero elements that arise from \texttt{SDDMM}. The outcome is then multiplied by $\mathbf{V}$ using an \texttt{SpMM} kernel. Note that, formally, this can be defined similarly to the masked attention mechanism discussed in Section~\ref{sec:background:dnn_basics:transformer_attention}, with $- \infty$ being the outcome of the operation in the zero mask elements. Then, we can apply the standard softmask function to the entire outcome. Sparse attention is a special case of masked attention, however, the typical masked attention  relies on dense operations.

\begin{figure}[ht!]
    \centering
    \includegraphics[scale = 0.55]{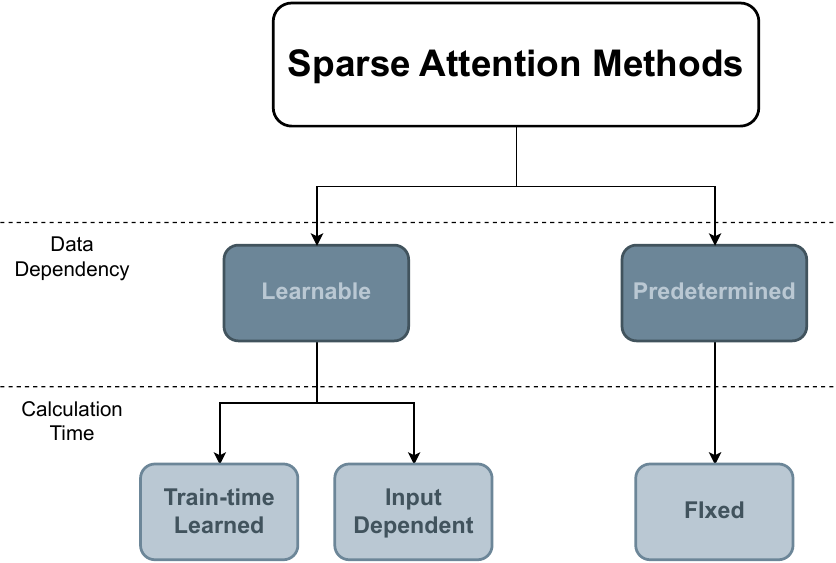}
    \caption{Taxonomy of sparse attention methods, organized by whether the sparsity pattern is fixed or dynamic, and when it is determined (before training, during training, or at inference). }
    \label{fig:sparse_attention_taxonomy}
\end{figure}

In sparse attention, some tokens do not interact directly, which can affect gradient propagation. This should be taken into account when designing the mask pattern. Obviously there is the usual performance vs accuracy trade-off, so caution is needed for choosing the proper mask pattern (usually a combination of more than one pattern is used). To systematically understand and compare alternative approaches to produce the sparse attention mask, we identify a first distinction, as to whether the mask is dependent on the model it will be applied to. We distinguish between \textbf{predetermined} and \textbf{learnable} masks. Predetermined masks are designed manually and remain unchanged regardless of the input or training process. This means that the mask is defined before training and remains fixed. In contrast, learnable masks are flexible and change based on training data or input content. This category is further divided into two main approaches: those where the mask is learned during training (train-time learned), and those where the mask is generated at inference time based on the input (input-dependent). This taxonomy is shown in Figure~\ref{fig:sparse_attention_taxonomy}.

\begin{figure}[tb]
    \centering
    \includegraphics[width=0.9\linewidth]{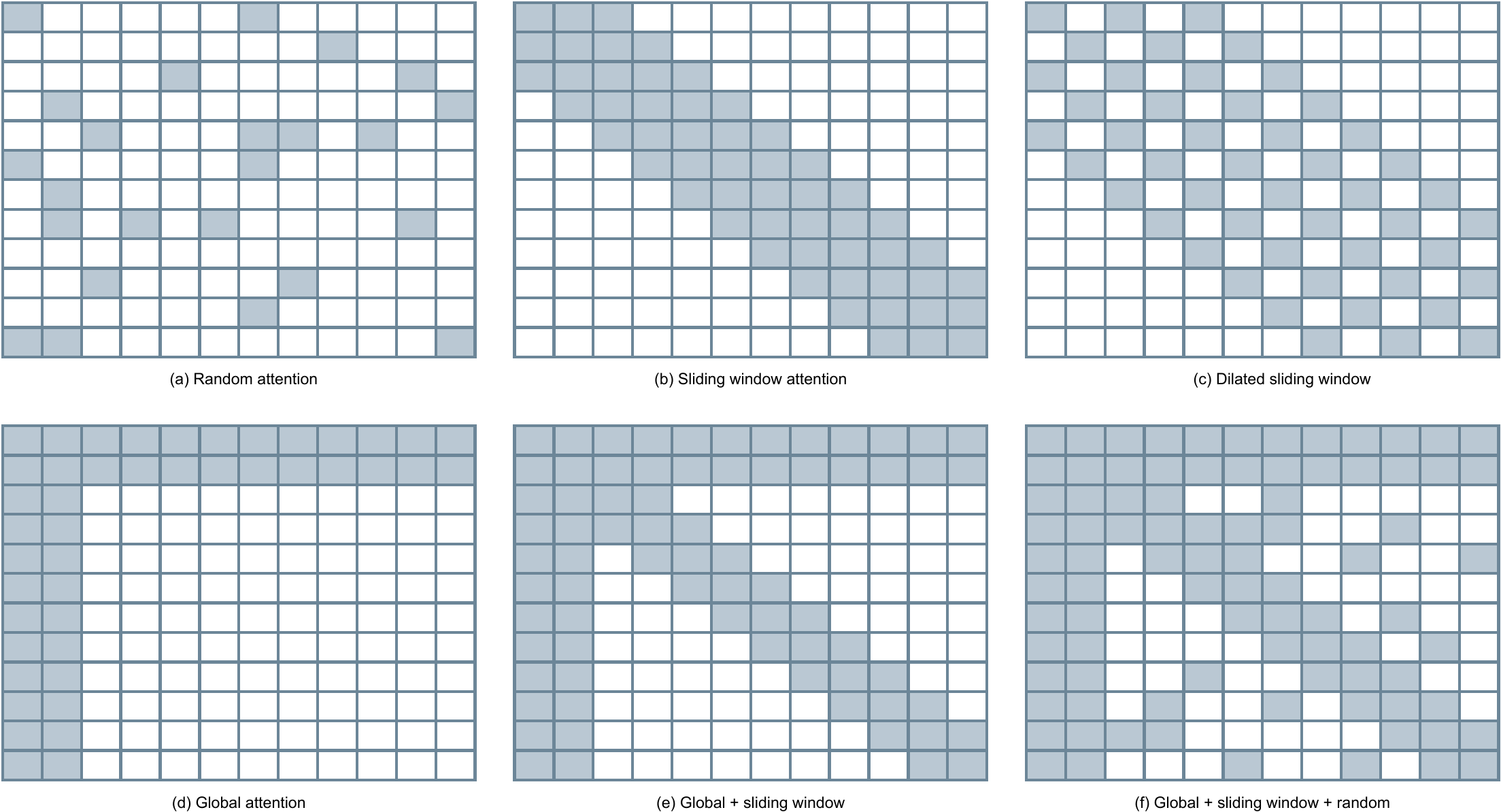}
    \caption{Examples of fixed sparse attention patterns. Top row: (a) Random, (b) Sliding Window, (c) Dilated Sliding Window. Bottom row: (d) Global, (e) Global + Sliding Window, (f) Global + Sliding Window + Random attention.}
    \label{fig:sparse_attention_masks}
\end{figure}

The most common predetermined patterns include structured approaches such as local attention, which confines attention computations to tokens within a fixed-size window. This significantly reduces complexity from $O\left( N^{2}\right)$ to $O\left(\operatorname{window\_size}^{2}\right)$, where $\operatorname{window\_size} \ll N$.  Blockwise patterns \cite{qiu2019blockwise,parmar2018image} represent a practical example of fixed patterns by dividing the input sequence into fixed-size chunks and computing attention only within these local blocks \cite{child2019generating}. Another fixed pattern method uses strided (or dilated) attention \cite{child2019generating,beltagy2020longformer}, where tokens attend to other tokens at fixed intervals, capturing a broader context efficiently. Combined patterns \cite{liu2018generating}, utilized by architectures such as Longformer and BigBird \cite{zaheer2020big}, integrate local attention with strategically placed global tokens to blend local context with selective global information. Additionally, some methods introduce random attention patterns, where each token is allowed to attend to a randomly selected subset of other tokens. While these patterns are technically produced for each instance, they are considered predetermined because they do not depend on the data or model dynamics during training or inference. This randomness serves to increase connectivity and mitigate information bottlenecks, but the sampling itself is independent of token content. 
Random attention is classified within the predetermined category due to its fixed structural pattern and data-agnostic design. Nevertheless, implementation-wise, the ad hoc computation of the mask hinders any code optimization approaches that could be applied considering the structure of the mask. Figure~\ref{fig:sparse_attention_masks} illustrates examples of the most common predetermined attention patterns and their combinations, including sliding window, global, dilated, and random attention.

\textbf{Learnable sparse attention} \cite{vyas2020fast,kitaev2020reformer,lou2024sparser} patterns adapt during training or inference, allowing the model to focus on the most relevant parts of the sequence. These masks are not predetermined, they instead emerge from training dynamics or are constructed on-the-fly during inference. This category can be further divided into two subtypes, based on when the mask is calculated.

The first subtype includes methods that learn the sparse attention mask during training, after which the mask remains fixed during inference. Examples of such approaches include the Routing Transformers \cite{roy2021efficient}, which clusters tokens into groups based on learned relationships to determine sparse attention patterns, and Layer-wise Attention Pruning \cite{shim2021layer}, which learns which attention heads are redundant and removes them permanently. Some adaptive sparse attention models also fall into this category, learning token-level importance scores during training and applying the resulting mask uniformly at inference. Hybrid models, such as Longformer, incorporate both fixed and learnable components—combining local attention, global tokens, and learnable sparse patterns to balance efficiency with contextual awareness.

The second category of learnable sparse attention methods involves computing the mask separately for each instance during inference. This includes methods like SAAP \cite{mazare2025inference} and SeerAttention-R \cite{gao2025seerattention}. The first method uses learned indexing networks to generate top$-k$ sparse attention patterns per query, with attention computed only over the selected keys using a standard softmax whereas the second dynamically constructs sparse attention masks by aligning with the attention structure of a dense teacher model for each input.

Another approach to achieving sparse attention goes beyond explicitly calculating a sparse matrix $\mathbf{M}$ and instead focuses on approximating the attention matrix using various techniques, such as low-rank approximations or kernel functions. For example methods such as Linformer \cite{wang2020linformer}, and Performer \cite{choromanski2020masked}, project keys and queries into lower-dimensional representations, reducing computational complexity to linear order. Performer and Random Feature Attention \cite{peng2021random} specifically employ kernel methods to approximate the softmax operation, drastically reducing both computational and memory requirements. These approaches collectively enable efficient scaling of transformer models and allow handling significantly longer sequences without substantial computational overhead.

Sparse attention methods also include recurrence-based models, such as Transformer-XL \cite{dai2019transformer}. These models process input sequences by splitting them into smaller segments and incorporating recurrent mechanisms, which maintain memory of previous segments, allowing models to retain information over longer distances. Additionally, downsampling techniques like Perceiver \cite{jaegle2021perceiver} and Funnel Transformers \cite{dai2020funnel} reduce the sequence length progressively by applying pooling operations, resulting in significant computational savings. Adaptive sparsity models, such as Switch Transformers \cite{fedus2022switch}, dynamically activate subsets of parameters based on the specific input context, further optimizing computational efficiency.

%%%%%%%%%%%%%%%%%%%%%%%%%%%%%%%%%%%%%%%%%%%%%%%%%%%%%%%%%%%%%%%%%%%%%%%%%%%%%%
\subsection{Combined sparsity} \label{sec:sparse:combined}
%%%%%%%%%%%%%%%%%%%%%%%%%%%%%%%%%%%%%%%%%%%%%%%%%%%%%%%%%%%%%%%%%%%%%%%%%%%%%%

One natural question that arises is whether combining the various forms of sparsity discussed in this section would be of additional benefit. For example, one could prune the trainable weights of a GNN \cite{peng2022towards, gurevin2024prunegnn} or utilize both activation and graph sparsity \cite{mukherji2023activity, mondal2022gnnie}. Although this is a clearly interesting field that can potentially lead to challenging accuracy-performance tradeoffs, and bring in the game additional sparse kernels like \texttt{SpGEMM} (sparse matrix by sparse matrix multiplication) \cite{zhang2023dynasparse}, we find that the research in the field is currently limited, and thus leave further investigation for future study. 

%%%%%%%%%%%%%%%%%%%%%%%%%%%%%%%%%%%%%%%%%%%%%%%%%%%%%%%%%%%%%%%%%%%%%%%%%%%%%%
\section{Sparse computational kernels for DNN inference} \label{sec:sparse_kernels}
%%%%%%%%%%%%%%%%%%%%%%%%%%%%%%%%%%%%%%%%%%%%%%%%%%%%%%%%%%%%%%%%%%%%%%%%%%%%%%

The performance of sparse computational kernels is decisive for the performance of DNN inference itself, and it is no surprise that their optimization has attracted vivid interest from vendors and researchers. In this section we review the state of the art of the sparse kernels involved in DNN inference, namely Sparse Matrix-Matrix Multiplication (\texttt{SpMM}), Sampled Matrix-Matrix Multiplication (\texttt{SDDMM}), Sparse Matrix-Vector Multiplication (\texttt{SpMV}) and Sparse Convolution. Common performance features of these kernels is their irregularity that leads to imbalance and memory bottlenecks. However, different and subtle instantiations of these generic performance bottlenecks may come forward depending on the kernel, matrix and architecture. The vast majority of proposed kernels are optimized for GPU execution, reflecting the dominant role of GPUs in high-throughput inference workloads. In the following paragraphs we present relevant work for each kernel.

%%%%%%%%%%%%%%%%%%%%%%%%%%%%%%%%%%%%%%%%%%%%%%%%%%%%%%%%%%%%%%%%%%%%%%%%%%%%%%
\subsection{SpMM and SDDMM} \label{sec:sparse_kernels:spmm_sddmm}
%%%%%%%%%%%%%%%%%%%%%%%%%%%%%%%%%%%%%%%%%%%%%%%%%%%%%%%%%%%%%%%%%%%%%%%%%%%%%%
As analyzed in Section~\ref{sec:background:sparsity_basics}, \texttt{SpMM} is a ubiquitous operation in sparse inference resident in all sources of sparsity. \texttt{SDDMM} on the other hand is used together with \texttt{SpMM} in GAT GNNs and sparse attention. Since researchers that work on \texttt{SDDMM} also work with \texttt{SpMM}, we include both kernels together in our analysis. The main performance challenges addressed for these kernels is the irregularity in the access of the dense vector(s) that prohibits the utilization of existing data locality, and load balance, again an artifact of sparsity. In Section~\ref{sec:eval} we provide evaluation results for these kernels in GPUs and CPUs where we validate that the performance of these kernels in modern compute devices is very low.

%%%%%%%%%%%%%%%%%%%%%%%%%%%%%%%%%%%%%%%%%%%%%%%%%%%%%%%%%%%%%%%%%%%%%%%%%%%%%%
\subsubsection{GPU implementations} \label{sec:sparse_kernels:spmm_sddmm:gpu}
%%%%%%%%%%%%%%%%%%%%%%%%%%%%%%%%%%%%%%%%%%%%%%%%%%%%%%%%%%%%%%%%%%%%%%%%%%%%%%
Several GPU-based \texttt{SpMM} and \texttt{SDDMM} approaches employ code-level optimizations such as tiling, row splitting or merging, and other low-level transformations to improve memory access patterns in line with GPU optimization guidelines (e.g., effective use of shared memory and memory coalescing). These techniques enhance memory locality and workload balance during execution. In addition, reordering the sparse matrix during preprocessing is a commonly used strategy to further improve locality.
ASpT \cite{hong2019adaptive} applies 2D tiling and column reordering to expose dense tiles and improve memory coalescing in both \texttt{SpMM} and \texttt{SDDMM}. 
ASpT-RR \cite{jiang2020novel} extends this by grouping rows based on access patterns and applying reordering to further enhance data locality. 
RoDe \cite{pang2024row} handles big rows in \texttt{SpMM} and \texttt{SDDMM} by splitting rows into regular and residual parts, enabling pipelined execution across GPU sub-blocks to improve load balance.
RS-SpMM \cite{hong2018efficient} partitions matrices into heavy and light segments and applies thread coarsening (i.e., increases the load per thread) and shared memory optimizations to each segment.
GNNAdvisor \cite{wang2021gnnadvisor} leverages neighbor partitioning, dimension partitioning for large embedding sizes, and warp-aware memory customizations to optimize workload balance, embedding-dimension parallelism, and memory access.
Merge-SpMM \cite{yang2018design} addresses \texttt{SpMM} through a merge-based row-splitting strategy guided by simple heuristics, combining thread- and instruction-level parallelism to hide latency and maintain balanced execution.

Many works opt for low-level GPU optimizations, such as memory alignment, warp scheduling, and caching, to improve workload balance.
Sputnik \cite{gale2020sparse} introduced \texttt{SpMM} and \texttt{SDDMM} kernel designs based on 1D subwarp tiling, incorporating reverse offset memory alignment (ROMA) to align sparse rows with vector memory instructions and a row swizzle technique to enhance load balancing across threads.
GE-SpMM \cite{huang2020ge} proposed a coalesced row caching mechanism along with coarse-grained warp merging, which improves memory access patterns and data reuse among warps by caching contiguous row segments in shared memory and grouping warps to reduce control divergence in CSR-based implementations.
More recently, GE-SpMM was incorporated into the dgSPARSE library \cite{dgsparse_lib} and was extended to provide an \texttt{SDDMM} implementation, with similar optimizations. The \texttt{SpMM} implementation evaluated in Section~\ref{sec:eval} under the name dgSPARSE is based on the GE-SpMM kernel.
MergePath-SpMM \cite{shan2023mergepath} aims in load balancing the sparse computation through nonzero-splitting and proposes a merge-path algorithm to explicitly track assigned matrix rows and avoid expensive atomic operations for complete row updates.

Existing approaches also utilize NVIDIA GPU Tensor Cores -- specialized hardware units designed to accelerate matrix operations using high-throughput, mixed-precision fused multiply-add instructions. These approaches adopt custom sparse formats and data restructuring techniques to expose denser submatrices, enabling efficient mapping of sparse workloads onto Tensor Cores and maximizing hardware utilization.
Acc-SpMM \cite{zhao2025acc} introduces the BitTCF compression format and data-affinity-based reordering in order to enhance the density of the blocks and pairs that with a high-throughput pipeline aiming to hide memory latency by overlapping memory access and sparsity-aware load balancing.
HC-SpMM \cite{li2025hc} uses hybrid core selection to dispatch dense submatrices to Tensor Cores and sparse ones to CUDA cores, employs kernel fusion to reduce kernel launch overhead and improve data reuse, and applies layout reformatting to expose denser segments suitable for Tensor Core processing.
Using the same logic, HR-SpMM \cite{wang2025hr} splits sparse computation by assigning short rows on regular CUDA cores and long rows on Tensor cores, combining load balancing and good hardware utilization.
Magicube \cite{li2022efficient} targets \texttt{SpMM} and \texttt{SDDMM} with the SR-BCRS format for 1D block sparsity and performs register-level online transposition of dense tiles to support low-precision Tensor Core execution.
VectorSparse \cite{chen2021efficient} also proposes 1D-blocking with column vector sparse encoding and supports single and half precision implementations for structured and unstructured sparsity.
DTC-SpMM \cite{fan2024dtc} proposes the ME-TCF format and integrates general matrix reordering, automatic code generation, and low-level kernel tuning.
Tiled-CSR \cite{xue2023releasing} presents a new compression format designed for unstructured sparsity, complemented by row shuffling, adaptive memory access, and 3D-grid tiling to enable fine-grained scheduling on Tensor Cores.
SMaT \cite{okanovic2024high} applies matrix permutation to create dense blocks for TCs and offers a BCSR implementation that overlaps data movement with computation.
FastSpMM \cite{wang2025fastspmm} proposes a new Compressed Sparse Tile (CST) format, virtual-row-based merging, and load-aware mapping, combined with runtime kernel optimizations like software pipelining and data prefetching, to overcome format conversion overheads, load imbalance, and irregular memory access.
Unlike the previous approaches, Flash-LLM \cite{xia2023flash} introduces a Load-as-Sparse, Compute-as-Dense approach and employs a tiled-CSL sparse format to mitigate memory bandwidth bottlenecks which is then unpacked to a dense matrix to fully utilize Tensor Cores.

Another line of work utilizes the NVIDIA Sparse Tensor Cores, which support a fixed 2:4 sparsity pattern -- requiring at least two zeros in every group of four elements. Several approaches focus on encoding sparse data into 2:4-compatible submatrices so that semi-structured sparsity can be efficiently executed on these specialized hardware units.
NM-SpMM \cite{ma2025nm} supports arbitrary vector-wise N:M sparsity without relying on hardware-specific features, using hierarchical blocking to improve locality across memory hierarchies and applying latency-hiding strategies.
nmSPARSE \cite{lin2023efficient} extends N:M sparsity to vector-wise and block-wise formats, eliminating shared memory bank conflicts through computation reorganization.
Jigsaw \cite{zhang2024jigsaw} reorganizes sparse matrices into 2:4 patterns using a hierarchical storage format and column reordering, allowing efficient mapping to Sparse Tensor Cores.
VENOM \cite{castro2023venom} generalizes N:M sparsity with a flexible V:N:M format, combining column-wise pruning and block sparsity, and maps these patterns onto Sparse Tensor Cores. 
Marlin \cite{frantar2025marlin} is also implemented for sparse tensor cores whilst introducing quantization and mixed precision support. 

Other works address the problem of tuning the sparse kernels over the large optimization space and propose code generation, heuristic tuning, and runtime adaptation to handle the diversity of sparsity patterns and hardware targets.
EC-SpMM \cite{lin2023ec} and LO-SpMM \cite{lin2024spmm} adopt code generation  approaches that apply 2D tiling and matrix reordering to optimize for pruned deep learning models. LO-SpMM extends this by introducing hierarchical tiling, constraint-based search space reduction to decrease the tuning time, and proxy-based performance evaluation.
SparseRT \cite{wang2020sparsert} uses an inspector-executor framework where sparse matrix data are stored in instruction cache and embedded directly in instruction operands, enabling efficient load balancing through compile-time specialization.
DA-SpMM \cite{dai2022heuristic} proposes a heuristic-driven design that selects appropriate kernel strategies based on input characteristics, using orthogonal design principles to adapt to input dynamics at runtime.

Several works integrate fused kernel designs and communication-aware scheduling to optimize sparse operations especially in GNN pipelines.
fuseGNN \cite{chen2020fusegnn} introduces a high-level message passing framework with fused \texttt{SDDMM}–\texttt{SpMM} execution, structured into five stages with register blocking to minimize kernel launch overhead and intermediate memory usage.
GNNOne \cite{gong2024gnnone} combines fused computation with inter-thread communication and scheduling strategies by organizing threads into groups within a warp and optimizing host-to-device data transfer, aiming to reduce synchronization and improve throughput during attention computation.
GNNPilot \cite{hu2025gnnpilot} enhances performance by combining neighbor packing for load balancing in sparse graphs, bin packing (BIN\_CSR format) for improved data locality in dense graphs, dynamic parallelization strategies, and a row-panel-based kernel fusion technique.
FeatGraph \cite{hu2020featgraph} extends the TVM compiler \cite{chen2018tvm} with GNN-specific optimizations, including feature-dimension tiling, hybrid graph partitioning and user-defined scheduling templates, to efficiently fuse and optimize graph traversal with feature computation.
Fused3s \cite{li2025fused3s} proposes one kernel that combines the three basic operations of sparse attention (\texttt{SDDMM}, Softmax, SpMM) in one, while utilizing CUDA Tensor Cores and minimizing data movement.

Finally, regarding \textbf{vendor} provided libraries, NVIDIA’s cuSPARSE offers optimized implementations for many sparse operations including \texttt{SpMM} and \texttt{SDDMM}, supporting different formats and access patterns. 
rocSPARSE \cite{amd_rocsparse} is a HIP-based library in the AMD-ROCm ecosystem that provides optimized sparse linear algebra routines for AMD GPUs.

%%%%%%%%%%%%%%%%%%%%%%%%%%%%%%%%%%%%%%%%%%%%%%%%%%%%%%%%%%%%%%%%%%%%%%%%%%%%%%
\subsubsection{CPU implementations} \label{sec:sparse_kernels:spmm_sddmm:cpu}
%%%%%%%%%%%%%%%%%%%%%%%%%%%%%%%%%%%%%%%%%%%%%%%%%%%%%%%%%%%%%%%%%%%%%%%%%%%%%%
To our knowledge there are very few works that propose sparse kernels needed in deep learning inference for CPU architectures.
TACO \cite{kjolstad2017tensor} is a library that provides optimized compiler generated kernels for sparse linear algebra including \texttt{SpMM} and \texttt{SDDMM}.
Intel’s MKL \cite{wang2014intel} is a widely used vendor library that provides optimized implementations for sparse kernels like \texttt{SpMM}, \texttt{SpMV}, \texttt{SpGEMM} but does not support \texttt{SDDMM}. 
Similarly, AMD's AOCL-Sparse \cite{amd_aocl} provides \texttt{SpMV} and \texttt{SpMM} implementations; however, it lacks \texttt{SDDMM} support.
From the aforementioned papers, ASpT and FeatGraph also provide CPU support.
SparseDNN \cite{wang2021sparsednn} introduces an x86 code generator that combines row reordering and tiling to optimize direct convolution on CPUs, tuned for cache utilization and SIMD execution.
FusedMM \cite{rahman2021fusedmm} fuses \texttt{SDDMM} and \texttt{SpMM} into a single multi-stage kernel with register blocking to minimize memory traffic and improve locality.
J-stream \cite{kurt2020efficient} proposes a tiling strategy based on matrix signatures and a streaming scheme along the output dimension to minimize memory movement. 

%%%%%%%%%%%%%%%%%%%%%%%%%%%%%%%%%%%%%%%%%%%%%%%%%%%%%%%%%%%%%%%%%%%%%%%%%%%%%%
\subsubsection{General observations} \label{sec:sparse_kernels:spmm_sddmm:other}
%%%%%%%%%%%%%%%%%%%%%%%%%%%%%%%%%%%%%%%%%%%%%%%%%%%%%%%%%%%%%%%%%%%%%%%%%%%%%%

Our analysis reveals that optimization strategies often align closely with the target domain of each paper.
Methods targeting graph attention, including GNNOne, FeatGraph, fuseGNN, fuded3s and GNNPilot, typically propose fused execution pipelines that combine \texttt{SDDMM} and \texttt{SpMM} into a single optimized workflow with tailored dataflow and scheduling strategies.
A distinct category of works -- such as ASpT, ASpT-RR, RS-SpMM, DA-SpMM, SMaT , FusedMM, and J-stream -- do not focus on a specific machine learning domain or sparsity type. Instead, they use generic matrices from the SuiteSparse collection and employ classical optimizations like 2D tiling, reordering, and shuffling, which are well-suited for the large, irregular structures commonly found in these datasets.
In contrast, works addressing attention sparsity -- such as Magicube, Sputnik, VectorSparse, and Tiled-CSR -- primarily rely on low-level GPU optimizations, including memory alignment, instruction-level tuning, and register-level transpositions, to improve throughput under structured or unstructured sparsity.

%%%%%%%%%%%%%%%%%%%%%%%%%%%%%%%%%%%%%%%%%%%%%%%%%%%%%%%%%%%%%%%%%%%%%%%%%%%%%%
\subsection{SpMV} \label{sec:sparse_kernels:spmv}
%%%%%%%%%%%%%%%%%%%%%%%%%%%%%%%%%%%%%%%%%%%%%%%%%%%%%%%%%%%%%%%%%%%%%%%%%%%%%%
The Sparse Matrix Vector Multiplication (\texttt{SpMV}) kernel has been widely studied in the HPC field. \texttt{SpMV} is a memory bound kernel and at the same time dominates the execution time of large system solvers that cover a wide range of applications like graph processing and scientific computing. Many different approaches \cite{kreutzer2014unified,liu2015csr5,elafrou2018sparsex} have addressed the bottlenecks of \texttt{SpMV} for both CPU and GPU architectures. These approaches are best suited for large matrices with a high unstructured sparsity of 99\% and therefore do not suit the smaller, denser and  more structured matrices that appear in LLMs for example. 
In deep learning, \texttt{SpMV} is typically found when the network layers process the input sequentially, one token at a time, i.e.,  \texttt{SpMV} is the primary computational kernel when running sparse RNNs and LSTMs. As observed by Hoefler et al. in \cite{hoefler2021sparsity}, most approaches targeting \texttt{SpMV} for recurrent neural network inference rely on specialized hardware accelerators that provide explicit support for sparse deep learning computations. In more modern network architectures \texttt{SpMV} can appear in the sequential, autoregressive decoding stage of Transformers when the batch size is one.
To accelerate LLM inference, EC-SpMV \cite{lin2025toward} proposes a novel approach that uses hierarchical block extraction and a compressed sparse format EC-CSR with delta indexing to enhance data locality and reduce storage overhead in GPUs. From the aforementioned papers, RS-SpMM and MergePath-SpMM also support an \texttt{SpMV} kernel implementation.

%%%%%%%%%%%%%%%%%%%%%%%%%%%%%%%%%%%%%%%%%%%%%%%%%%%%%%%%%%%%%%%%%%%%%%%%%%%%%%
\subsection{Sparse Convolution} \label{sec:sparse_kernels:spconv}
%%%%%%%%%%%%%%%%%%%%%%%%%%%%%%%%%%%%%%%%%%%%%%%%%%%%%%%%%%%%%%%%%%%%%%%%%%%%%%

As discussed in Section~\ref{sec:background:sparsity_basics:unstructured}, sparse convolution is a key computational kernel in CNNs, that are particularly important in computer vision tasks where sparsity arises from structured inputs (e.g., 3D point clouds) or pruning techniques used to accelerate inference. Many approaches have been proposed that cover the different sparse convolutions in both CPU and GPU architectures. 
TACO-UCF \cite{won2023unified} is an extension of the TACO library, offering a generalized approach that tackles all the different types of sparse convolutions. 
Escoin \cite{chen2018escoin} proposes a direct approach for filter sparse convolution that avoids the im2col lowering of the kernel into \texttt{SpMM} and tries to maximize data reuse in GPUs by exploiting the inherent spatial locality of convolution operations. 
XNNPACK \cite{elsen2020fast} and SparseDNN also provide im2col sparse convolution implementations. 
In \cite{kurtz2020inducing} Kurtz et al. propose a CPU-tailored sparse convolution algorithm that exploits activation sparsity by efficiently integrating run-time SIMD compression of sparse input with low-level kernel optimizations, including vectorization, memory access pipelining, and data blocking.
Lastly, several papers propose optimized implementations of submanifold sparse convolution and are discussed in detail in Appendix~\ref{app:spconv}.

%%%%%%%%%%%%%%%%%%%%%%%%%%%%%%%%%%%%%%%%%%%%%%%%%%%%%%%%%%%%%%%%%%%%%%%%%%%%%%
%%%%%%%%%%%%%%%%%%%%%%%%%%%%%%%%%%%%%%%%%%%%%%%%%%%%%%%%%%%%%%%%%%%%%%%%%%%%%%
%%%%%%%%%%%%%%%%%%%%%%%%%%%%%%%%%%%%%%%%%%%%%%%%%%%%%%%%%%%%%%%%%%%%%%%%%%%%%%

\begin{sidewaystable}
    \centering
    \caption{Summary of \texttt{SpMM}, \texttt{SDDMM} and \texttt{SpConv} implementations discussed in Section~\ref{sec:sparse_kernels}}
    \label{tab:wide-data}
\resizebox{\textwidth}{!}{%
    \begin{tabular}{lllllll}
        \toprule
        \textbf{Name} & \textbf{Platform} & \textbf{Kernels} & \textbf{Dataset} & \textbf{Sparsity Type} & \textbf{Optimization} & \textbf{Domain} \\
        \midrule

cuSPARSE \cite{nvidia_cusparse} & GPU & \texttt{SpMM}, \texttt{SDDMM} & - & unstructured & Closed-source NVIDIA library & - \\
rocSPARSE \cite{amd_rocsparse} & GPU (AMD) & \texttt{SpMM}, \texttt{SDDMM} & - & unstructured & Open-source AMD library (GPU) & - \\
ASPT \cite{hong2019adaptive} & GPU, CPU & \texttt{SpMM}, \texttt{SDDMM} & SuiteSparse & unstructured & 2D tiling, column reordering & HPC \\
ASPT-RR \cite{jiang2020novel} & GPU & \texttt{SpMM}, \texttt{SDDMM} & SuiteSparse & unstructured & Clustering-based row reordering & HPC \\
dgSPARSE \cite{huang2020ge,dgsparse_lib} & GPU & \texttt{SpMM}, \texttt{SDDMM} & SuiteSparse & unstructured & CSR-based, coalesced row caching, coarse-grained warp merging & GNN \\
Sputnik \cite{gale2020sparse} & GPU & \texttt{SpMM}, \texttt{SDDMM} & DLMC & unstructured & 1D subwarp tiling, ROMA, row swizzle & CNN, Transformers \\
EC-SpMM \cite{lin2023ec} & GPU & \texttt{SpMM} & custom pruned weights & unstructured & Codegen, 2D tiling, matrix reordering & DNN \\
LO-SpMM \cite{lin2024spmm} & GPU & \texttt{SpMM} & custom pruned weights & unstructured & Codegen, hierarchical tiling, reordering, prefetching & DNN \\
SparseRT \cite{wang2020sparsert} & GPU & \texttt{SpMM}, \texttt{SpConv} & DLMC & unstructured & Codegen, inspector-executor framework, load balancing & CNN, Transformers \\
RoDe \cite{pang2024row} & GPU & \texttt{SpMM}, \texttt{SDDMM} & SuiteSparse, DLMC & unstructured & Row splitting (regular/residual), pipelined execution, ROMA & HPC, Transformers, CNN \\
DTC-SpMM \cite{fan2024dtc} & GPU (TC) & \texttt{SpMM} & custom pruned weights, SuiteSparse & unstructured & ME-TCF format, reordering, codegen & GNN \\
Acc-SpMM \cite{zhao2025acc} & GPU (TC) & \texttt{SpMM} & custom pruned weights, SuiteSparse & unstructured & BitTCF format, data-affinity reordering, load balancing & GNN \\
RS-SpMM \cite{hong2018efficient} & GPU & \texttt{SpMM}, \texttt{SpMV} & SuiteSparse & unstructured & Split matrix (heavy/light segments), streaming, thread coarsening & HPC \\
VectorSparse \cite{chen2021efficient} & GPU (TC) & \texttt{SpMM}, \texttt{SDDMM} & DLMC, ResNet & (un)structured & 1D Octet Tiling, column-vector sparse encoding & Transformers, CNN \\
Escoin \cite{chen2018escoin} & GPU & \texttt{SpConv} & custom pruned weights, SuiteSparse & unstructured & Direct sparse convolution (avoids im2col) & CNN \\
GNNOne \cite{gong2024gnnone} & GPU & \texttt{SpMM}, \texttt{SDDMM} & custom pruned weights, SuiteSparse & unstructured & Fused computations, inter-thread communication, scheduling & GAT \\
HC-SpMM \cite{li2025hc} & GPU & \texttt{SpMM} & SNAP, TUDataset, KONECT & unstructured & Hybrid core selection (TC/CUDA), kernel fusion & GNN \\
DA-SpMM \cite{dai2022heuristic} & GPU & \texttt{SpMM} & SuiteSparse & unstructured & Heuristic-driven algorithm selection, orthogonal design principles & HPC \\
Merge-SpMM \cite{yang2018design} & GPU & \texttt{SpMM} & SuiteSparse & unstructured & Merge-based row-splitting, load balancing & HPC \\
Magicube \cite{li2022efficient} & GPU (TC) & \texttt{SpMM}, \texttt{SDDMM} & DLMC, LRA transformer benchmark & structured & SR-BCRS format for 1D block sparsity, online transposition & Transformers \\
fuseGNN \cite{chen2020fusegnn} & GPU & \texttt{SpMM}, \texttt{SDDMM} & graph dataset & unstructured & Kernel fusion, kernel launch overhead minimization & GAT \\
FastSpMM \cite{wang2025fastspmm} & GPU (TC) & \texttt{SpMM} & SuiteSparse & unstructured & CST format, virtual-row-based merging, runtime optimizations & GNN \\
Flash-LLM \cite{xia2023flash} & GPU (TC) & \texttt{SpMM} & LLM models & unstructured & Tiled-CSL format, load-as-sparse/compute-as-dense approach & Transformers \\
SMaT \cite{okanovic2024high} & GPU (TC) & \texttt{SpMM} & SuiteSparse, synthetic banded matrices & unstructured & BCSR format, matrix permutation, computation/communication overlapping & HPC \\
Tiled-CSR \cite{xue2023releasing} & GPU (TC) & \texttt{SpMM} & DLMC-transformers & unstructured & Tiled-CSR format, row shuffling, 3D-grid tiling & Transformers \\
Activation Spconv \cite{kurtz2020inducing} & CPU & \texttt{SpConv} & ImageNet & unstructured & SIMD compression, vectorization, pipelining, blocking & CNN\\
GNNPilot \cite{hu2025gnnpilot} & GPU & \texttt{SpMM}, \texttt{SDDMM} & Suitesparse & unstructured & Neighbor/Bin packing (BIN\_CSR), dynamic parallelization, kernel fusion & GAT \\
FeatGraph \cite{hu2020featgraph} & GPU, CPU & \texttt{SpMM}, \texttt{SDDMM} & graph dataset & unstructured & TVM extension, feature-dimension tiling, hybrid partitioning & GAT \\
NM-SpMM \cite{ma2025nm} & GPU (STC) & \texttt{SpMM} & Llama (N:M‑pruned weight matrices) & semi-structured & Hierarchical blocking, sparsity-aware latency hiding & HPC \\
GNNAdvisor \cite{wang2021gnnadvisor} & GPU & \texttt{SpMM}, \texttt{SDDMM} & graph dataset & unstructured & Neighbor/dimension partitioning, warp-aware memory customization & GNN \\
nmSPARSE \cite{lin2023efficient} & GPU (STC) & \texttt{SpMM}, \texttt{SpMV} & BERT‑Large, OPT, synthetic & semi-structured & Vector-Wise/Block-Wise N:M formats, bank conflict elimination & Transformers, HPC \\
Jigsaw \cite{zhang2024jigsaw} & GPU (STC) & \texttt{SpMM} & DLMC & semi-structured & Hierarchical reorder-aware format, column reordering & Transformers, DNN \\
VENOM \cite{castro2023venom} & GPU (STC) & \texttt{SpMM} & HuggingFace & semi-structured & V:N:M format, column-wise pruning, Sparse TC & Transformers, DNN \\
FusedMM \cite{rahman2021fusedmm} & CPU & \texttt{SpMM}, \texttt{SDDMM} & SuiteSparse, NetworkRepository & unstructured & Kernel fusion, memory traffic minimization, locality awareness & GNN \\
TACO \cite{kjolstad2017tensor} & CPU & \texttt{SpMM}, \texttt{SDDMM} & - & unstructured & Compiler generated kernels, OpenMP directives & - \\
TACO-UCF \cite{won2023unified} & CPU, GPU & \texttt{SpConv} & DLMC, S3DIS \cite{armeni20163d}, SemanticKITTI \cite{behley2019semantickitti} & (un)structured & TACO extension for \texttt{SpConv} & CNN \\
MKL \cite{wang2014intel} & CPU & \texttt{SpMM} & - & unstructured & Closed-source Intel library & - \\
AOCL-Sparse \cite{amd_aocl} & CPU & \texttt{SpMM} & - & unstructured & Open-source AMD library (CPU) & - \\
SparseDNN \cite{wang2021sparsednn} & CPU & \texttt{SpMM}, \texttt{SpConv} & custom pruned matrices & unstructured & x86 codegen, row reordering, z-curve tiling & CNN \\
XNNPACK \cite{elsen2020fast} & CPU & \texttt{SpConv} & ImageNet & unstructured & Im2col sparse convolution, cache prefetching& CNN \\
J-stream \cite{kurt2020efficient} & CPU & \texttt{SpMM}, \texttt{SDDMM} & SuiteSparse & unstructured & Tiling based on matrix signatures, memory traffic minimization & HPC \\
MergePath-SpMM \cite{shan2023mergepath} & GPU & \texttt{SpMM}, \texttt{SpMV} & graph dataset & unstructured & Merge-path algorithm, nonzero-splitting load balancing & GNN \\
fused3S \cite{li2025fused3s} & GPU (TC) & \texttt{SpMM}, \texttt{SDDMM} & graph dataset & unstructured & MET-TCF based format, Kernel fusion (\texttt{SDDMM}-Softmax-\texttt{SpMM})& GAT \\
HR-SpMM \cite{wang2025hr} & GPU (CC, TC) & \texttt{SpMM} & DLMC, SuiteSparse & unstructured & Row splitting (short rows on CCs, long on TCs) & Transformers, HPC \\
Marlin \cite{frantar2025marlin} & GPU (STC) & \texttt{SpMM}, \texttt{SDDMM} & integrated in vLLM & semi-structured & Sparse TC, 4-bit weight quantization, mixed precision & Transformers \\
EC-SpMV \cite{lin2025toward} & GPU & \texttt{SpMV} & SparseGPT & unstructured & EC-CSR format, hierarchical blocking, delta indexing & Transformers \\
        \bottomrule
    \end{tabular}
}
\end{sidewaystable}
        
\newpage
%%%%%%%%%%%%%%%%%%%%%%%%%%%%%%%%%%%%%%%%%%%%%%%%%%%%%%%%%%%%%%%%%%%%%%%%%%%%%%
\section{Datasets for sparse inference research} \label{sec:data}
%%%%%%%%%%%%%%%%%%%%%%%%%%%%%%%%%%%%%%%%%%%%%%%%%%%%%%%%%%%%%%%%%%%%%%%%%%%%%%
In the previous section we presented numerous works that involve designing new hand-crafted or generated kernels to speedup sparse neural network inference. To evaluate their results, these works either incorporate their kernels into an existing framework like PyTorch and run a model end-to-end, or isolate parts of the model—for example, a single layer or operation—and test their kernel’s effectiveness specifically on that part of the model. Applying this to the most time consuming parts of the model allows them to generalize their findings to the full network. A critical issue in this process is to identify appropriate data for conducting experiments. Works that evaluate end-to-end performance over a model need to test their work on fully trained sparse models with acceptable accuracy to ensure that the shape and characteristics of the model’s sparsity are representative of real-world scenarios. On the other hand, works that focus on the performance of isolated kernels require internal data from the neural network, such as the sparse weight matrices and activations of particular layers, to test their approach.

In this section we examine the availability and characteristics of data that can support research and experimental assessment of performance engineering methods that target sparse DNNs. We present the characteristics of the DLMC (Deep Learning Matrix Collection) \cite{gale2020sparse} which serves as the standard matrix suite of pruned model layers, custom approaches for model pruning, the availability of sparse models in public repositories, sparsification approaches for semi-structured pruning, graphs that represent DNNs, attention masks and datasets relevant to sparse activation.

%%%%%%%%%%%%%%%%%%%%%%%%%%%%%%%%%%%%%%%%%%%%%%%%%%%%%%%%%%%%%%%%%%%%%%%%%%%%%%
\subsection{Deep Learning Matrix Collection (DLMC)} \label{sec:data:dlmc}
%%%%%%%%%%%%%%%%%%%%%%%%%%%%%%%%%%%%%%%%%%%%%%%%%%%%%%%%%%%%%%%%%%%%%%%%%%%%%%
The Deep Learning Matrix Collection (DLMC) is a curated dataset of sparse matrices designed specifically for benchmarking and evaluating sparse kernels in deep learning workloads introduced in \cite{gale2020sparse}. Unlike general-purpose collections such as SuiteSparse \cite{davis2011university}, DLMC consists of matrices extracted directly from the weight tensors of deep neural networks, resulting in sparsity patterns that closely reflect those found in real-world models. The matrices are generated using a variety of unstructured pruning techniques, including magnitude pruning, random pruning, $L_0$ regularization \cite{louizos2017learning}, and variational dropout \cite{molchanov2017variational}. The latter two are examples of training-time pruning methods.

The models used are a classic Transformer \cite{vaswani2017attention}, the foundational architecture for most Natural Language Processing tasks, and a ResNet-50, a widely adopted convolutional neural network in the field of Computer Vision. These two baseline models are well-established and typically sufficient for conducting basic experiments on DNN kernels. Additionally, the dataset creators provide essential metadata and statistics for each matrix, allowing researchers to analyze the underlying data characteristics. DLMC has become the staple benchmark for research on unstructured sparsity \cite{gale2020sparse, pang2024row, chen2021efficient, li2022efficient, xue2023releasing, zhang2024jigsaw}, serving as a reference point for evaluating sparse kernel performance as it is currently the only structured and publicly available dataset that can be readily used with custom kernels—eliminating the need for researchers to generate weight matrices and perform pruning themselves.

However, DLMC’s utility for modern kernel experiments could be considered more limited due to its age and scope. The models in the collection, such as the original Transformer and ResNet-50, date back to 2019 and 2020, failing to represent the exponential growth in scale seen in contemporary architectures. For instance, the original `Attention Is All You Need' Transformer featured a model dimension of 512, a feed-forward dimension of 2,048, and 8 attention heads—yielding weight matrices with sizes in the order of 106 parameters. In contrast, modern large language models (LLMs) such as LLaMA 3 405B employ a model dimension of 16,384 and a feed-forward dimension of 53,248 across 126 layers, resulting in individual feed-forward matrices containing in the order of 109 parameters. These models therefore operate at scales several orders of magnitude larger, with total parameter counts extending into the hundreds of billions, far beyond the representational range captured in DLMC.

As mentioned above, modern LLMs like Llama 3 \cite{dubey2024llama}, Qwen3 \cite{yang2025qwen3} and DeepSeek V3 \cite{liu2024deepseek} operate on projection and FFN matrices that dwarf those in the original Transformer. Apart from the different size, modern models also integrate techniques like Grouped‑Query Attention (GQA) \cite{ainslie2023gqa}, Multi‑Head Latent Attention (MLA) \cite{meng2025transmla}, MoE routing \cite{shazeer2017outrageously}, FlashAttention’s fused I/O‑aware kernels \cite{dao2022flashattention}, rotary/relative embeddings \cite{su2024roformer}, and speculative decoding \cite{leviathan2023fast}.
These innovations alter both the shape and execution flow—for example, GQA changes KV matrix dimensions and cache patterns, MLA applies low-rank compressions and runtime up‑sampling, and FlashAttention fuses \(QK^\top\), softmax, and matrix multiplication into streaming kernels. This means that we cannot treat inference as a vanilla sequence of dense \texttt{GEMM}s. Current DLMC benchmarks, built around small, sparsified Transformer and ResNet matrices, miss these real-world inference characteristics. If DLMC is to remain relevant for kernel designers and researchers, it must evolve; including sparse, pruned matrices sampled from these mega-scale LLMs and reflecting their grouped attention, compressed latent KV, fused execution, and dynamic sparsity patterns during inference—so that benchmarks match the reality of how state‑of‑the‑art models are actually computed. Furthermore, the dataset lacks matrices from key modern architectures like Diffusion Models, State-Space Models and MLP-Mixers and GNNs.

%%%%%%%%%%%%%%%%%%%%%%%%%%%%%%%%%%%%%%%%%%%%%%%%%%%%%%%%%%%%%%%%%%%%%%%%%%%%%%
\subsection{Customly pruned layers} \label{sec:data:custom_pruned_layers}
%%%%%%%%%%%%%%%%%%%%%%%%%%%%%%%%%%%%%%%%%%%%%%%%%%%%%%%%%%%%%%%%%%%%%%%%%%%%%%
Apart from the DLMC dataset, many studies rely on their own custom benchmarks and matrix collections. The methods used to create these diverse datasets can be broadly categorized into three approaches:
\begin{enumerate}
    \item \textbf{Pruned weights from real models:}  This is a popular trend, as it provides the most realistic sparsity patterns. Researchers take well-known models like ResNet-50, Transformer, BERT, MobileNet, EfficientNet, and various LLMs (e.g., OPT-30B) and apply specific pruning algorithms to their weight layers. The key benefit of this approach is that the resulting sparse matrices reflect the characteristics of a real-world neural network layer, including the distribution of nonzero values and the overall matrix dimensions. Papers like SparseRT \cite{wang2020sparsert}, Faster CNNs \cite{park2016faster}, Escoin \cite{chen2018escoin}, and Flash-LLM \cite{xia2023flash} all follow this strategy. These datasets are not just matrices; they are direct outputs of a model compression process, making them ideal for testing the efficiency of sparse kernels on realistic workloads.

    \item \textbf{Synthetic matrices:} This approach involves generating matrices from scratch with controlled parameters like size and sparsity level. While not derived from a pruned model, these matrices are often designed to mimic the structural characteristics of DNN layers. For example, EC-SpMM \cite{lin2023ec}  and LO-SpMM \cite{lin2024spmm} create synthetic matrices that are `DNN-inspired', meaning they have shapes and sparsity patterns that are representative of what you would find in deep learning models. This method offers the advantage of reproducibility and control, allowing researchers to isolate variables like matrix dimension, sparsity, and block size to understand their impact on kernel performance. SparseDNN
    \cite{wang2021sparsednn} is a clear example of this trend, where matrices are generated with specific sizes and sparsity percentages for systematic testing.

    \item \textbf{Hybrid and specialized approaches:}  Some research combines these methods or focuses on highly specific model types. For instance, SparTA \cite{zheng2022sparta} applies structured sparsity to models like BERT and HuBERT, which is a form of pruning that results in a more regular sparsity pattern than the unstructured variety found in DLMC. Similarly, Minuet \cite{yang2024minuet} focuses on sparse 3D CNNs and voxelized data, which introduces a different type of sparsity pattern relevant to 3D vision tasks. This shows a trend toward increasingly specialized datasets tailored to niche applications.

\end{enumerate}

In essence, the DLMC serves as a robust and standardized starting point, but the trend of creating custom datasets allows researchers to push the boundaries of sparse kernel research by exploring new models, algorithms, and sparsity patterns that are not yet covered by public collections. However, the lack of a standardized benchmarking approach may pose a challenge for progress in the field, as it complicates fair and consistent comparisons between sparse kernel implementations. Furthermore, many of these custom benchmarks fail to accurately recreate weight matrices or models that reflect real-world deep learning workloads, thereby limiting the practical relevance and generalizability of their results. 

%%%%%%%%%%%%%%%%%%%%%%%%%%%%%%%%%%%%%%%%%%%%%%%%%%%%%%%%%%%%%%%%%%%%%%%%%%%%%%
\subsection{Sparse models in public repositories} \label{sec:data:public_sparse_models}
%%%%%%%%%%%%%%%%%%%%%%%%%%%%%%%%%%%%%%%%%%%%%%%%%%%%%%%%%%%%%%%%%%%%%%%%%%%%%%
Public repositories such as Neural Magic's SparseZoo \cite{SparseZoo} and the Hugging Face Hub \cite{HuggingFaceHub} can serve as an ideal source of sparse datasets. However, it is worth noting that finding already sparse networks in model repositories is not very common. 
SparseZoo has historically been the most prominent resource, but as of 2025, it is no longer being actively updated, with the most recent state-of-the-art contribution being Llama3-8B-2:4. By contrast, platforms like Hugging Face Hub host a vast number of pre-trained models, but these are typically dense. 

In contrast, the Hugging Face Hub contains a much larger variety of pre-trained models, but dense models are the default: few models explicitly enforce sparsity. Notable exceptions include cases like: \texttt{Intel/bert-base-uncased-sparse-/1\_2}, which employs structured 1:2 sparsity. Also, the DeepSparse Sparse LLMs collection—hosted via Neural Magic’s presence on Hugging Face—includes some models where at least 50\% of the weights have been pruned.

One reason for the rarity of sparse models is that many model hubs prioritize broad compatibility and ease of use. Dense models are heavily supported by existing frameworks, hardware, and inference pipelines. Sparse inference runtimes like DeepSparse exist and can leverage pruned models, but their adoption is still limited by compatibility, performance trade-offs, and toolchain complexity.

%%%%%%%%%%%%%%%%%%%%%%%%%%%%%%%%%%%%%%%%%%%%%%%%%%%%%%%%%%%%%%%%%%%%%%%%%%%%%%
\subsection{Data for semi-structured sparsity} \label{sec:data:semistructured}
%%%%%%%%%%%%%%%%%%%%%%%%%%%%%%%%%%%%%%%%%%%%%%%%%%%%%%%%%%%%%%%%%%%%%%%%%%%%%%
N:M pruning has become a key technique for optimizing large-scale deep learning models, particularly where matrix multiplication operations dominate the computational time. Transformer-based architectures, especially in NLP, are the most common targets due to their heavy reliance on linear layers, with weight matrices from models such as \cite{devlin2019bert}, GPT \cite{mann2020language}, OPT \cite{zhang2022opt}, and LLaMA \cite{touvron2023llama} frequently used to construct N:M sparse datasets. Beyond these real-world models, researchers also consider broader model suites, including CNNs like ResNet \cite{he2016deep} and VGG \cite{simonyan2014very}, object detection models such as MaskRCNN \cite{dollar2017mask}, and even GANs \cite{goodfellow2020generative}, to demonstrate the general applicability of pruning methods. 

The standard approach for producing N:M sparse models follows a train, prune, finetune workflow: a dense model is first trained to convergence on the target task, its weights are then pruned to satisfy a specified N:M sparsity pattern, and finally the pruned model is fine-tuned or fully retrained to recover any lost accuracy. Research has also focused on moving beyond NVIDIA’s hardware-native 2:4 sparsity pattern (50\%) to explore more aggressive, arbitrary N:M ratios, including 2:8 (75\%), 2:16 (87.5\%), and even greater than 95\% sparsity, as investigated in 
\cite{lin2023efficient}, VENOM \cite{castro2023venom}, and NM-SpMM \cite{ma2025nm}. Additionally, different pruning granularities are being explored, from element-wise to vector-wise and block-wise N:M sparsity, with coarser granularities often yielding more regular memory access and higher performance. To support these experiments, frameworks such as NVIDIA’s Automatic Sparse Pruning (ASP) \cite{mishra2021accelerating} are frequently extended to handle arbitrary N:M patterns and granularities, enabling systematic evaluation of both sparsity and computational efficiency.

For controlled micro-benchmarking, synthetically generated matrices are often employed, allowing evaluation of kernel efficiency independently of any specific model architecture, as seen in studies like \cite{lin2023efficient} and \cite{castro2023venom}. We consider this a solid approach, since the actual distribution of nonzero elements in semi-structured sparsity does not influence performance in this case. What actually matters is the size of the matrix and the N, M factors. 

%%%%%%%%%%%%%%%%%%%%%%%%%%%%%%%%%%%%%%%%%%%%%%%%%%%%%%%%%%%%%%%%%%%%%%%%%%%%%%
\subsection{Graphs for GNNs} \label{sec:data:gnn_graphs}
%%%%%%%%%%%%%%%%%%%%%%%%%%%%%%%%%%%%%%%%%%%%%%%%%%%%%%%%%%%%%%%%%%%%%%%%%%%%%%
Evaluating performance on GNNs is rather straightforward. Since both GNN models and their input data are inherently sparse, there is no need for additional modifications such as pruning or retraining/fine-tuning. Moreover, most studies rely on well-established benchmark graph datasets that are already widely used, such as SuiteSparse \cite{davis2011university}, SNAP \cite{leskovec2016snap}, OGB \cite{hu2020open}, TUD \cite{morris2020tudataset}, allowing models to be run end-to-end and their performance measured directly. 

The sizes and densities of the graphs in the most popular datasets vary a lot. Measuring these two quantities is important since different sparse matrix techniques have additional costs for the sparsification of matrices and show different performance for different input densities. The sizes range from a few tens of nodes for graph classification (in molecular sciences), to a few thousands of nodes for node classification and link prediction (in citation datasets), to hundreds of thousands of nodes in knowledge graphs. In all datasets, the average number of edges per node is rarely above 10, resulting in densities of less than 1\% in most cases.

An important factor in graph-based learning tasks is whether the task involves processing a single large graph or many smaller, disjoint graphs:
\begin{enumerate}
    \item Single large graph scenarios (e.g., citation networks, social networks, web graphs): the dataset is represented as one connected or partially connected graph, with message passing occurring globally across the node space.
    \item Many small graph scenarios (e.g., bioinformatics, molecular graphs, protein structures): consist of multiple graphs that serve as independent connected components. Each graph corresponds to a distinct entity (e.g., a molecule), and message passing is restricted within each graph. During inference, these graphs are processed in batches, allowing the model to handle many graphs in parallel. Vectorized computations are still employed, with indexing ensuring message passing remains confined to each respective graph. In benchmark reporting, node and edge counts are typically averaged per graph.
\end{enumerate}

Another key feature in real-world datasets is the distribution of nonzero elements. There are two main cases: a) power-law distribution, where arbitrarily long `evil' rows are present, leading to workload imbalance and irregular memory accesses and b) structured distribution, where the number of nonzero elements per row does not vary significantly.

Some graph-related parameters here are the average and maximum node degree. When their values have small variance (i.e., max degree is relatively small), this indicates that the nonzeros are evenly distributed. For example, from the table of datasets (Table~\ref{data:gnn_datasets}), Cora, PubMed, Citeseer, artist, and amazon0505 are considered type 1 (with artist ranking first), while OVCAR-8H and PROTEINS are considered type 2.

While the input feature size provides a useful overview of the aggregation workload, it does not directly impact the actual computations involved. This is because very large initial feature sizes in a GNN are computationally expensive, so a projection layer (also called a combiner, feature projector, or encoder) is typically used to reduce the input feature dimension to a manageable hidden size before computation. Alternatively, some GNN architectures perform this projection implicitly within the first GNN layer, where the learnable weight matrix simultaneously handles dimensionality reduction and feature transformation. In both cases, the dimensionality relevant for \texttt{SpMM} is determined by the projected hidden size, rather than the raw input feature size. Typical hidden dimensions for GNN layers range from 16, 32, 64, 128, 512, and occasionally up to 1024, depending on model design, computational resources, and task complexity. The hidden dimension (embedding size) is treated as a hyperparameter during optimization, often selected through empirical tuning on validation performance, and is then used in the central Sparse Matrix-Matrix multiplication (\texttt{SpMM}) operation. This dimension determines the output size of intermediate GNN layers, balancing computational cost with model capacity, and serves as the key parameter for optimizing the \texttt{SpMM} kernel. The direct use of sparse, well-defined matrices along with their associated feature dimensions simplifies both performance evaluation and kernel design.

\begin{table}[t]
  \caption{Graph datasets used in bibliography}
  \label{data:gnn_datasets}
  \centering
  \resizebox{\textwidth}{!}{\begin{tabular}{llrrlll}
    \toprule
    \textbf{Title} & \textbf{Structure} & \textbf{Nodes} & \textbf{Edges} & \textbf{Feature size(initial)} & \textbf{Category} & \textbf{Source}\\
    \midrule
    Cora & Single graph & 2,708 & 5,278 & 1,433 (node) & citation & Planetoid \\
    PubMed & Single graph & 19.717 & 88.648 & 500 (node) & bioinformatics & Planetoid \\
    Reddit & Single graph & 232,965 & 114,615,892 & 602 (node) & internet-forum & - \\ 
    Reddit Binary & Multiple subgraphs & $\sim$429.6 & $\sim$995.5 & - & internet-forum & TUDortmund \\
    AmazonProducts & Single graph & 1,569,960 & 264,339,468 & 200 & e-shop & GraphSAINT \\
    ogbn-proteins & Multiple subgraphs & 132,534 & 39,561,252 & 8 (edge) & bioinformatics & OGB \\
    ogbn-mag & Single graph & 1,939,743 & 21,111,007 & 128 (node) & citation & OGB \\
    ogbn-arxiv & Single graph & 169,343 & 1,166,243 & 128 (node) & citation & OGB \\
    com-Youtube & Single graph & 1,138,499 & 2,990,443 & - & web-network & SNAP \\
    com-Orkut & Single graph & 3,072,441 & 117,185,083 & - & web-network & SNAP \\
    com-Amazon & Single graph & 334,863 & 925,872 & - & e-shop & SNAP \\
    Flickr & Single graph & 8,925 & 899,765 & - & web-network & GraphSAINT \\
    Harvard & Multiple subgraphs & 15,126 & 824,617 & - & citation & -  \\ 
    Citeseer & Single graph & 3.327 & 9.103 & 3,703 (node) & citation & SNAP \\
    sx-stackoverflow & Single graph & 2,601,977 & 63,497,050 & - & internet-forum & SNAP \\
    roadNet-CA & Single graph & 1,965,206 & 2,766,607 & - & road-network & SNAP \\
    web-BerkStan & Single graph & 685,230 & 7,600,595 & - & citation & SNAP \\
    artist & Single graph & 50,515 & 1,638,396 & - & web-network (?) & Kaggle \\
    amazon0505 & Single graph & 410,236	& 3,356,824 & - & e-shop & SNAP \\
    OVCAR-8H & Multiple subgraphs (200) & 46.67 & 48.70 & - & bioinformatics & TUDortmund \\
    PROTEINS & Multiple subgraphs (1,113) & 39.1 & 145.6 & 3 & bioinformatics & TUDortmund \\
    PPI & Multiple subgraphs & 2,245.3 & 61,318.4 & - & - & - \\ 
    ogbl-collab & Single graph & 235,868 & 1,285,465 & 128 (node) & citation & OGB \\
    ogbl-ppa & Single graph & 576,289 & 30,326,273 & 58 (node) & bioinformatics & OGB \\
    ogbl-ddi & Single graph & 4,267 & 1,334,889 & - & bioinformatics & OGB \\
    ogbn-products & Single graph & 2,449,029 & 61,859,140 & 100 (node PCA) & e-shop & OGB \\
    \bottomrule
  \end{tabular}}
\end{table}

%%%%%%%%%%%%%%%%%%%%%%%%%%%%%%%%%%%%%%%%%%%%%%%%%%%%%%%%%%%%%%%%%%%%%%%%%%%%%%
\subsection{Attention masks} \label{sec:data:attention_masks}
%%%%%%%%%%%%%%%%%%%%%%%%%%%%%%%%%%%%%%%%%%%%%%%%%%%%%%%%%%%%%%%%%%%%%%%%%%%%%%
As discussed in Section~\ref{sec:sparse:attention}, masks are either predetermined or learnable. In general, the scientific community has increasingly shifted its focus towards learnable and dynamic attention masks rather than predetermined ones. This trend reflects the growing demand for models that can adaptively modulate attention based on input context, achieving better trade-offs between efficiency and accuracy. Despite this growing interest, relatively little research has systematically investigated how these masks impact kernel performance, and consequently, there are few publicly available datasets or ready-to-use masks for testing sparse attention kernels.

Several prominent repositories accompany leading papers, offering either ready-to-use libraries or reference implementations that support dynamic sparse attention during both training and inference. Researchers typically rely on these tools to apply sparse attention schemes and measure model performance. These implementations are often integrated with popular deep learning frameworks such as PyTorch or TensorFlow, facilitating experimentation and model adaptation. For example, SpargeAttn \cite{zhang2025spargeattn} provides a modular framework for dynamically generating sparse masks during inference, using heuristics like cosine similarity and mean pooling. This toolkit allows plug-and-play integration with pre-trained models, enabling sparsity without retraining from scratch. A primary limitation is that achieving competitive accuracy generally requires relatively low sparsity ratios (around 50\%). Similarly, MInference \cite{jiang2024minference} integrates dynamic sparse attention into fine-tuning pipelines for large language models, reflecting a trend toward scalable, resource-efficient adaptation of foundation models.

Finally, it is important to note that these activations are typically applied in groups—semi-structured or structured—rather than in an unstructured, element-wise manner. As a result, the exact computations performed can vary depending on the chosen sparsity pattern.

%%%%%%%%%%%%%%%%%%%%%%%%%%%%%%%%%%%%%%%%%%%%%%%%%%%%%%%%%%%%%%%%%%%%%%%%%%%%%%
\subsection{Sparse Activations} \label{sec:data:activations}
%%%%%%%%%%%%%%%%%%%%%%%%%%%%%%%%%%%%%%%%%%%%%%%%%%%%%%%%%%%%%%%%%%%%%%%%%%%%%%
Recent work showcases that sparse activations tend to compress mostly into binary masks that extract which neurons are active (i.e., nonzero) at inference. To the best of our knowledge, no generic tool specifically crafted for revealing activation masks from numerous model architectures and numerous methods that create sparsity exists. That said, workarounds to extract sparse activations are possible. A universal technique is to intercept the outputs of particular layers during the forward pass and apply a threshold to the intermediate activations to generate binary masks. This technique is most suitable for sparsity that is induced by activation functions like ReLU \cite{mirzadeh2023relu} or by learned gate control \cite{song2024prosparse}.

In cases where sparsity is spatial or structure-induced at the level of convolutional layers,
masks are revealed by finding regions with zero-valued activations. If sparsity is implicitly exposed, such as with the `lazy neuron' effect with transformers \cite{li2022lazy}, the practical solution is profiling activation statistics across different inputs and thresholding their mean or variance. Likewise, in approaches employing regularization or noise injection \cite{kurtz2020inducing}, one could estimate sparse patterns by looking at activations across different batches and finding stably inactive components. These methods, however, depend on specific prerequisites, namely, access to the pre-trained model or to the training pipeline, including the source code, so that instrumentation during either training or inference can be implemented.

%%%%%%%%%%%%%%%%%%%%%%%%%%%%%%%%%%%%%%%%%%%%%%%%%%%%%%%%%%%%%%%%%%%%%%%%%%%%%%
\section{State of practice} \label{sec:sop}
%%%%%%%%%%%%%%%%%%%%%%%%%%%%%%%%%%%%%%%%%%%%%%%%%%%%%%%%%%%%%%%%%%%%%%%%%%%%%%

Deploying a sparse neural network is a complex engineering task that requires careful consideration of performance goals, target hardware, and model architecture. The process of model sparsification—including the selection of appropriate pruning techniques and activation schemes—is intrinsically linked to the deployment strategy. The primary objectives of introducing sparsity are almost always to reduce latency, memory footprint, and energy consumption. Towards these objectives, this section attempts to outline a selection of tools, libraries, and frameworks that support sparsity in DNN inference. It is important to acknowledge that the domain is extremely broad and fast evolving and therefore, this cannot be considered a complete or fixed list. Following Section \ref{sec:sparse} we divide our analysis into four subsections based on the source of sparsity, in order to determine which frameworks offer support for each category.

%%%%%%%%%%%%%%%%%%%%%%%%%%%%%%%%%%%%%%%%%%%%%%%%%%%%%%%%%%%%%%%%%%%%%%%%%%%%%%
\subsection{Utilizing sparsity in weights}
%%%%%%%%%%%%%%%%%%%%%%%%%%%%%%%%%%%%%%%%%%%%%%%%%%%%%%%%%%%%%%%%%%%%%%%%%%%%%%

%%%%%%%%%%%%%%%%%%%%%%%%%%%%%%%%%%%%%%%%%%%%%%%%%%%%%%%%%%%%%%%%%%%%%%%%%%%%%%
\subsubsection{Unstructured sparsity}
%%%%%%%%%%%%%%%%%%%%%%%%%%%%%%%%%%%%%%%%%%%%%%%%%%%%%%%%%%%%%%%%%%%%%%%%%%%%%%
Pruning a DNN model is quite straightforward with the support provided by  popular frameworks like PyTorch or TensorFlow. Once an unstructured sparse model has been produced, deployment needs to incorporate the proper sparse data structures and optimized kernels. Unlike the case of semi-structured sparsity discussed next, unstructured sparsity has no predictable pattern as the zeros are scattered across the weight tensors.
These irregular unstructured sparsity patterns require specifically designed sparse kernels to realize the potential performance gains. However, the specialized kernels provided by researchers are typically highly complex and not easily integrated into the current production-ready deep learning frameworks, which are primarily optimized for dense or highly regular data formats. Therefore the support for unstructured sparsity in deep learning inference is currently limited. In the following paragraphs we split our analysis into three sections: we first cover the mature, optimized kernel implementations in both CPUs and GPUs (see Section \ref{sec:sparse_kernels} for detailed analysis regarding research implementations), and then examine the high-level, end-to-end frameworks that offer comprehensive support for deploying sparsity.

\paragraph{Support for unstructured sparsity in CPUs}

The nature of unstructured sparsity is best suited for CPUs, yet due to the current dominance of GPUs and their overall performance benefits, the available support for sparsity in CPUs is limited. Key production and research tools in this area include the following:

\begin{itemize}
    \item While often outperformed by more specialized kernels, vendor libraries provide an easily integrated solution that still offers decent performance gains. Intel’s MKL \cite{wang2014intel} is a widely used vendor library that provides optimized implementations for sparse kernels like \texttt{SpMM}, \texttt{SpMV}, \texttt{SpGEMM}. The library provides highly optimized implementations that take into account the input matrix' structure using inpector-executor logic and are based on the CSR storage format, thus making them easily adjustable to any framework. Similarly, ARM performance libraries \cite{arm_pl} and  AMD's AOCL-Sparse \cite{amd_aocl} provide sparse linear algebra support for ARM and AMD CPUs respectively.
    \item LIBXSMM \cite{heinecke2016libxsmm} is an optimized, Just-In-Time (JIT) code generation library for specialized dense and sparse matrix operations, including small matrix multiplications and deep learning primitives like convolutions (with fusions like GEMM + activation).        
    \item Specialized research inference engines (e.g., SparseDNN) propose advanced techniques, such as code generation and optimized scheduling, specifically to accelerate unstructured sparse inference on CPUs. While generally not yet production-ready, these engines serve as valuable resources for studying new sparse algorithms and often provide highly improved prototype kernels.
    
\end{itemize}

\paragraph{Support for unstructured sparsity in GPUs}
Unstructured-sparse GPU acceleration presents a greater challenge because GPUs fundamentally favor massively parallel applications with structured and coalesced memory access. Thus, support is often more research-driven, though several industry-grade options exist:

\begin{itemize}
    \item NVIDIA cuSPARSE \cite{nvidia_cusparse} is the standard vendor library for sparse linear algebra on NVIDIA GPUs, offering optimized baseline implentations that are essential for many inference pipelines that rely on unstructured sparsity.
    \item ROCm stack (AMD) provides rocSPARSE \cite{amd_rocsparse} for sparse linear algebra on AMD GPUs.
    \item Triton (OpenAI) \cite{Triton} provides a standard for writing custom sparse kernels, allowing developers to produce highly performant CUDA kernels for unstructured sparsity without the need for extensive knowledge of CUDA programming or specific hardware details.
\end{itemize}

\paragraph{Support  for unstructured sparsity in high level frameworks}
While hardware support for unstructured sparsity is provided by specialized libraries and low-level kernels, high-level frameworks need to offer APIs and runtimes to deploy these techniques effectively:

\begin{itemize}
    \item PyTorch \cite{paszke2019pytorch} offers torch.sparse APIs and some semi-structured support, but general-purpose, high-performance unstructured sparse acceleration (especially on GPUs) is limited and often requires custom kernels or external runtimes.
    \item TensorFlow / Keras \cite{TensorFlow} provides tf.sparse.SparseTensor and sparse operations, but GPU inference acceleration for unstructured sparse neural network weights is limited in practice. Most TensorFlow production sparse workflows use structured patterns or CPU-specific optimizations.
    \item The OpenVINO Toolkit \cite{OpenVINO} offers an end-to-end approach for training, pruning and deploying deep learning models. It supports quantization and handles sparse weights with the Neural Network Compression Framework (NNCF), leveraging Sparse Weight Decompression in its runtime to reduce memory bandwidth requirements on compatible Intel hardware.
    \item ONNX \cite{ONNX} defines the SparseTensor prototype, but practical, widely-supported unstructured-sparse kernels are limited. The ONNX Runtime has started adding block-sparse kernels and special-case support, but general unstructured-sparse acceleration often requires custom operators or engine-specific extensions.
    \item DeepSparse (Neural Magic) \cite{DeepSparse} is an industrial approach for executing unstructured sparse models. DeepSparse implements a proprietary runtime specifically designed to skip zero-computations effectively on x86 architectures (AVX-512/AVX2). It executes unstructured pruned models significantly faster than dense baselines by keeping the active dataset entirely within the CPU caches hierarchy, effectively turning memory-bound workloads into compute-bound ones on commodity hardware.
    \item Hugging Face Transformers \cite{wolf2020transformers}  focus primarily on dense computations. Sparse acceleration is enabled via integration with external toolchains (TVM \cite{chen2018tvm}, DeepSparse, OpenVINO, or vendor runtimes) rather than native unstructured-sparse kernels.
\end{itemize}

%%%%%%%%%%%%%%%%%%%%%%%%%%%%%%%%%%%%%%%%%%%%%%%%%%%%%%%%%%%%%%%%%%%%%%%%%%%%%%
\subsubsection{Semi-structured sparsity}
%%%%%%%%%%%%%%%%%%%%%%%%%%%%%%%%%%%%%%%%%%%%%%%%%%%%%%%%%%%%%%%%%%%%%%%%%%%%%%
\paragraph{N:M Sparsity}

Once a model has been pruned to an N:M pattern, deployment becomes primarily a matter of aligning the model’s format and runtime execution with the capabilities of the target hardware. While training and fine-tuning determine the quality and stability of the sparse model, deployment determines whether that sparsity actually translates into measurable efficiency gains.

Modern GPUs, including NVIDIA’s Ampere and Hopper architectures as well as AMD’s RDNA 4 generation, are currently the main execution targets for N:M sparse models. NVIDIA platforms provide native support for 2:4 sparsity through Sparse Tensor Cores, while AMD supports 4:2 structured sparsity. The approaches are practically the same. On such hardware, the runtime recognition of fixed patterns allows execution units to skip zero multiplications efficiently, achieving up to a 2× improvement in throughput for matrix multiplications and roughly a 50\% reduction in memory footprint without additional software overhead.

In practice, deployment requires exporting the model with its block masks or compressed sparse representation intact, often using framework-level support such as PyTorch’s torch.sparse utilities or NVIDIA’s Automatic Sparse Pruning (ASP) extensions. The key step is ensuring that the sparse format used during training matches the runtime format expected by the inference engine, typically a compressed layout optimized for the specific 2:4 or 4:2 pattern supported by the target GPU as discussed in Section~\ref{sec:background:sparsity_basics:semi_structured}. For research-oriented or custom accelerators, alternative ratios like 2:8 or 2:16 can be used, provided that corresponding sparse kernels are implemented in CUDA, Triton, or specialized runtime backends.

Standard CPUs lack specific support for this kind of sparsity. Consequently, when an N:M model is deployed on a CPU, the runtime must handle it either as unstructured sparsity with the generic representations and kernels of this case, or as dense computations that can utilize the resident SIMD support. Therefore, outside of aforementioned GPUs environments, N:M sparsity currently serves primarily as a storage and memory bandwidth optimization rather than a computational accelerator.

\subparagraph{Support for N:M sparsity in GPUs}
\begin{itemize}
    \item TensorRT \cite{TensorRT} (and TensorRT-LLM \cite{TensorRT_LLM}) is NVIDIA's high-performance inference optimizer. The TensorRT Model Optimizer includes modules like modelopt.torch.sparsity that enable sparsification of PyTorch models into the 2:4 N:M format and other patterns. The resulting optimized model is then run by TensorRT-LLM which uses highly-fused CUDA kernels to leverage the Sparse Tensor Cores for peak throughput.
    \item Core PyTorch APIs lack native N:M kernel acceleration. However, high-performance sparse workflows are enabled through extensions like NVIDIA's TensorRT Model Optimizer (which works as a PyTorch extension) or torch.ao (PyTorch's Acceleration Optimizer), which is building towards a unified sparse inference framework. The common workflow involves training and pruning in PyTorch, then converting and optimizing with TensorRT tools.
    \item vLLM \cite{kwon2023efficient} is a highly efficient LLM serving engine. It is not natively built around N:M sparsity like TensorRT-LLM. However, there exist specialized versions or integrations. For instance, Neural Magic's nm-vllm package \cite{nm_vllm} extends vLLM with custom CUDA kernels to support sparse models (including N:M pruned models) and deliver acceleration, combining vLLM's efficient attention mechanisms (PagedAttention) with sparse computation.
\end{itemize}

\paragraph{Block sparsity}
Block sparsity splits weight matrices into fixed-size blocks and prunes whole blocks, preserving local structure and improving memory locality compared to unstructured sparsity. A common workflow is to apply a block mask, and convert the masked weights to a Block Sparse Row (BSR) layout such as PyTorch’s Tensor.to\_sparse\_bsr(), which stores only nonzero blocks and their indices. During inference, frameworks with BSR-aware kernels can execute these weights efficiently, either via specialized block-sparse kernels (i.e., DeepSpeed’s Sparse Attention Kernels \cite{deepspeed_sparse_attention}) or by breaking the work into many small dense matrix multiplications.

%%%%%%%%%%%%%%%%%%%%%%%%%%%%%%%%%%%%%%%%%%%%%%%%%%%%%%%%%%%%%%%%%%%%%%%%%%%%%%
\subsubsection{Structured sparsity}
%%%%%%%%%%%%%%%%%%%%%%%%%%%%%%%%%%%%%%%%%%%%%%%%%%%%%%%%%%%%%%%%%%%%%%%%%%%%%%
Structured sparsity offers the simplest deployment pathway.
While the pruning procedure itself may be complex, deployment is trivial: after pruning, the model is restructured into a smaller dense network. For example, pruning 128 out of 512 convolutional filters produces a new layer with 384 filters that operates like any standard dense layer.
From an engineering standpoint, structured sparsity is universally compatible. The resulting model can be deployed on any hardware platform, CPUs, GPUs, TPUs, mobile, or edge devices, and runs without modification in standard inference frameworks such as PyTorch, TensorFlow, or ONNX runtime.
Structured sparsity is widely used when deploying models on resource-constrained devices or whenever a straightforward, low-effort performance improvement is needed without altering the inference pipeline.

%%%%%%%%%%%%%%%%%%%%%%%%%%%%%%%%%%%%%%%%%%%%%%%%%%%%%%%%%%%%%%%%%%%%%%%%%%%%%%
\subsection{Activation sparsity}
%%%%%%%%%%%%%%%%%%%%%%%%%%%%%%%%%%%%%%%%%%%%%%%%%%%%%%%%%%%%%%%%%%%%%%%%%%%%%%
Since the zero elements in activation sparsity are input-dependent, skipping zero operations must occur dynamically at runtime. This typically requires hardware or kernels capable of detecting and bypassing inactive neurons on the fly, a capability available mainly in specialized ASICs and FPGA accelerators rather than general-purpose CPUs or GPUs.
Nevertheless, recent research has shown that dynamic activation sparsity can still be partially exploited on standard hardware. Prototype systems  demonstrate that early CPU-oriented sparse-convolution libraries have achieved measurable speedups by skipping zero activations \cite{kurtz2020inducing}, while recent work on large transformers (e.g., ReLU-based LLM variants \cite{mirzadeh2023relu}, activation-predictive pruning, and lightweight sparsity predictors \cite{liu2023deja, shin2025sparseinfer, zhang2025rank}) has shown that dynamic sparsity patterns can be estimated or enforced well enough to accelerate inference with negligible accuracy impact. Although these approaches are not yet mainstream or universally supported, they indicate a growing trajectory toward practical activation-sparsity-aware inference on commodity hardware.

%%%%%%%%%%%%%%%%%%%%%%%%%%%%%%%%%%%%%%%%%%%%%%%%%%%%%%%%%%%%%%%%%%%%%%%%%%%%%%
\subsection{Attention sparsity}
%%%%%%%%%%%%%%%%%%%%%%%%%%%%%%%%%%%%%%%%%%%%%%%%%%%%%%%%%%%%%%%%%%%%%%%%%%%%%%
In Transformer models, sparsity can be introduced into the attention mechanism to reduce its quadratic complexity. Architectures like Longformer \cite{beltagy2020longformer} and BigBird \cite{zaheer2020bigbird} use structured attention patterns (e.g., sliding-window, global tokens, random blocks) to approximate dense attention efficiently. 
These models are natively supported in Hugging Face Transformers \cite{wolf2020transformers} and can be deployed with minimal changes. 
For high-performance inference, tools like DeepSpeed \cite{aminabadi2022deepspeed} (with Triton-based block-sparse kernels \cite{deepspeed_sparse_attention}), xFormers \cite{lefaudeux2022xformers} and vLLM offer optimized support for various sparsity patterns. These libraries reduce memory usage and latency, enabling scalable deployment of sparse attention models on modern hardware.

%%%%%%%%%%%%%%%%%%%%%%%%%%%%%%%%%%%%%%%%%%%%%%%%%%%%%%%%%%%%%%%%%%%%%%%%%%%%%%
\subsection{GNN sparsity}
%%%%%%%%%%%%%%%%%%%%%%%%%%%%%%%%%%%%%%%%%%%%%%%%%%%%%%%%%%%%%%%%%%%%%%%%%%%%%%
In practice, deploying GNNs is heavily dependent on specialized frameworks such as PyTorch Geometric (PyG) \cite{fey2019fast} and Deep Graph Library (DGL) \cite{wang2019deep}, which provide optimized CUDA and CPU kernels tailored for graph operations. From an engineering perspective, the key responsibilities are ensuring that the production system uses these optimized kernels and that graph data is represented efficiently in a sparse format. However, there are trade-offs: not all GNN layers have fused kernels, and for very large graphs or latency-critical systems, you may need custom tuning (e.g., JIT compilation, multi-core runtime systems, or distributed GNN frameworks).

%%%%%%%%%%%%%%%%%%%%%%%%%%%%%%%%%%%%%%%%%%%%%%%%%%%%%%%%%%%%%%%%%%%%%%%%%%%%%%
\section{Evaluation} \label{sec:eval}
%%%%%%%%%%%%%%%%%%%%%%%%%%%%%%%%%%%%%%%%%%%%%%%%%%%%%%%%%%%%%%%%%%%%%%%%%%%%%%

%%%%%%%%%%%%%%%%%%%%%%%%%%%%%%%%%%%%%%%%%%%%%%%%%%%%%%%%%%%%%%%%%%%%%%%%%%%%%%
%\subsection{Introduction} \label{sec:eval:intro}
%%%%%%%%%%%%%%%%%%%%%%%%%%%%%%%%%%%%%%%%%%%%%%%%%%%%%%%%%%%%%%%%%%%%%%%%%%%%%%
In this section, we evaluate the performance of the previously discussed \texttt{SpMM} and \texttt{SDDMM} kernels, focusing exclusively on those that are both published and open-source\footnote{The repository, including all kernel implementations and benchmarking scripts, is available at \url{https://github.com/pmpakos/sparse_survey_evaluation}.}. We focus on those kernels as they currently dominate efforts as discussed in Section~\ref{sec:sparse_kernels}. The interested reader can find a recent evaluation for \texttt{SpMV} in HPC setups in \cite{mpakos2023feature}. The primary goal of this evaluation is to benchmark all kernels within a unified experimental framework, enabling a fair and consistent comparison of their performance. Beyond that, we aim to explore the computational capabilities of both GPU and CPU platforms by examining the extent to which each kernel approaches peak performance on its respective architecture. Another objective is to highlight the performance variability across different types of matrix datasets; particularly comparing graph-based matrices from the SuiteSparse collection against dense-like matrices from the DLMC repository. We also investigate how structural properties of the matrices influence kernel performance, seeking to identify any strong correlations between matrix characteristics and computational efficiency. Lastly, this section reflects on the practical aspects of the evaluation process itself. We assess the setup complexity involved in running each kernel, noting that some implementations require minimal effort, while others demand more extensive preparation and tuning.

%%%%%%%%%%%%%%%%%%%%%%%%%%%%%%%%%%%%%%%%%%%%%%%%%%%%%%%%%%%%%%%%%%%%%%%%%%%%%%
\subsection{Experimental setup} \label{sec:eval:setup}
%%%%%%%%%%%%%%%%%%%%%%%%%%%%%%%%%%%%%%%%%%%%%%%%%%%%%%%%%%%%%%%%%%%%%%%%%%%%%%
In our experiments, we use server-class CPU and GPU; a 24-core AMD EPYC 7402 CPU and an NVIDIA A100 GPU. A detailed presentation of each platform is presented in Table~\ref{eval:experimental_setup}. The 24-core EPYC processor is composed of multiple chiplets, referred to as Core Complex Dies (CCDs), with each CCD containing 3 cores that share the L3 cache. Additionally, the CPU is configured with a single NUMA domain (NPS1). Threads are pinned to cores using OpenMP environment variables, and matrices are initialized in parallel using the Linux first-touch policy, ensuring that data is allocated in the NUMA node local to the thread accessing it. For each testbed, we calculate the memory bandwidth using the STREAM benchmark and report it alongside the nominal bandwidth. For each configuration (testbed, matrix, and format), we execute 128 iterations of \texttt{SpMM} and \texttt{SDDMM}, and record the average performance in GFLOPs. Subsequently, we calculate the arithmetic mean across three independent experiments to ensure robustness. 

\begin{table}[t]
    \caption{Experimental setup}
    \label{eval:experimental_setup}
    \begin{center}
        \begin{tabular}{ccc}
        \toprule
            \textbf{Testbed}     & \textbf{NVIDIA A100 (GPU)} & \textbf{AMD EPYC 7402 (CPU)} \\
        \midrule
            \textbf{Cores}       & 6912 CUDA cores            & 24 cores \\
        \midrule
            \textbf{Memory}      & 40 GB HBM2                 & 128 MB LLC (L3) \\
                                 & (BW 1555 GB/s)             & 256 GB DDR4 \\
        \midrule
            \textbf{Measured BW} & 1350 GB/s                  & DRAM 77 GB/s \\
        \midrule
            \textbf{Compiler}    & cuda-12.5                  & gcc 12.2.0 \\
        \bottomrule
        \end{tabular}
    \end{center}
\end{table}

%%%%%%%%%%%%%%%%%%%%%%%%%%%%%%%%%%%%%%%%%%%%%%%%%%%%%%%%%%%%%%%%%%%%%%%%%%%%%%
\subsection{Datasets} \label{sec:eval:datasets}
%%%%%%%%%%%%%%%%%%%%%%%%%%%%%%%%%%%%%%%%%%%%%%%%%%%%%%%%%%%%%%%%%%%%%%%%%%%%%%
The first dataset comprises the graph collection described in Section~\ref{sec:data:gnn_graphs}, Table~\ref{data:gnn_datasets}. This dataset spans a wide range of graph sizes, from a few thousand to several hundred thousand nodes, and from approximately 9K up to 40M edges. Due to hardware limitations, not all graphs could be included in the experiments, as some exceeded the memory capacity of a single device. Ultimately, 18 representative graphs were selected, as listed in Table~\ref{eval:graph_dataset}. These graphs originate from diverse domains such as bioinformatics, social and communication networks, and citation graphs. The majority are extremely sparse (99.9\% sparsity), with only one exception exhibiting 94\% sparsity. We utilize these graphs to assess the performance of the available \texttt{SpMM} and \texttt{SDDMM} kernels.

\begin{table}[t]
    \caption{Graph matrices used for evaluation}
    \label{eval:graph_dataset}
    \begin{center}
        \begin{tabular}{crrrr}
        \toprule
            \textbf{Matrix} & \textbf{Rows} & \textbf{Nonzeros} & \textbf{Sparsity (\%)} & \textbf{Size (MB)}\\
        \midrule
            citeseer & 3327 & 9340 & 99.9156 & 0.08 \\
            cora & 2708 & 10858 & 99.8519 & 0.09 \\
            pubmed & 19717 & 88673 & 99.9772 & 0.75 \\
            PROTEINS & 43471 & 162088 & 99.9914 & 1.40 \\
            ogbl-ddi & 4267 & 1067911 & 94.1347 & 8.16 \\
            ogbl-collab & 235868 & 967632 & 99.9983 & 8.28 \\
            ogbn-arxiv & 169343 & 1166243 & 99.9959 & 9.54 \\
            harvard & 15126 & 1649234 & 99.2792 & 12.64 \\
            com-Amazon & 334863 & 1851744 & 99.9983 & 15.41 \\
            REDDIT-BINARY & 859254 & 1991016 & 99.9997 & 18.47 \\
            amazon0505 & 410236 & 3356824 & 99.9980 & 27.18 \\
            OVCAR-8H & 1890931 & 3946402 & 99.9999 & 37.32 \\
            wiki-Talk & 2394385 & 5021410 & 99.9999 & 47.44 \\
            roadNet-CA & 1971281 & 5533214 & 99.9999 & 49.73 \\
            com-Youtube & 1134890 & 5975248 & 99.9995 & 49.92 \\
            web-BerkStan & 685230 & 7600595 & 99.9984 & 60.60 \\
            sx-stackoverflow & 2601977 & 36233450 & 99.9995 & 286.36 \\
            ogbn-proteins & 132534 & 39561252 & 99.7748 & 302.33 \\
        \bottomrule
        \end{tabular}
    \end{center}
\end{table}

The second dataset is drawn from the DLMC benchmark suite, presented in Section~\ref{sec:data:dlmc} and originally introduced in \cite{gale2020sparse}. These matrices are derived from deep learning workloads, specifically from the weight tensors of neural networks subjected to unstructured pruning. The pruning strategies include magnitude pruning, random pruning, L0 regularization, and variational dropout, all of which induce sparsity patterns representative of modern deep learning models. The characteristics of these matrices are summarized in Table~\ref{eval:dlmc_dataset}, with a total of 2680 matrices being available. For the \texttt{SpMM} kernels, we directly employ the DLMC matrices. For the \texttt{SDDMM} kernels, however, we adopt a different strategy. Instead of weight-derived sparsity, we generate attention masks following the methodology described in \cite{zaheer2020big} and Section~\ref{sec:sparse:attention}. Specifically, we construct random, windowed, and global attention masks, as well as the BigBird mask that combines these three patterns. The defining features of these masks, such as the number of rows and target sparsity levels, are used as input to the generator and correspond to the values reported in the DLMC dataset, leading to the generation of 84 masks.

\begin{table}[t]
    \caption{DLMC dataset used for evaluation}
    \label{eval:dlmc_dataset}
    \begin{center}
        \begin{tabular}{cc}
        \toprule
            \textbf{Features} & \textbf{Values}\\
        \midrule
            \textbf{Rows} & 512, 2048 \\
        \midrule
            \textbf{Nonzeros} & 250 - 673K \\
        \midrule
            \multirow{2}{*}{\textbf{Sparsity (\%)}} & 50, 60, 70, 80,\\
                                                    & 90, 95, 98\\
        \bottomrule
        \end{tabular}
    \end{center}
\end{table}

%%%%%%%%%%%%%%%%%%%%%%%%%%%%%%%%%%%%%%%%%%%%%%%%%%%%%%%%%%%%%%%%%%%%%%%%%%%%%%
\subsection{Formats and algorithms examined} \label{sec:eval:formats}
%%%%%%%%%%%%%%%%%%%%%%%%%%%%%%%%%%%%%%%%%%%%%%%%%%%%%%%%%%%%%%%%%%%%%%%%%%%%%%
Our performance evaluation contains publicly available implementations provided both by hardware vendors (Intel, AMD, and NVIDIA) and by the broader research community.

All experiments are conducted using single-precision floating point (FP32), as this is the most universally supported precision level across the selected formats. Formats operating on alternative precisions—such as bfloat16, FP16, or INT8—are excluded from this study due to limited and inconsistent support across platforms and implementations. 

To ensure a fair comparison of computational performance, we deliberately exclude any preprocessing overheads associated with specific formats. This decision allows us to focus purely on the runtime performance of \texttt{SpMM} and \texttt{SDDMM} operations, without skewing results due to format-specific setup or conversion stages.

In terms of kernel support across platforms, we observe a significant imbalance in format availability as discussed in Section~\ref{sec:sparse_kernels}. On GPUs, a wide variety of \texttt{SpMM} implementations are available, offering flexibility in both format and optimization strategy. In contrast, CPU-based \texttt{SpMM} support is more limited, with only four available. We note that for three GPU \texttt{SpMM} kernels (dgSPARSE, GNNPilot, DTC) multiple versions are available. After testing with both datasets, we select and present results for the best performing version of each format.

The disparity in kernel support is even more pronounced for \texttt{SDDMM} kernels. While several GPU-based implementations exist, CPU support is extremely limited. In fact, only one usable \texttt{SDDMM} kernel is available on CPU, the implementation from ASpT \cite{hong2019adaptive}. Although the FusedMM library \cite{rahman2021fusedmm} includes an \texttt{SDDMM} operator, it is tightly integrated with its \texttt{SpMM} routine, preventing us from isolating and benchmarking the \texttt{SDDMM} kernel independently. As a result, FusedMM's \texttt{SDDMM} component is excluded from our evaluation. 
The complete mapping of supported formats for each kernel and platform is summarized in Table~\ref{eval:formats_tested}.

\begin{table}[t]
    \caption{Formats and implementations tested on each testbed}
    \label{eval:formats_tested}
    \begin{center}
        \begin{tabular}{cccc}
        \toprule
            \multicolumn{2}{c}{\textbf{NVIDIA A100 (GPU)}} & \multicolumn{2}{c}{\textbf{AMD EPYC 7402 (CPU)}} \\
            \texttt{SpMM} & \texttt{SDDMM} & \texttt{SpMM} & \texttt{SDDMM} \\
        \midrule
            cuSPARSE \cite{nvidia_cusparse} & cuSPARSE \cite{nvidia_cusparse} & Intel-MKL \cite{wang2014intel} & ASpT \cite{hong2019adaptive} \\
            ASpT \cite{hong2019adaptive} & ASpT \cite{hong2019adaptive} & AOCL-Sparse \cite{amd_aocl} & \\
            RoDe \cite{pang2024row} & dgSPARSE \cite{dgsparse_lib} & ASpT \cite{hong2019adaptive} & \\
            dgSPARSE \cite{huang2020ge,dgsparse_lib} & GNNPilot \cite{hu2025gnnpilot} & FusedMM \cite{rahman2021fusedmm} & \\
            GNNPilot \cite{hu2025gnnpilot} & & & \\
            DTC-SpMM \cite{fan2024dtc} & & & \\
            Sputnik \cite{gale2020sparse} & & & \\
        \bottomrule
        \end{tabular}
    \end{center}
\end{table}

Our evaluation is carried out across multiple inner dimensions of the matrices involved in the computations, specifically the sequence length, with three representative values examined: 32, 128, and 1024.

%%%%%%%%%%%%%%%%%%%%%%%%%%%%%%%%%%%%%%%%%%%%%%%%%%%%%%%%%%%%%%%%%%%%%%%%%%%%%%
\subsection{Results and analysis} \label{sec:eval:results}
%%%%%%%%%%%%%%%%%%%%%%%%%%%%%%%%%%%%%%%%%%%%%%%%%%%%%%%%%%%%%%%%%%%%%%%%%%%%%%
Before proceeding with the analysis of the evaluation results, we note how the results are presented and interpreted. All figures in this section are shown as boxplots, which aggregate the performance measurements of all matrices belonging to each category shown on the x-axis. These plots provide an overview of the performance trends, while also illustrating the median performance and the variability across different matrices within each category.

In addition, it is important to highlight a key consideration regarding the CPU measurements. The performance of the CPU kernels is strongly influenced by the total problem size in relation to the capacity of the Last-Level Cache (LLC). By problem size, we refer to the combined memory footprint of the sparse matrix and the dense matrices involved in the computations. When the working set fits within the LLC, data reuse is maximized and memory access overheads are minimized, enabling higher performance. However, once the working set exceeds the LLC capacity, frequent accesses to main memory dominate execution, leading to a significant drop in performance. This effect is particularly relevant for the graph dataset, where the total problem size may fall either below or above the LLC threshold depending on the embedding dimension (k). In contrast, the DLMC matrices are relatively small, resulting in problem sizes that consistently remain within the LLC across all cases.

%%%%%%%%%%%%%%%%%%%%%%%%%%%%%%%%%%%%%%%%%%%%%%%%%%%%%%%%%%%%%%%%%%%%%%%%%%%%%%
\subsection{SpMM} \label{sec:eval:spmm}
\begin{figure}[tb]
    \centering
    \includegraphics[width=\linewidth]{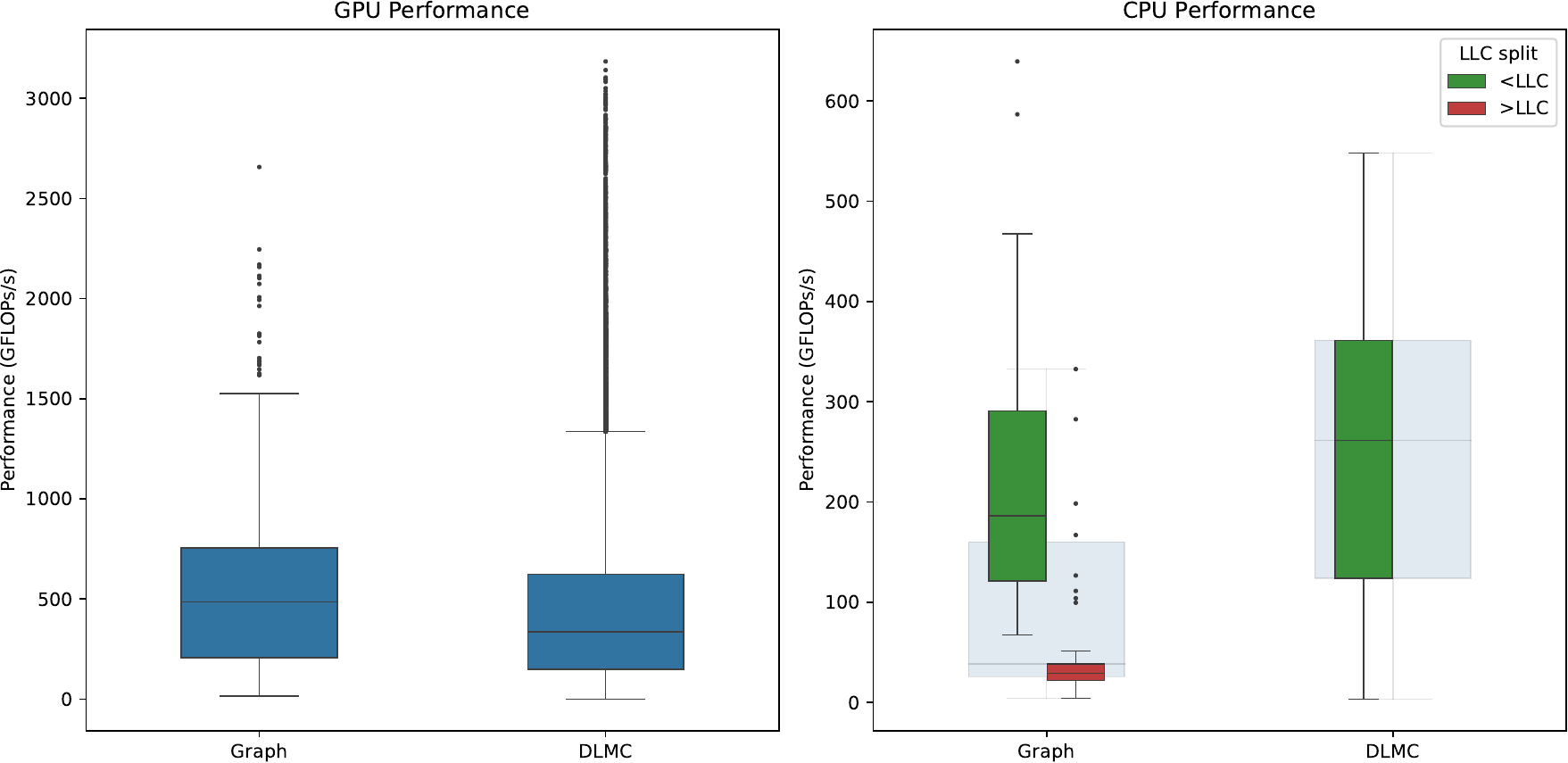}
    \caption{Performance of the \texttt{SpMM} kernel for the two different datasets on GPU and CPU (embedding dimension k=128). The CPU results are split in two boxplots, for smaller- and larger-than-LLC problem sizes, as LLC affects results of CPU performance by a great factor.}
    \label{fig:eval_spmm}
\end{figure}

%%%%%%%%%%%%%%%%%%%%%%%%%%%%%%%%%%%%%%%%%%%%%%%%%%%%%%%%%%%%%%%%%%%%%%%%%%%%%%
Figure~\ref{fig:eval_spmm} presents the performance of the \texttt{SpMM} kernel across the two datasets. On the GPU, we observe comparable average performance for both graph and DLMC matrices, with only a small fraction of DLMC cases (fewer than 5\%) achieving noticeably higher performance. It is worth emphasizing that the theoretical peak performance of the \texttt{GEMM} kernel on the evaluated GPU is 18 TFLOPs/s. In comparison, the \texttt{SpMM} kernel attains on average only about 2.5\% of this peak, with rare cases reaching up to 20\%.

The CPU exhibits a different trend. Here, \texttt{SpMM} performance on the DLMC matrices is, on average, 1.5$\times$ higher than on graph matrices, for problem sizes smaller than the LLC. Regarding larger than LLC problem sizes, which are observed in graph matrices only, we observe a large drop in performance, 6$\times$ on average, which shows the large effect that the LLC size has on the attainable performance. The theoretical peak performance of the \texttt{GEMM} kernel on the CPU is 1.4 TFLOPs/s. Relative to this bound, the DLMC matrices achieve approximately 20\% of peak performance on average, with some matrices approaching the upper limit. In contrast, the graph matrices remain far behind, reaching about 13\% of peak performance on average.

%%%%%%%%%%%%%%%%%%%%%%%%%%%%%%%%%%%%%%%%%%%%%%%%%%%%%%%%%%%%%%%%%%%%%%%%%%%%%%
\subsubsection{Format evaluation} \label{sec:eval:spmm:format}
\begin{figure}[tb]
    \centering
    \includegraphics[width=\linewidth]{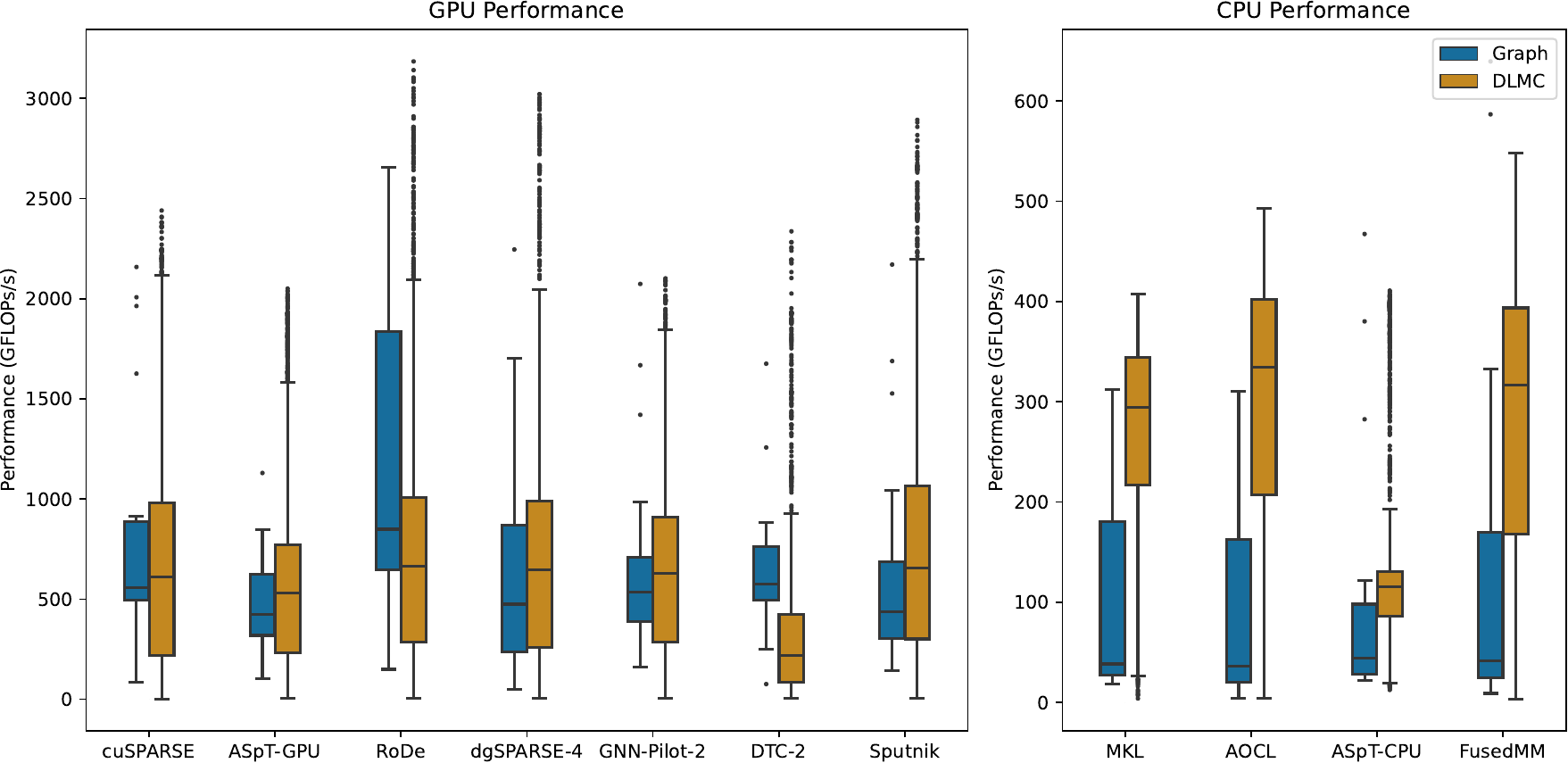}
    \caption{Performance of the evaluated \texttt{SpMM} formats for the two different datasets on GPU and CPU (embedding dimension k=128).}
    \label{fig:eval_spmm_format}
\end{figure}

%%%%%%%%%%%%%%%%%%%%%%%%%%%%%%%%%%%%%%%%%%%%%%%%%%%%%%%%%%%%%%%%%%%%%%%%%%%%%%
Figure~\ref{fig:eval_spmm_format} presents the performance of the various \texttt{SpMM} formats and implementations evaluated in this study. Across both matrix datasets, similar performance trends are observed. Vendor-provided formats, such as cuSPARSE (NVIDIA), MKL (Intel), and AOCL (AMD), deliver stable and consistent performance across all cases. However, these implementations do not achieve the highest attainable performance, exhibiting relatively limited performance ceilings compared to other evaluated formats. All formats exhibit considerable performance variability, with RoDe, dgSPARSE, DTC, and Sputnik for the GPU showing particularly large fluctuations on the DLMC matrices. We also observe that no format is particularly dominant in performance compared to the others, with the exception of the RoDe format for the graph dataset. 

%%%%%%%%%%%%%%%%%%%%%%%%%%%%%%%%%%%%%%%%%%%%%%%%%%%%%%%%%%%%%%%%%%%%%%%%%%%%%%
\subsubsection{Sparsity effect on performance} \label{sec:eval:spmm:sparsity}
\begin{figure}[tb]
    \centering
    \includegraphics[width=\linewidth]{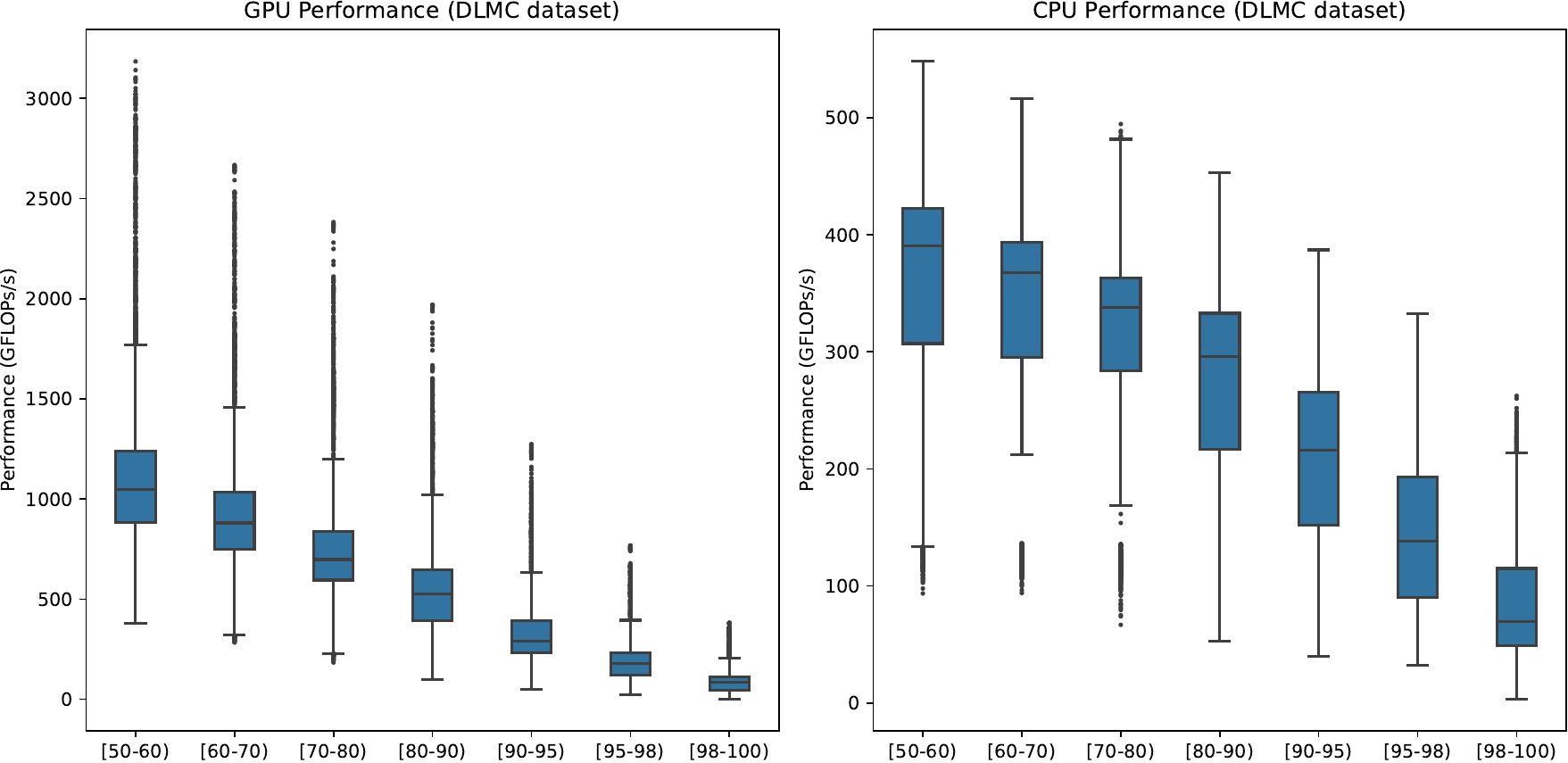}
    \caption{Effect of sparsity levels of the DLMC matrices on the performance of \texttt{SpMM} on the GPU and the CPU (embedding dimension k=128).}
    \label{fig:eval_spmm_sparsity}
\end{figure}

%%%%%%%%%%%%%%%%%%%%%%%%%%%%%%%%%%%%%%%%%%%%%%%%%%%%%%%%%%%%%%%%%%%%%%%%%%%%%%
The impact of matrix sparsity can be analyzed using the DLMC dataset, where sparsity levels range from 50\% to 99\%. The graph matrices, in contrast, are predominantly extremely sparse, with roughly 99\% sparsity in most cases. Figure~\ref{fig:eval_spmm_sparsity} presents the effect of sparsity on \texttt{SpMM} performance for both GPU and CPU. As expected, performance decreases substantially on both devices as sparsity increases from 50\% toward 100\%, approximately a 12$\times$ drop on the GPU and a 5.5$\times$ drop on the CPU. On the GPU, the decline in performance is nearly linear with increasing sparsity, because GPUs rely on a large number of concurrent computations to maintain high throughput; when the number of computations decreases due to higher sparsity, the GPU is underutilized and performance drops sharply. On the CPU, performance remains relatively stable for sparsity levels up to 80\%, since CPUs can still exploit vectorized operations and efficiently schedule the smaller number of computations. Beyond this point, however, the reduced workload leads to a steep drop in performance, as vectorization and instruction-level parallelism can no longer be efficiently utilized. We need to note, however, that the performance reported here expresses throughput, i.e., operations per time unit. The overall latency of the entire kernel is affected by both throughput and number of operations, with sparsity having a direct and opposite impact on both. This constitutes a significant tradeoff when tuning an inference engine, especially when one needs to take into account the impact of sparsity on accuracy as a third important parameter. 

%%%%%%%%%%%%%%%%%%%%%%%%%%%%%%%%%%%%%%%%%%%%%%%%%%%%%%%%%%%%%%%%%%%%%%%%%%%%%%
\subsubsection{Embedding dimension (k) effect on performance}  \label{sec:eval:spmm:k}
\begin{figure}[tb]
    \centering
    \includegraphics[width=\linewidth]{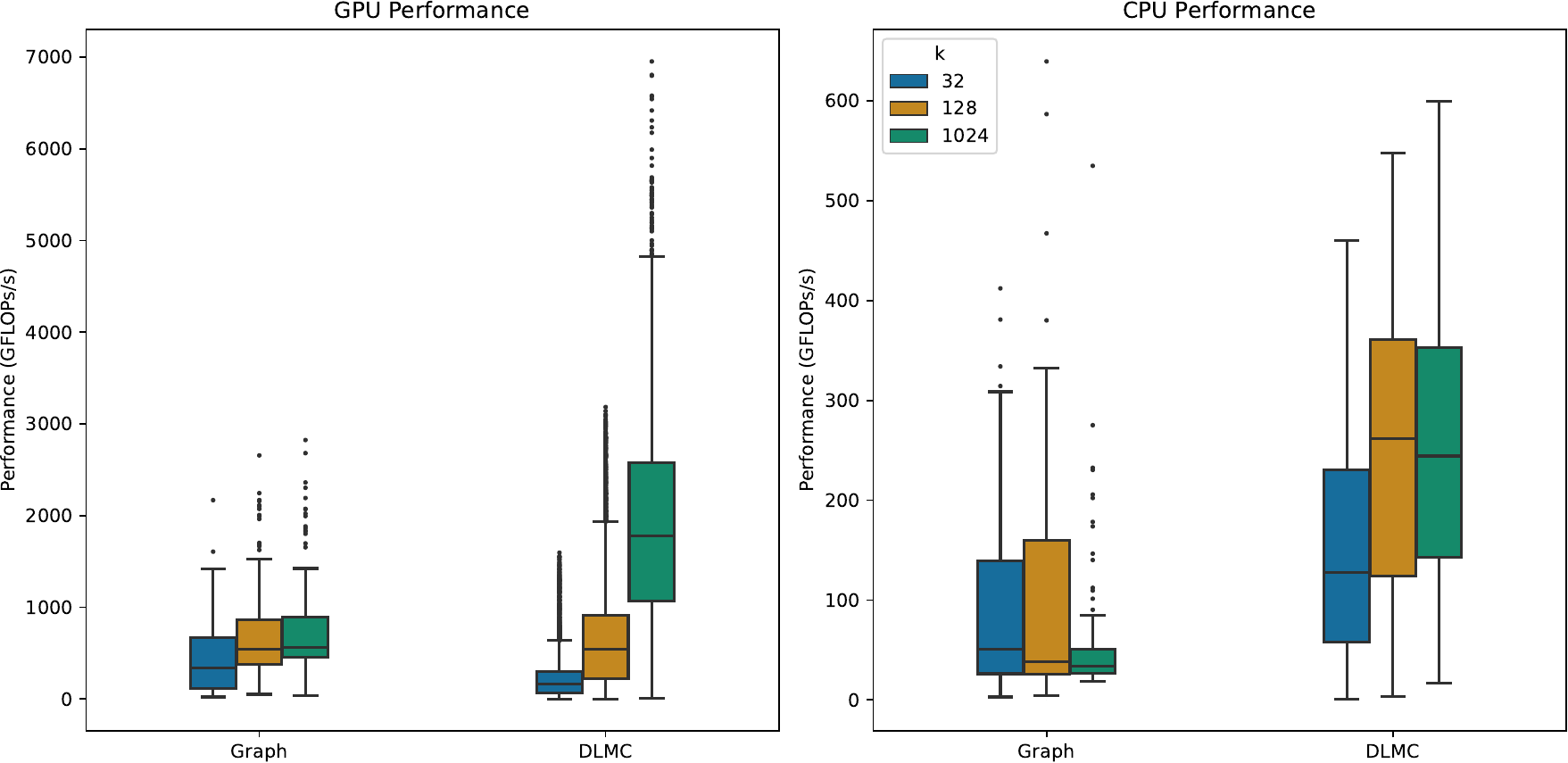}
    \caption{Effect of embedding dimension on the performance of \texttt{SpMM} for the two different datasets on GPU and CPU.}
    \label{fig:eval_spmm_k}
\end{figure}

%%%%%%%%%%%%%%%%%%%%%%%%%%%%%%%%%%%%%%%%%%%%%%%%%%%%%%%%%%%%%%%%%%%%%%%%%%%%%%
Figure~\ref{fig:eval_spmm_k} presents the effect of the different k values on the performance of the \texttt{SpMM} kernel. We observe distinct patterns for the graph and DLMC datasets. On the GPU, changing the embedding dimension does not significantly affect performance on graph matrices. In contrast, DLMC matrices, which exhibit higher density and therefore greater computational requirements, show performance scaling by up to 10$\times$ as k increases from 32 to 1024. This behavior can be explained by the interaction between computational intensity and memory access patterns on each platform. The higher density of the DLMC matrices provides sufficient computational workload to exploit the massive parallelism of the hardware, and increasing k directly increases the arithmetic intensity, leading to proportional performance gains. On the other hand, graph matrices are too sparse, and their memory-bound nature limits the benefit of scaling k. 

\begin{figure}[tb]
    \centering
    \includegraphics[width=\linewidth]{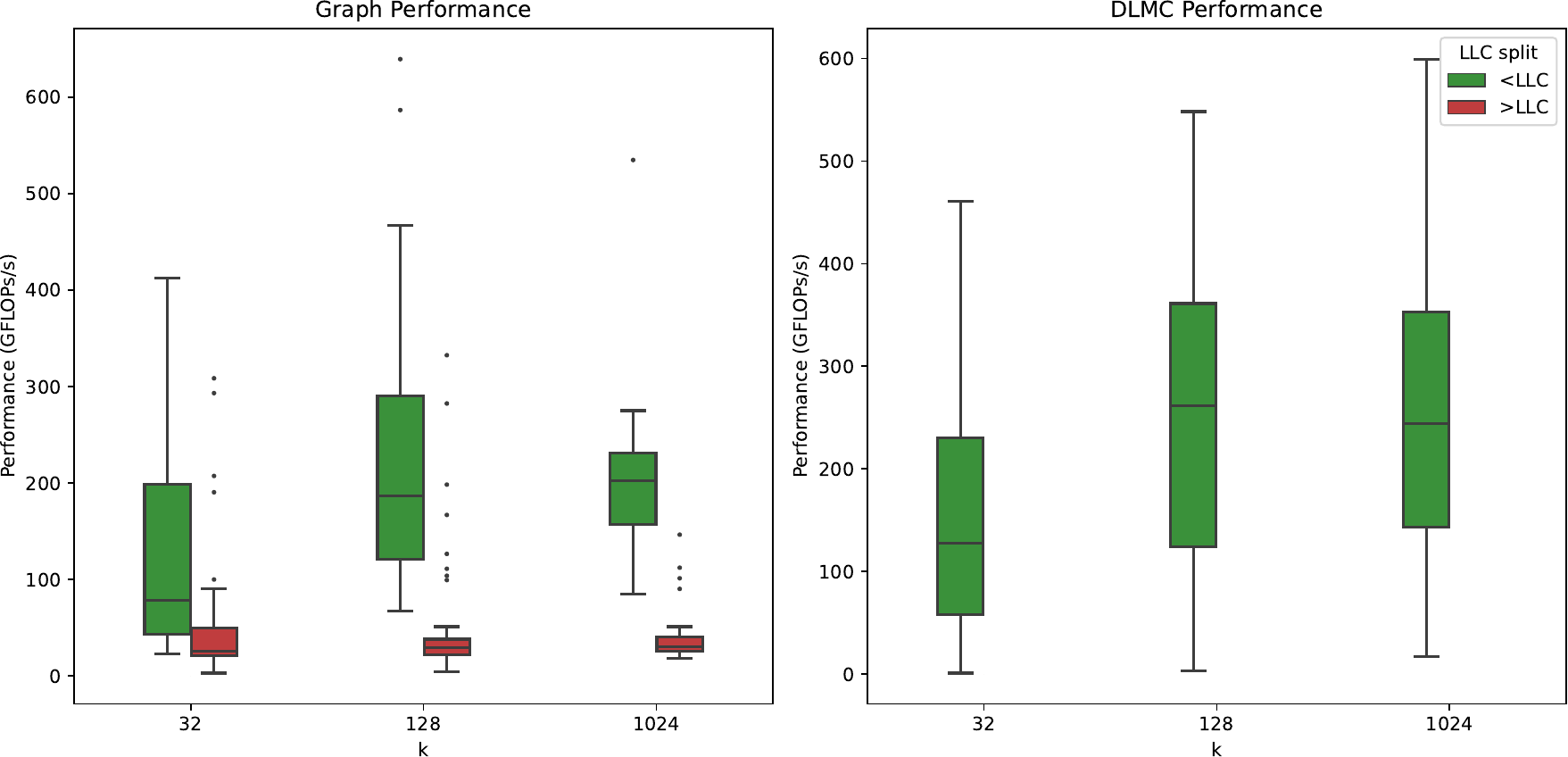}
    \caption{Performance of \texttt{SpMM} on CPU for the graph and DLMC datasets, with splitting between smaller- and larger-than-LLC problem sizes.}
    \label{fig:eval_spmm_k_cpu}
\end{figure}

On the CPU, we must take the total problem size into account before analyzing the results. Figure~\ref{fig:eval_spmm_k_cpu} illustrates the different performance patterns as k varies for problem sizes smaller and larger than the LLC. For both datasets, when the embedding dimension is large (1024), performance does not improve substantially compared to the medium size (128) for problem sizes smaller than the LLC. For the larger-than-LLC cases, which occur only for graph matrices, we observe that the embedding dimension has no impact on performance, as the memory bandwidth becomes the dominant bottleneck limiting CPU performance, and additional computation from larger k values cannot be effectively utilized.

%%%%%%%%%%%%%%%%%%%%%%%%%%%%%%%%%%%%%%%%%%%%%%%%%%%%%%%%%%%%%%%%%%%%%%%%%%%%%%
\subsection{SDDMM}  \label{sec:eval:sddmm}
%%%%%%%%%%%%%%%%%%%%%%%%%%%%%%%%%%%%%%%%%%%%%%%%%%%%%%%%%%%%%%%%%%%%%%%%%%%%%%
Before presenting the \texttt{SDDMM} results, it is important to clarify that the matrices referred to as part of the DLMC dataset are not directly taken from DLMC, but are instead generated matrices that combine feature values originating from the DLMC dataset with the structural sparsity patterns of matrices from the BigBird \cite{zaheer2020big} paper.

\begin{figure}[tb]
    \centering
    \includegraphics[width=\linewidth]{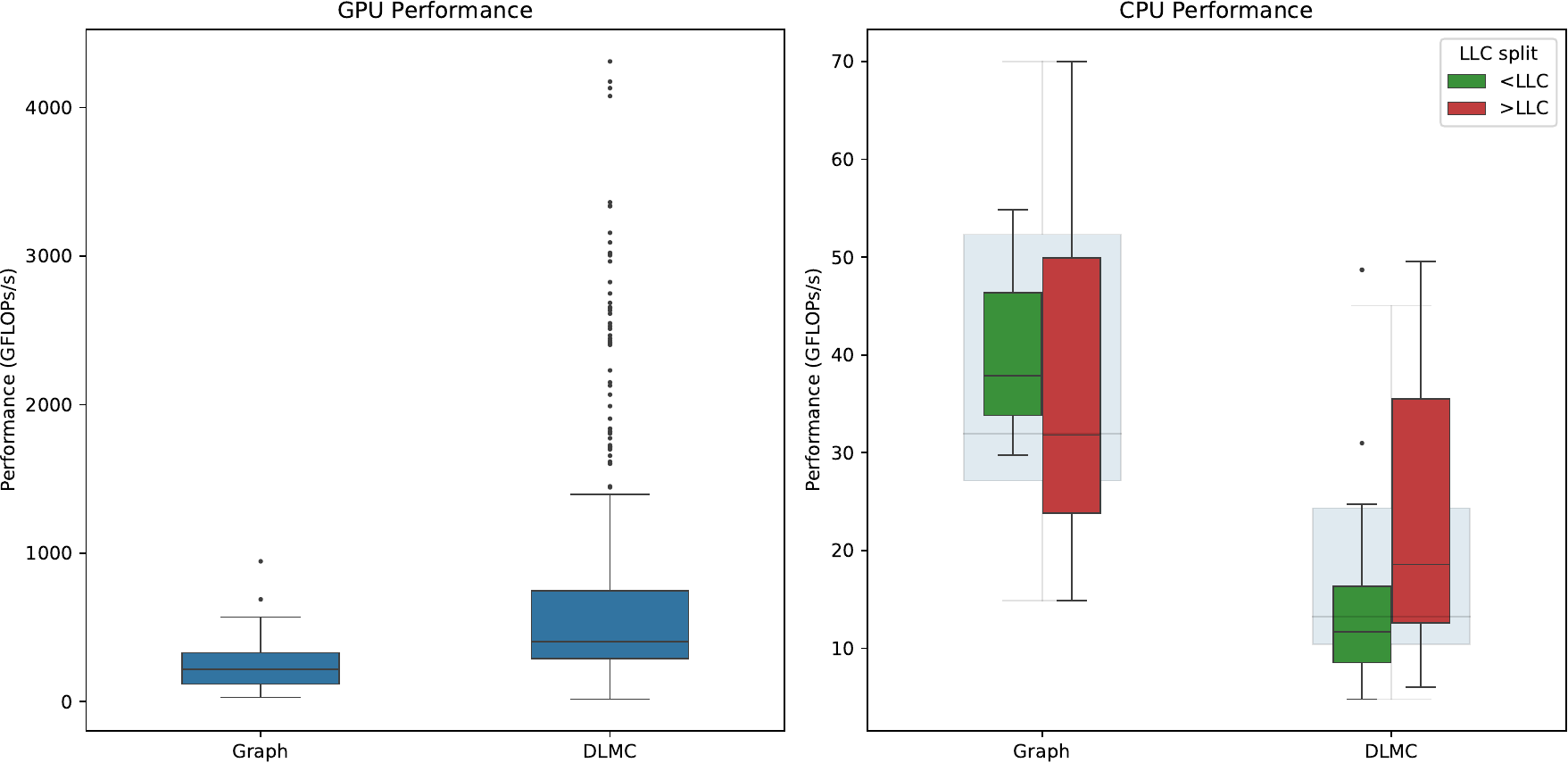}
    \caption{Performance of the \texttt{SDDMM} kernel for the two different datasets on GPU and CPU (embedding dimension k=128).}
    \label{fig:eval_sddmm}
\end{figure}

Figure~\ref{fig:eval_sddmm} presents the performance of the \texttt{SDDMM} kernel across the two datasets. On the GPU, the two datasets behave quite differently. The very high sparsity levels of the graph matrices prevent the \texttt{SDDMM} kernel from achieving high performance. Conversely, the DLMC matrices contain some rare cases of higher throughput and, on average, deliver 1.44$\times$ better performance than the graph matrices. Overall, the \texttt{SDDMM} kernel reaches only 2.3\% of the peak GPU performance, with a few outliers achieving up to 22\%. On the CPU, where only a single implementation (ASpT-CPU) is available, the picture changes. Contrary to the GPU results, graph matrices consistently achieve higher performance than DLMC matrices. Interestingly, and unlike our previous observations, problem sizes that exceed the LLC capacity show better performance than those that fit within it. Nevertheless, the absolute performance levels remain low: only 2.3\% of peak performance on average for graph matrices and about 1\% for DLMC matrices.

%%%%%%%%%%%%%%%%%%%%%%%%%%%%%%%%%%%%%%%%%%%%%%%%%%%%%%%%%%%%%%%%%%%%%%%%%%%%%%
\subsubsection{Format evaluation}  \label{sec:eval:sddmm:format}
\begin{figure}[tb]
    \centering
    \includegraphics[width=\linewidth]{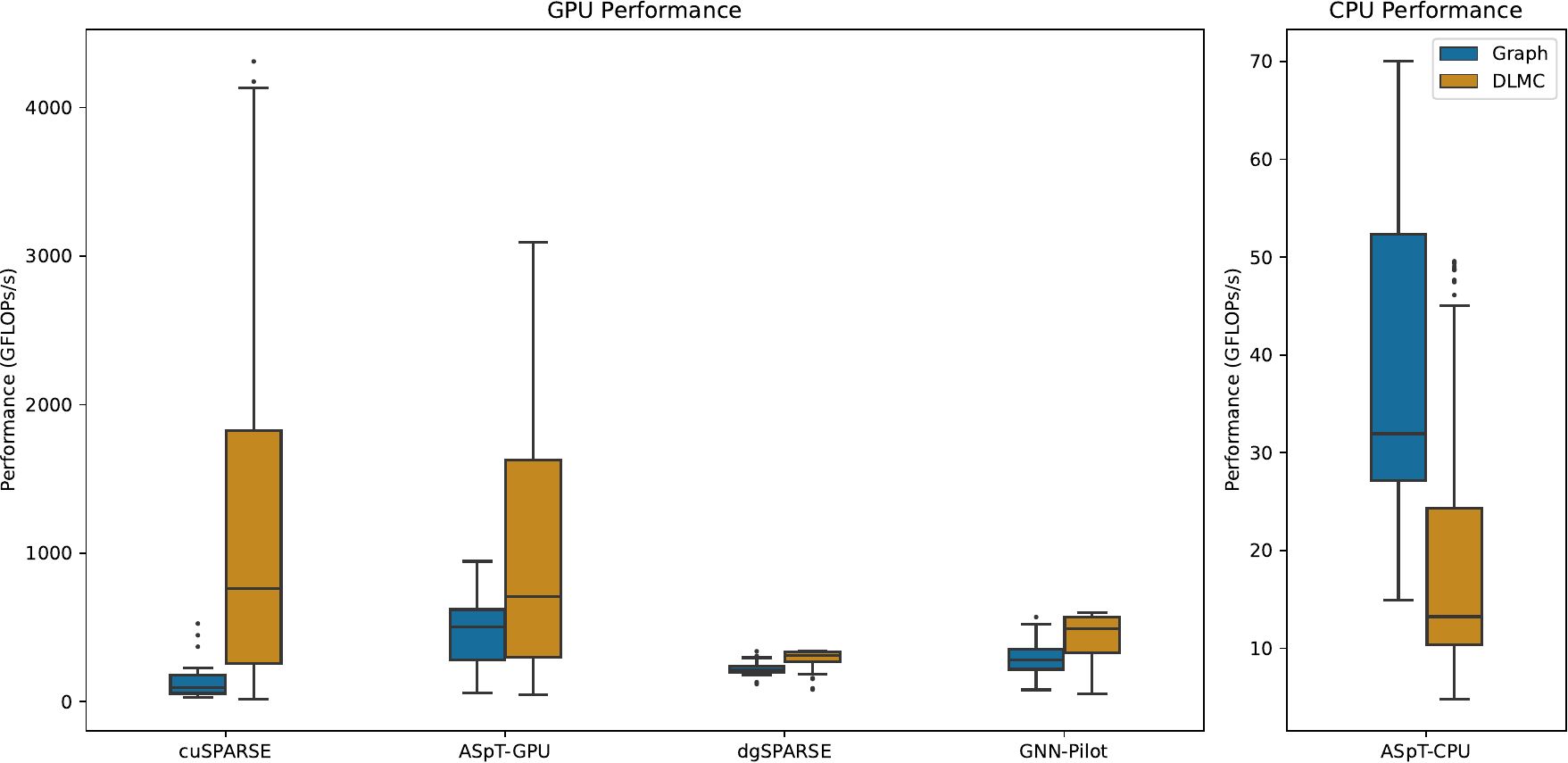}
    \caption{Performance of the evaluated \texttt{SDDMM} formats for the two different datasets on GPU and CPU (embedding dimension k=128).}
    \label{fig:eval_sddmm_format}
\end{figure}

%%%%%%%%%%%%%%%%%%%%%%%%%%%%%%%%%%%%%%%%%%%%%%%%%%%%%%%%%%%%%%%%%%%%%%%%%%%%%%
Figure~\ref{fig:eval_sddmm_format} presents the performance of the different \texttt{SDDMM} formats and implementations evaluated. On the GPU, the large performance gap between the graph and DLMC datasets is primarily driven by the cuSPARSE and ASpT-GPU implementations, which perform significantly better on the DLMC matrices. In contrast, the other two formats deliver consistently lower and comparable performance. On the CPU side, with only a single available format, the observations remain unchanged compared to the previous analysis.

%%%%%%%%%%%%%%%%%%%%%%%%%%%%%%%%%%%%%%%%%%%%%%%%%%%%%%%%%%%%%%%%%%%%%%%%%%%%%%
\subsubsection{Sparsity effect on performance} \label{sec:eval:sddmm:sparsity}
\begin{figure}[tb]
    \centering
    \includegraphics[width=\linewidth]{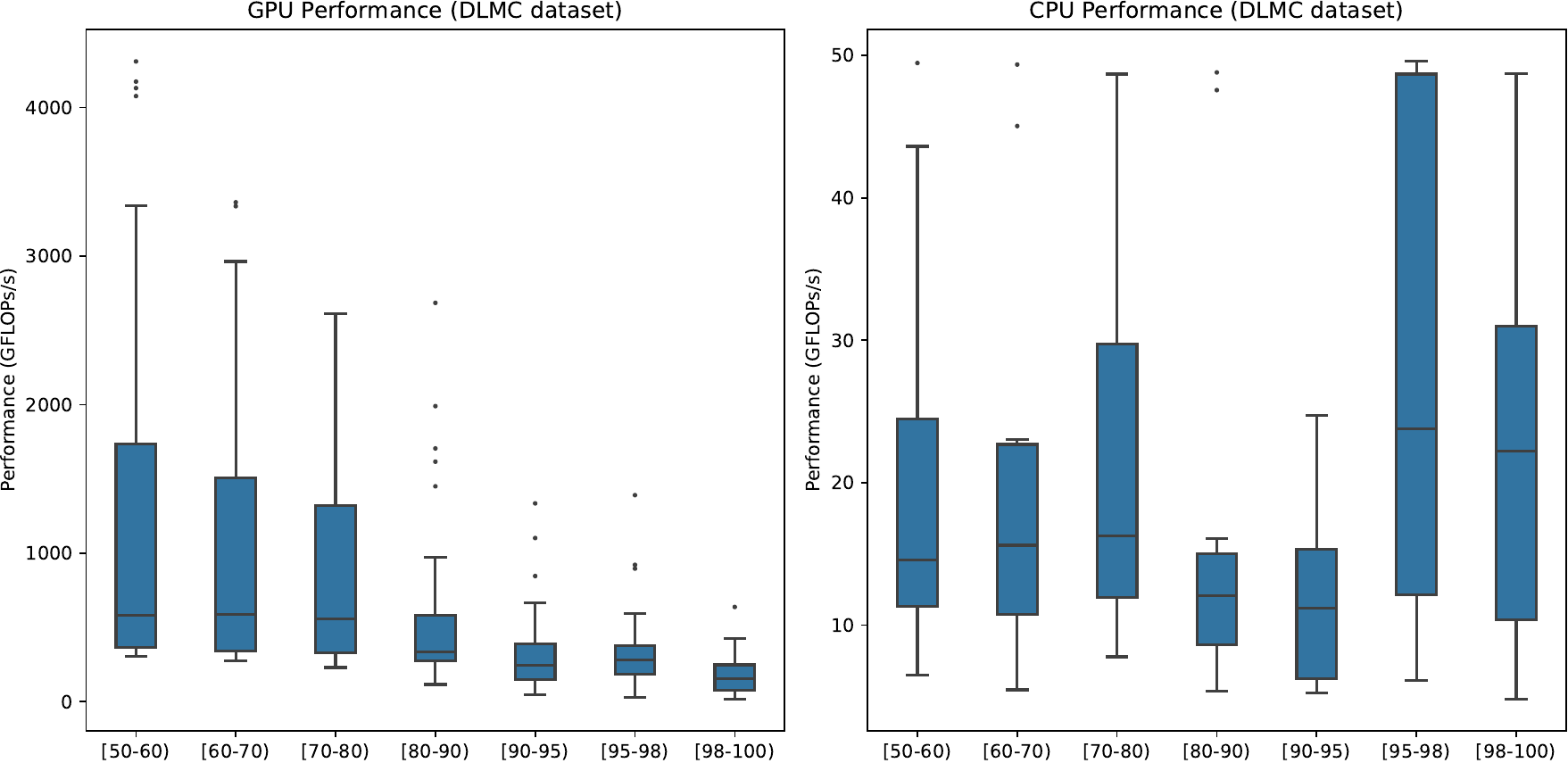}
    \caption{Effect of sparsity levels of the DLMC matrices on the performance of \texttt{SDDMM} on the GPU and the CPU (embedding dimension k=128).}
    \label{fig:eval_sddmm_sparsity}
\end{figure}

%%%%%%%%%%%%%%%%%%%%%%%%%%%%%%%%%%%%%%%%%%%%%%%%%%%%%%%%%%%%%%%%%%%%%%%%%%%%%%
Figure~\ref{fig:eval_sddmm_sparsity} shows the effect of mask sparsity on \texttt{SDDMM} performance for both GPU and CPU. On the GPU, we observe a clear sensitivity to sparsity: once sparsity exceeds 80\%, performance drops steeply. Overall, as sparsity increases from 50\% to nearly 100\%, average performance decreases by a factor of 3.8x, and the GPU is, similarly to the \texttt{SpMM} kernel, underutilized. On the CPU, performance patterns are less stable across sparsity levels. Given the already low performance of the available format, no firm conclusions can be drawn regarding the impact of sparsity in this case.

%%%%%%%%%%%%%%%%%%%%%%%%%%%%%%%%%%%%%%%%%%%%%%%%%%%%%%%%%%%%%%%%%%%%%%%%%%%%%%
\subsubsection{Embedding dimension (k) effect on performance} \label{sec:eval:sddmm:k}
\begin{figure}[tb]
    \centering
    \includegraphics[width=\linewidth]{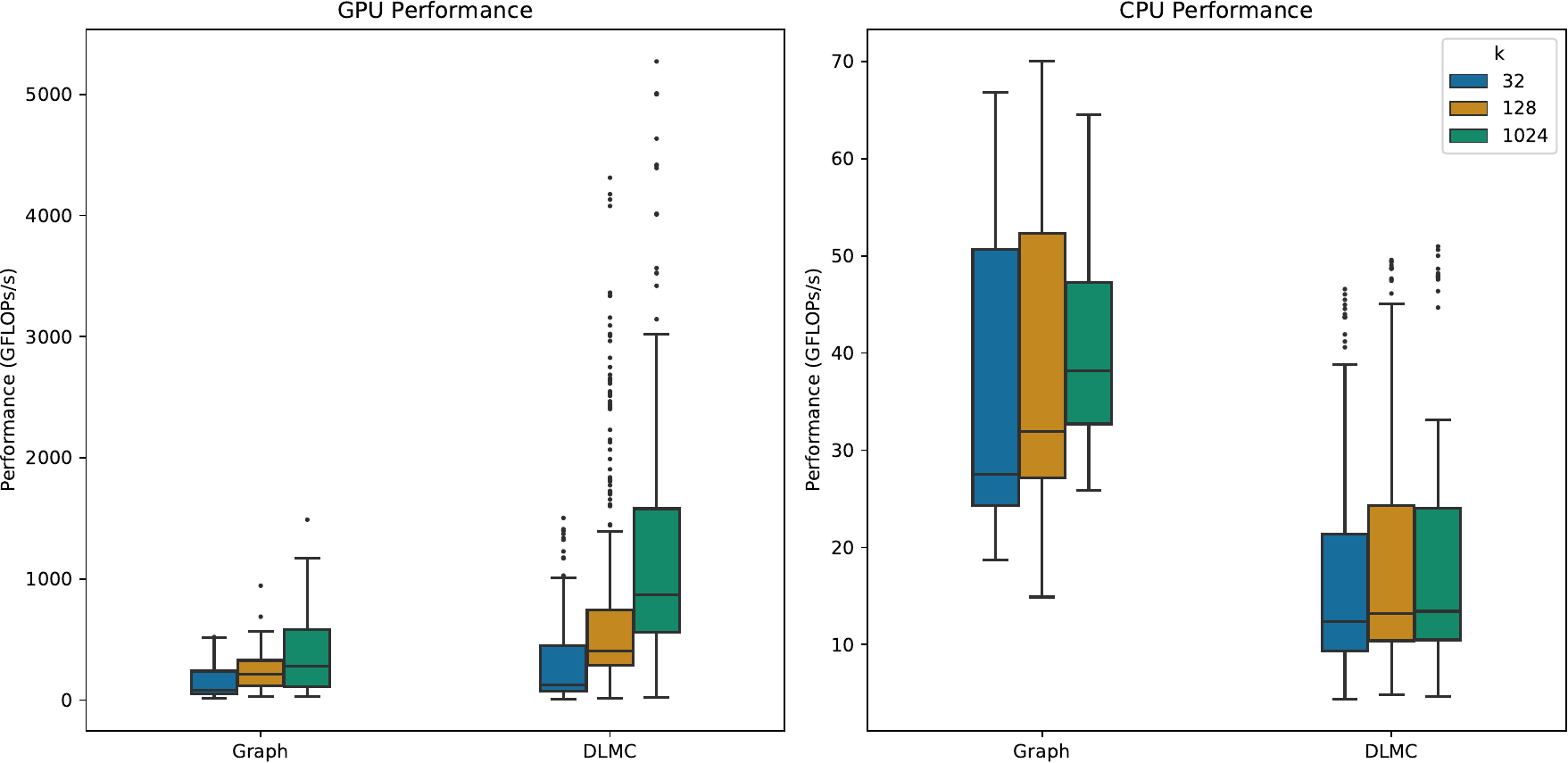}
    \caption{Effect of embedding dimension on the performance of \texttt{SDDMM} for the two different datasets on GPU and CPU.}
    \label{fig:eval_sddmm_k}
\end{figure}

\begin{figure}[tb]
    \centering
    \includegraphics[width=\linewidth]{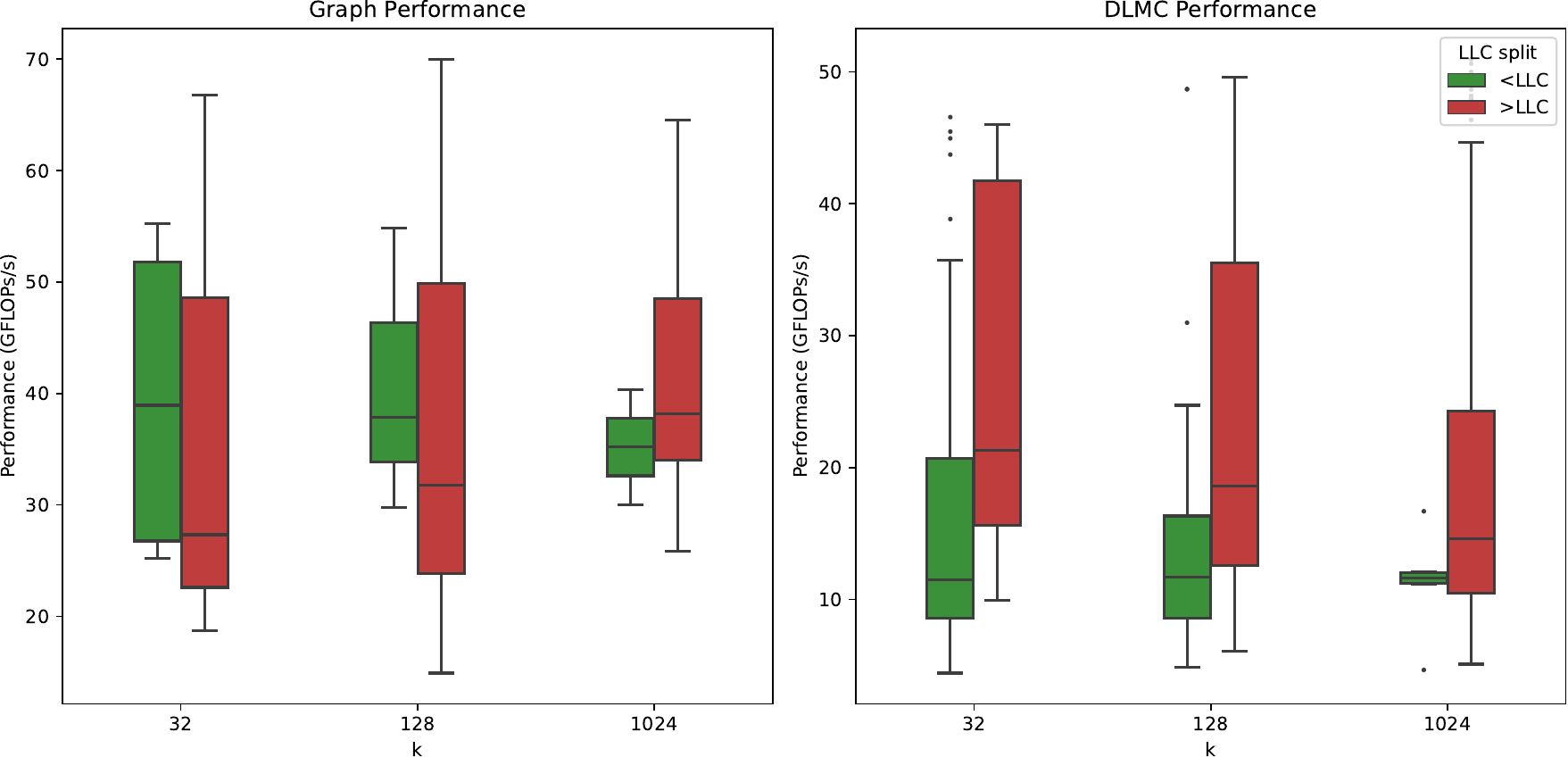}
    \caption{Performance of \texttt{SDDMM} on CPU for the graph and DLMC datasets, with splitting between smaller- and larger-than-LLC problem sizes.}
    \label{fig:eval_sddmm_k_cpu}
\end{figure}

%%%%%%%%%%%%%%%%%%%%%%%%%%%%%%%%%%%%%%%%%%%%%%%%%%%%%%%%%%%%%%%%%%%%%%%%%%%%%%
Figure~\ref{fig:eval_sddmm_k} presents the effect of the embedding dimension (k) on the performance of the \texttt{SDDMM} kernel, revealing distinct patterns for the graph and DLMC datasets. On the GPU, varying the embedding dimension has little impact on graph matrices. In contrast, DLMC matrices, being denser and thus more computationally demanding, show performance scaling by a factor of 6.7$\times$ as k increases from 32 to 1024. This trend is consistent with what we observed for \texttt{SpMM}, though the scaling effect here is less pronounced. On the CPU, the analysis must again take into account the total problem size and the LLC capacity, with the results presented in Figure~\ref{fig:eval_sddmm_k_cpu}. For problem sizes that fit within the LLC, performance decreases on average as the embedding dimension grows. For problem sizes exceeding the LLC, however, the two datasets diverge: graph matrices experience a modest performance improvement, while DLMC matrices exhibit a drop in performance. However, given the limited computational capabilities of the only available CPU format, it is difficult to draw firm conclusions regarding the true effect of embedding dimension on CPU performance.

%%%%%%%%%%%%%%%%%%%%%%%%%%%%%%%%%%%%%%%%%%%%%%%%%%%%%%%%%%%%%%%%%%%%%%%%%%%%%%
\subsection{Porting effort} \label{sec:eval:porting}
%%%%%%%%%%%%%%%%%%%%%%%%%%%%%%%%%%%%%%%%%%%%%%%%%%%%%%%%%%%%%%%%%%%%%%%%%%%%%%
The process of building a common benchmarking framework and integrating the various \texttt{SpMM} and \texttt{SDDMM} implementations proved far from being trivial. We categorize the challenges along two axes: ease of integration and functional validation.

%%%%%%%%%%%%%%%%%%%%%%%%%%%%%%%%%%%%%%%%%%%%%%%%%%%%%%%%%%%%%%%%%%%%%%%%%%%%%%
\subsubsection{Ease of integration} \label{sec:eval:porting:ease}
%%%%%%%%%%%%%%%%%%%%%%%%%%%%%%%%%%%%%%%%%%%%%%%%%%%%%%%%%%%%%%%%%%%%%%%%%%%%%%

Vendor-provided libraries such as NVIDIA cuSPARSE \cite{nvidia_cusparse}, Intel MKL \cite{wang2014intel}, and AMD AOCL-Sparse \cite{amd_aocl} were the most straightforward to integrate. These libraries are well-documented and generally worked out-of-the-box, with only minor adjustments required. For instance, cuSPARSE required matrices to be stored in row-major layout with explicitly specified leading dimensions, while AOCL-Sparse required switching to the development branch of their repository to access the multi-threaded \texttt{SpMM} variant.
In contrast, research-oriented formats lacked the same plug-and-play convenience. Many required substantial code modifications, and documentation was not detailed in most cases, making the integration process more challenging. A few representative cases include:
\begin{itemize}
    \item ASpT \cite{hong2019adaptive} (GPU): Required extensive editing to replace deprecated CUDA primitives (e.g., replacing shfl with shfl\_sync). For the CPU implementation, we had to replace Intel’s icpc with gcc, while several compiler pragmas were unavailable.
    \item FusedMM \cite{rahman2021fusedmm} (CPU): Required detailed inspection of internal functions to isolate the \texttt{SpMM} operator. Unfortunately, the \texttt{SDDMM} operator could not be extracted in a standalone manner.
    \item DTC-SpMM \cite{fan2024dtc}, HC-SpMM \cite{li2025hc}, and GNNPilot \cite{hu2025gnnpilot} (GPU): Designed for PyTorch integration. To isolate their \texttt{SpMM} and \texttt{SDDMM} kernels, we first had to compile them within the PyTorch framework to obtain the correct compilation and linking instructions, before adapting the computational kernels for standalone benchmarking.
    \item RoDe \cite{pang2024row} (GPU): Required removal of several external dependencies, including absl (Abseil Common Libraries) and glog (Google logging).
    \item Sputnik \cite{gale2020sparse} and dgSPARSE \cite{dgsparse_lib} (GPU): Could be compiled successfully but required manual reconfiguration of build and linking procedures.
\end{itemize}

%%%%%%%%%%%%%%%%%%%%%%%%%%%%%%%%%%%%%%%%%%%%%%%%%%%%%%%%%%%%%%%%%%%%%%%%%%%%%%
\subsubsection{Functional validation} \label{sec:eval:validation}
%%%%%%%%%%%%%%%%%%%%%%%%%%%%%%%%%%%%%%%%%%%%%%%%%%%%%%%%%%%%%%%%%%%%%%%%%%%%%%
Successful compilation did not always guarantee correct execution. While vendor libraries ran reliably, several research implementations encountered issues:
\begin{itemize}
    \item Acc-SpMM \cite{zhao2025acc} and HC-SpMM \cite{li2025hc} (GPU): Ran successfully for only a single graph matrix, preventing the collection of meaningful results.
    \item Sputnik \cite{gale2020sparse} (GPU): Produced correct results for both \texttt{SpMM} and \texttt{SDDMM}, but the \texttt{SDDMM} kernel was prohibitively slow and, thus, excluded from the final evaluation.
    \item DTC-SpMM \cite{fan2024dtc} (GPU): Offered multiple variants, with some requiring specific embedding dimensions (k) to execute correctly. Similar restrictions on k were observed in HC and ACC implementations.
    \item RoDe \cite{pang2024row} (GPU): The \texttt{SDDMM} implementation produced incorrect results with no straightforward fix.
    \item VectorSparse \cite{chen2021efficient}, Magicube \cite{li2022efficient} and SMaT \cite{okanovic2024high} (GPU): Supported only FP16 precision, whereas our evaluation focused on FP32, making them unsuitable for inclusion.
\end{itemize}

%%%%%%%%%%%%%%%%%%%%%%%%%%%%%%%%%%%%%%%%%%%%%%%%%%%%%%%%%%%%%%%%%%%%%%%%%%%%%%
\subsubsection{Summary} \label{sec:eval:porting:summary}
%%%%%%%%%%%%%%%%%%%%%%%%%%%%%%%%%%%%%%%%%%%%%%%%%%%%%%%%%%%%%%%%%%%%%%%%%%%%%%
Overall, the integration effort revealed a clear divide: vendor libraries were reliable, easy to use, and stable, but limited in flexibility and peak performance potential. Research kernels, while often promising higher performance, typically required extensive code modifications and debugging to become operational, with many exhibiting incomplete or unstable functionality.

%%%%%%%%%%%%%%%%%%%%%%%%%%%%%%%%%%%%%%%%%%%%%%%%%%%%%%%%%%%%%%%%%%%%%%%%%%%%%%
\section{Conclusions} \label{sec:conclusions}
%%%%%%%%%%%%%%%%%%%%%%%%%%%%%%%%%%%%%%%%%%%%%%%%%%%%%%%%%%%%%%%%%%%%%%%%%%%%%%
To summarize our conclusions, we return to the initial questions posed in Section~\ref{sec:introduction} and discuss our findings below.

\paragraph{In which forms does sparsity arise in DNN inference?} We found that sparsity is discussed in several forms, as summarized in Figure~\ref{fig:dnn_sparsity}. Sparsity can be enforced, i.e., explicitly injected in the process as is the case of weight pruning and sparse attention in LLMs. The sparsity can also be natural in the case of GNNs that represent a naturally sparse domain. Finally, sparsity can be ephemeral, that happens exactly due to the nature of several activation functions that zero out the outcome of DNNs layers. All aforementioned forms of sparsity ultimately resort to actual sparse computations in their implementations. On the other hand, 3D convolutions for point clouds and Mixture of Experts (MoE) models, although frequently categorized as sparse approaches, rely on dense computations. 

We distinguish activation sparsity as a more challenging form in the sense that sparse data are dynamically generated during inference for each distinct inference operation. This means that sparse matrices need to be generated on the fly and any computational optimization (e.g., matrix reordering) must seriously consider the preprocessing cost. 

In the case of weight pruning, there exist three general types of sparsity: structured sparsity that removes large parts of the network and leads to smaller dense networks, unstructured sparsity that freely removes any weight of the network according to some optimization criterion, and semi-structured sparsity, a hybrid of the above that is allowed to remove a specific ratio of nonzero elements (e.g., 2 out of 4) within an area of the original weight matrix. These lead to different performance/efficiency/accuracy tradeoffs that differ across computational devices. 

\paragraph{How do the original dense computations transform to sparse computational kernels?} Original DNN inference is dominated by dense matrix operations, specifically matrix multiplication,  matrix-vector multiplication and convolution. Starting from the initial dense computation in each type of DNN layer (Section~\ref{sec:background:dnn_basics}), we show in Section~\ref{sec:sparse} that the involved kernels in sparse DNN inference are primarily sparse matrix by dense matrix multiplication (\texttt{SpMM}), sampled matrix-matrix multiplication (\texttt{SDDMM}), sparse matrix by dense vector multiplication (\texttt{SpMV}) and sparse convolution.

\paragraph{What is the current state-of-the-art in the implementation of these kernels?} Sparse kernels, especially those that work on unstructured sparsity, are well known for their very low efficiency, as they suffer from memory bottlenecks and imbalance. There is a vivid research interest in optimizing the \texttt{SpMM} and \texttt{SDDMM} kernels on GPUs applying optimizations that target specifically the improvement of locality and load balance. \texttt{SpMV}, convolutions, and in general CPU implementations seem to be lagging behind. We argue that there is a lot of room for performance improvement in all kernels across computational devices. 

\paragraph{What sparse datasets can be used to assist research and development in sparse DNNs?} We studied the datasets used in the various research papers that deal with sparsity in the field and note the following: 
\begin{itemize}
    \item Sparse matrices that represent graphs are collected from well-known databases like Suitesparse \cite{davis2011university}, SNAP \cite{leskovec2016snap}, OGB \cite{hu2020open}, TUD \cite{morris2020tudataset}, and seem to capture well the properties of the domain.
    \item Researchers rely heavily on the valuable DLMC \cite{gale2020sparse} dataset for their experiments, which includes matrices from sparsified layers of the original Transformer and ResNet-50 networks. 
    We argue, however, that these networks are rather old, and a new dataset reflecting the sizes and properties of current architectures needs to be generated.
    \item Evaluating end-to-end performance in state-of-the-art works relies on ad-hoc sparsification approaches of existing networks, typically distinct in each research work. We find that this is an approach that can be followed primarily by domain experts. Access to publicly available entirely sparsified networks could be of value to performance engineers. 
    \item For experimentation with fixed attention masks in Transformers, Big Bird \cite{zaheer2020big} provides a solid base. However, datasets to support experimentation with learnable sparse attention seem to be missing.
    \item Experimenting with activation sparsity can only be done by incorporating the approach in an end-to-end network, a workflow well understood by domain experts but less familiar to the broader community. Datasets that reflect the outputs of different activation mechanisms in modern networks seem to be missing. These could also enable more intense involvement of performance engineers. 
\end{itemize}

\paragraph{What is the current software support to deploy production systems that rely on sparse DNNs?} Vendors and researchers have invested significant effort to provide libraries and tools that manage sparsity. Semi-structured, N:M sparsity seems to be more mature, it has solid software and hardware support, and can be more easily incorporated in existing deployment pipelines. Sparsity in GNNs also enjoys very good support and is operational. Regarding unstructured sparsity in weights, activation, and attention, a large number of libraries for optimized kernels is available, together with other supporting tools like pruning software. However, an end-to-end framework or methodology to deploy a unstructured sparse DNN with production quality seems to be missing, to the best of our knowledge. 

\paragraph{What is the performance behavior of sparse kernels in modern CPUs and GPUs?} In Section~\ref{sec:eval} we performed a preliminary performance evaluation of the state-of-the-art \texttt{SpMM} and \texttt{SDDMM} kernels that work with unstructured sparsity. We summarize our findings below:

\begin{itemize}
    \item As expected, GPU performance for both kernels greatly exceeds CPU performance.
    \item In most of the cases, however, the efficiency of the kernels is extremely low, with the GPU not surpassing 3\% of peak. The CPU was able to reach a decent 20\% over peak only for the \texttt{SpMM} kernel. 
    \item In all cases, the sparsity of the matrix and the embedding dimension, significantly affect performance. In several cases, computations on graph matrices led to different performance behavior compared to that of DLMC matrices. 
    \item Storage formats and optimizations influence performance. There does not seem to be a clear winner and further investigation is required on this aspect. 
\end{itemize}

Our evaluation effort also served a second purpose, i.e., to assess how straightforward it is to perform comparison of related work. We found that a lot needs to be done in order to have a solid and straightforward benchmarking process.  We summarize our thoughts below:

\begin{itemize}
    \item As mentioned above, there are several gaps in the available datasets.
    \item Profiling, benchmarking and performance evaluation over complex, deeply stacked frameworks like PyTorch may be cumbersome and counter-intuitive.
    \item We seem to be missing a common, simple and indicative benchmark or benchmarking methodology that can be used to assess the end-to-end impact of sparsity in modern DNNs, with an additional focus on LLMs. 
\end{itemize}

\newpage
\appendix
{\Large \textbf{\sffamily{Appendix}}}
%%%%%%%%%%%%%%%%%%%%%%%%%%%%%%%%%%%%%%%%%%%%%%%%%%%%%%%%%%%%%%%%%%%%%%%%%%%%%%
\section{Weight pruning for model sparsification} \label{app:pruning}
%%%%%%%%%%%%%%%%%%%%%%%%%%%%%%%%%%%%%%%%%%%%%%%%%%%%%%%%%%%%%%%%%%%%%%%%%%%%%%

There exist various approaches that apply weight pruning and differ in terms of when, how, and what to sparsify. These are analyzed in the following sections.

%%%%%%%%%%%%%%%%%%%%%%%%%%%%%%%%%%%%%%%%%%%%%%%%%%%%%%%%%%%%%%%%%%%%%%%%%%%%%%
\subsection{When to sparsify}
%%%%%%%%%%%%%%%%%%%%%%%%%%%%%%%%%%%%%%%%%%%%%%%%%%%%%%%%%%%%%%%%%%%%%%%%%%%%%%
Sparsification can take place \textit{before}, \textit{during} or \textit{after} training:

\begin{enumerate}
    \item \textbf{Sparsify before training}: 
    The model is initialized with a fixed sparse connectivity pattern and trained from scratch without modifying the mask. Such static sparse initializations are simple and avoid the pruning cost, but generally lead to reduced accuracy. Examples include random sparse initialization and fixed-topology sparse training methods~\cite{mocanu2018scalable} and recent analyses showing that randomly pruned networks can sometimes train effectively from scratch~\cite{liu2022unreasonable}.
    
    \item \textbf{Sparsify during training}: 
    Sparsity can also be introduced while the model is being optimized, and training may start from either a dense or a sparsely initialized network. 
    First, such approaches may be \emph{sparsity-enforcing}, where explicit top-$k$ or magnitude-based rules maintain a desired sparsity level (either globally or per-layer), or \emph{sparsity-inducing}, where regularizers such as L1, group-lasso, or L0 gates encourage weights to become zero without enforcing a specific target\cite{louizos2017learning}.
    Second, the sparsity pattern may remain \emph{static}, follow a \emph{gradual pruning schedule} that increases sparsity over training~\cite{zhu2017prune}, or evolve dynamically through \emph{pruning-and-regrowth} mechanisms, as in dynamic sparse training (DST) approaches~\cite{mocanu2018scalable, evci2020rigging, vanderschueren2023straight}.
    These mechanisms introduce or maintain sparsity during learning, forming the basis of \emph{sparsity-aware training}, where sparsification is integrated into the optimization process rather than applied only after convergence.
    \item \textbf{Sparsify after training}: The train-then-sparsify is the most common approach and the one implied in Figure~\ref{fig:depl}. In this method, also known as post-training pruning, the network is first trained densely to convergence, making it easier to assess which weights are genuinely redundant. The trained model is then analyzed to identify low-importance parameters—typically those with small magnitudes or low sensitivity—and these weights are removed.
    Post-training pruning provides a natural baseline for evaluating sparsification methods, since the performance of the pruned model can be directly compared against that of the original dense network~\cite{lecun1989optimal, han2015learning, han2015deep}. However, despite its intuitive nature and simplicity, a dense model that was not trained with sparsity in mind may not be optimally positioned for extreme sparsification, which motivates methods that incorporate sparsity constraints during training~\cite{evci2020rigging, gale2019state}.
\end{enumerate}

%%%%%%%%%%%%%%%%%%%%%%%%%%%%%%%%%%%%%%%%%%%%%%%%%%%%%%%%%%%%%%%%%%%%%%%%%%%%%%
\subsection{How to sparsify}
%%%%%%%%%%%%%%%%%%%%%%%%%%%%%%%%%%%%%%%%%%%%%%%%%%%%%%%%%%%%%%%%%%%%%%%%%%%%%%

%%%%%%%%%%%%%%%%%%%%%%%%%%%%%%%%%%%%%%%%%%%%%%%%%%%%%%%%%%%%%%%%%%%%%%%%%%%%%%
\subsubsection{Structured sparsity}
%%%%%%%%%%%%%%%%%%%%%%%%%%%%%%%%%%%%%%%%%%%%%%%%%%%%%%%%%%%%%%%%%%%%%%%%%%%%%%
Structured pruning targets entire filters or neurons and achieves significantly better performance on current hardware due to its compatibility with dense computations~\cite{li2016pruning, han2015learning, wen2016learning}. It also reduces  the representational overhead, as fewer indices need to be stored. Structured pruning tends to perform well even when applied randomly at initialization, making it useful for early-stage architecture search~\cite{luo2017thinet}. However, it typically leads to lower accuracy per pruned element and may require more training iterations or computational resources to achieve similar final results \cite{molchanov2016pruning, hoefler2021sparsity}.

%%%%%%%%%%%%%%%%%%%%%%%%%%%%%%%%%%%%%%%%%%%%%%%%%%%%%%%%%%%%%%%%%%%%%%%%%%%%%%
\subsubsection{Unstructured sparsity}
%%%%%%%%%%%%%%%%%%%%%%%%%%%%%%%%%%%%%%%%%%%%%%%%%%%%%%%%%%%%%%%%%%%%%%%%%%%%%%
Unstructured pruning, often called unstructured sparsity, is the most common type of sparsity investigated. This method, exemplified by weight-level pruning, generally preserves higher accuracy per pruned element compared to structured methods \cite{han2015learning, gale2019state}. Its fine-grained approach allows for precise control over which parameters are removed, and the resulting sparse matrices can be efficiently represented using standard sparse matrix formats (see Section~\ref{sec:background:sparsity_basics}), which is why these neural networks can retain their accuracy even at high sparsity ratios.
Despite the advantages of unstructured pruning, it comes with severe tradeoffs as it suffers from lower computational efficiency on modern hardware and incurs higher memory overhead because it needs to store indices for each individual pruned weight. In addition, unstructured pruning typically degrades accuracy when applied randomly at initialization, since the specific connections in the network are more important at that level of granularity. Although this type of sparsity can be induced in multiple ways, the resulting weight matrices can have distinct characteristics. While these properties could potentially be exploited to improve sparse kernel performance, such exploration is beyond the scope of this discussion. The following section will explore some of the most common and state-of-the-art pruning algorithms used by the community.

\paragraph{Random pruning} The simplest way to introduce unstructured sparsity is, of course, randomly. This method is not used in practice, as the accuracy of the resulting model is significantly degraded most of the time~\cite{frankle2018lottery}. 
Nevertheless, random pruning provides a convenient baseline for experimentation, serving as a control condition for evaluating more principled pruning methods.

\paragraph{Magnitude pruning} 
This is the most intuitive, widely used, and often surprisingly effective method. It is based on the simple idea that the most influential weights of the model are those with the largest absolute values, so the rest can be removed with little effect on accuracy. In the post-training setting, this typically involves ranking weights by magnitude, pruning the smallest ones (either globally or per layer), and then fine-tuning the model to recover accuracy \cite{han2015deep, liu2018rethinking}. 
Although one-shot magnitude pruning can lead to noticeable accuracy degradation at high sparsity levels, its performance after retraining is often competitive with more sophisticated methods, which, together with its minimal computational overhead, explains its widespread adoption.
Beyond post-training pruning, magnitude-based criteria also play an important role in sparse-aware training methods where masks are learned jointly with model parameters. Such dynamic sparse training approaches rely on magnitude-based pruning steps due to their efficiency and simplicity, while focusing on how to enable better joint optimization of weights and masks~\cite{retsinas2021online, jayakumar2020top, georgoulakis2023feather, vanderschueren2023straight}.

\paragraph{More elaborate pruning methods}
More elaborate pruning methods \cite{lecun1989optimal, hassibi1993optimal} use second-order information, typically approximations of the Hessian or Fisher matrix, to estimate the sensitivity of the weights to be removed. These approaches aim to eliminate parameters that minimally affect the loss, often achieving higher accuracy than magnitude pruning at comparable sparsity levels. Although computing exact second-order information is expensive, modern scalable variants \cite{singh2020woodfisher, kurtic2022optimal, wang2020neural} employ block-diagonal, low-rank, or Hessian-free approximations, making curvature-based pruning practical for contemporary architectures.

\paragraph{Pruning methods for LLMs} Based on the same idea, pruning LLMs is considerably more difficult due to the increased number of parameters and the prohibitive retraining requirements. This is why recent state-of-the-art unstructured pruning methods for these models such as SparseGPT \cite{frantar2023sparsegpt}, Wanda \cite{sun2023simple} and FISTAPruner \cite{zhao2025fistapruner}, pivot from traditional fine-tuning to highly efficient `one-shot' post-training approaches. These advanced techniques address the challenge in different ways: SparseGPT prunes weights layer by layer while locally adjusting them so the model's output remains close to the original, Wanda evaluates both the weight values and their activation responses to decide which ones matter most, and FISTAPruner frames pruning as an optimization problem that automatically balances model compression with error correction. All are designed to achieve significant model compression and hardware efficiency while preserving high accuracy without the need for costly post-pruning retraining. To balance accuracy and compression though these methods usually do not reach very high sparsity ratios.

\paragraph{Other pruning methods} Beyond simple magnitude pruning and its more advanced Hessian-based descendants, several other sophisticated approaches have been explored for post-training unstructured pruning without extensive retraining. These include gradient-based methods like SNIP \cite{lee2018snip}, and GraSP \cite{wang2020picking} which evaluate weight importance by their impact on the loss's gradients, movement pruning \cite{sanh2020movement} which assesses importance by how much weights `move' during a brief fine-tuning phase, favoring those that actively adapt. An emerging and theoretically grounded approach is spectral graph sparsification \cite{suzuki2018spectral} which views the neural network as a graph and prunes connections while aiming to preserve the graph's fundamental spectral properties, thereby maintaining critical information flow and connectivity. While these methods offer more sophisticated ways to identify and remove redundancies with high accuracy retention, they are generally more prevalent in research settings and less commonly adopted in widespread industrial deployments compared to magnitude pruning, largely due to the complexities of implementation, hardware compatibility challenges with highly sparse models, and the ongoing need for further validation on diverse large-scale tasks.

%%%%%%%%%%%%%%%%%%%%%%%%%%%%%%%%%%%%%%%%%%%%%%%%%%%%%%%%%%%%%%%%%%%%%%%%%%%%%%
\subsubsection{Semi-structured sparsity}
%%%%%%%%%%%%%%%%%%%%%%%%%%%%%%%%%%%%%%%%%%%%%%%%%%%%%%%%%%%%%%%%%%%%%%%%%%%%%%
Semi-structured pruning has recently gained significant attention as it strikes a balance between the flexibility of unstructured pruning and the computational efficiency of structured approaches. It imposes sparsity patterns with limited regularity, such as N:M sparsity—where exactly $N$ out of every $M$ weights remain nonzero within each predefined block \cite{mishra2021accelerating}. This constrained pattern makes semi-structured sparsity more amenable to hardware acceleration than fully unstructured sparsity, as the regular grouping enables partially optimized kernels and lighter indexing overhead. At the same time, semi-structured sparsity retains considerably more flexibility than coarse structured pruning, often allowing for higher compression rates with minimal accuracy degradation.

Although semi-structured sparsity is not as hardware-friendly as fully dense computation or coarse structured pruning, it provides a practical middle ground for balancing model expressivity, computational efficiency, and memory savings. The growing support for these patterns in modern accelerators—most notably NVIDIA’s 2:4 sparsity support on Ampere and Hopper GPUs—has accelerated their adoption, making semi-structured sparsity a compelling middle ground in the neural network compression landscape.

\paragraph{Basic N:M pruning} Simple N:M pruning strategies serve as foundational baselines for comparison with more sophisticated methods, defined by their simplicity and computational efficiency~\cite{mishra2021accelerating}. Random Pruning is a basic control method that enforces a fixed N:M sparsity pattern by randomly selecting N weights to retain within each block of M weights, without regard to weight importance. A more effective, yet still simple, approach is Magnitude-Based Pruning. This method adapts the common magnitude pruning heuristic to the N:M constraint by selecting the N weights with the largest absolute magnitudes within each block of M. This strategy is easy to implement and is compatible with hardware that supports N:M sparsity.

\paragraph{Advanced N:M pruning} 
Beyond blockwise heuristics, recent work has introduced more sophisticated learning-based methods that aim to identify optimal N:M sparsity patterns, particularly for transformer architectures and large language models (LLMs). Most of these approaches operate during training or retraining, where the sparsity mask is learned jointly with the model parameters rather than imposed post hoc. While earlier post-training pruning approaches such as SparseGPT~\cite{frantar2023sparsegpt} and Wanda~\cite{sun2023simple} established the feasibility of N:M sparsity in the LLM setting, newer methods pursue end-to-end optimization to learn masks that better preserve model performance.
The Adaptive Sparse Trainer (AST) \cite{huang2025pruning} is an efficient retraining framework that learns optimal masks during the weight update process. It integrates knowledge distillation from a dense teacher model to accelerate convergence and preserve performance, achieving near-dense model performance on LLaMA2-7B with minimal retraining cost. Similarly, MaskLLM \cite{fang2024maskllm}and MaskPro \cite{sun2025maskpro} reframe pruning as a learnable problem. MaskLLM uses Gumbel Softmax sampling to model N:M patterns as a learnable distribution, allowing for end-to-end training that directly optimizes the language modeling loss. MaskPro employs a linear-space probabilistic framework to learn a categorical distribution for each weight group, using a moving average of loss residuals to stabilize training. ProxSparse \cite{liu2025proxsparse} further advances this paradigm with a regularized optimization framework. By transforming the non-differentiable mask selection into a gradual search process, it enables end-to-end learning with global gradient feedback, consistently outperforming previous baselines and demonstrating the clear benefits of a learned, global approach to pruning.

%%%%%%%%%%%%%%%%%%%%%%%%%%%%%%%%%%%%%%%%%%%%%%%%%%%%%%%%%%%%%%%%%%%%%%%%%%%%%%
\section{Sparse Convolutions for Multidimensional Sparse Input} \label{app:spconv}
%%%%%%%%%%%%%%%%%%%%%%%%%%%%%%%%%%%%%%%%%%%%%%%%%%%%%%%%%%%%%%%%%%%%%%%%%%%%%%

Multidimensional sparse input refers to data that exist in a high-dimensional space (e.g., a 3D volume, but also 4D spatio-temporal sequences \cite{choy20194d}) and are naturally sparsely populated, meaning that most of the space remains empty. A common example is 3D point clouds, i.e., unordered sets of 3D points, each associated with a feature vector, i.e., there are the point coordinates (e.g., $x$, $y$, $z$ for a 3D space) and a feature vector at each point (e.g., R, G, B values for each nonzero point in the 3D space). The data is collected through sensors like LiDAR scanners and depth cameras \cite{choy20194d}.

Unlike 2D image pixels, this representation is irregularly distributed, making it challenging to apply standard convolutions for computer vision tasks \cite{wu2019pointconv}. Point cloud convolution approaches that first convert point clouds to a dense volumetric representation and then apply dense CNNs or point-based methods, that directly perform convolution on the k-nearest neighbor (spherical nearest neighbor of each point), are less efficient in large outdoor scenes \cite{lin2021pointacc, tang2022torchsparse}. Thus, many state-of-the-art deep networks for point cloud segmentation and detection \cite{graham20183d, choy20194d, wu2024point, wang2023insmos} voxelize the data \cite{graham20183d} and primarily use a different kind of convolution, coined `sparse convolution' \cite{tang2022torchsparse}. Sparse convolution avoids the irregular or large memory footprint of the other methods and is empirically proven to be able to scale up to large point clouds \cite{tang2022torchsparse}.

This category of point cloud convolution is different from the sparsified convolution layer discussed in Section~\ref{sec:background:sparsity_basics}, where there is some sparsity either in the weights (pruning) and/or the input (from previous ReLU activations) \cite{tang2022torchsparse}. Both operations share the same name in literature (e.g., `SparseRT' \cite{wang2020sparsert}, which refers to convolution on pruned weights as sparse convolution), potentially causing some confusion. Sparse Convolution in the context of 3D point clouds aims to maintain the inherent sparse pattern of the input by computing features only at specific output coordinates \cite{yang2024minuet} and therefore it can be viewed as a form of masked convolution \cite{won2023unified}.

A sparse convolution layer of dimension $D$ and kernel size $K$ has weights of shape $K^{D}\times C_{in}\times C_{out}$, where $ C_{in}$ and $ C_{out}$ are the input and output feature sizes. Kernel offsets can be defined as the $K^{D}$ vectors that describe the relative positions around the center of the kernel window (e.g for $\text{kernel size} = 3$ the kernel offsets are the $3^{D}$ vectors of the form $\{ -1, 0, 1\}^{D}$). The weights can be broken down into $K^{D}$ matrices of shape $ C_{in}\times C_{out}$, where each $W_{\delta}$ corresponds to the weights of a particular offset $\delta$, thus called weight offset matrices.

For a convolution of stride $s$, an output point with coordinates $q_{i}$ can be linked through an offset $\delta$ with an input point with coordinates $p_{j}$ if $p_{j} = sq_{i} + \delta$. From this perspective, the features $x_{i}$ of an output point $q_{i}$ can be calculated as $x_{i} = \sum\limits_{\delta}\sum\limits_{j}1\left[p_{j} = sq_{i} + \delta \right](x_{j} {\cdot} W_{\delta})$, where $1[{\cdot}]$ is the binary indicator of the condition  $p_{j} = sq_{i} + \delta$ \cite{tang2022torchsparse}.

Following this observation, state-of-the-art implementations of sparse convolution compute as a first step an input-output map $M$, called kernel map, that has the mapping $\left\{p_{j}, q_{i}, \delta \right\}$ if  $p_{j} = sq_{i} + \delta$ (\textit{Map Step}) \cite{tang2022torchsparse}. To check whether $sq_{i} + \delta$ is an input point, most frameworks use a hash table to store the input points and iteratively perform queries \cite{tang2022torchsparse,hong2023exploiting,choy20194d}. More recently, Minuet \cite{yang2024minuet}, introduced an innovative binary search-like approach to optimize the locality of the query operations.

It is important to note that this input-output mapping differs from the adjacency matrix used in graph convolution, as the latter uses the same weight matrix of shape $C_{in}\times C_{out}$ for all the different neighbors. On the contrary, in sparse convolution, an output point can be associated with different `neighboring' input points through different offset and thus the computations would need different weight offset matrices. As a result, sparse libraries that accelerate graph convolution cannot be directly applied for sparse convolution \cite{lin2021pointacc, tang2022torchsparse}.

Sparse Convolution engines for 3D point cloud inference use the map step to result in dense computations. Each $\left\{p_{j}, q_{i}, \delta \right\}$ corresponds to the matrix multiplication of $x_{i} += x_{j}W_{\delta}$, so iterating over all mappings will calculate the output features. However, matrix-vector multiplication tends to exhibit low utilization on GPUs. To address this, early implementations adopted the \textit{Gather-Matmul-Scatter} dataflow \cite{tang2022torchsparse}. In this process, all input feature vectors that share the same weight offset matrix are first gathered and concatenated into a contiguous feature matrix. A dense \texttt{GEMM} is then performed between this feature matrix and the weight offset matrix to produce partial sums. Finally, these partial sums are scattered and accumulated into their corresponding output feature vectors \cite{tang2022torchsparse, yang2024minuet}.

An alternative dataflow is \textit{Fetch-on-Demand} \cite{lin2021pointacc,hong2023exploiting}, where instead of materializing the input feature matrix to a gather buffer in DRAM and the partial sums in a scatter buffer, it fetches on demand feature vectors on chip, performs a matrix multiplication and then directly writes the partial sum to the corresponding output. Thus, this dataflow achieves overlap between computations and memory accesses and can be viewed as a kernel fusion version of \textit{Gather-Matmul-Scatter} \cite{yang2024minuet}.

Finally, a third dataflow introduced in \cite{spconv2022} is \textit{Implicit-GEMM}. By performing a im2col-like transformation on the input features, which is based on kernel map, sparse convolution can be viewed as a dense implicit \texttt{GEMM} of $X_{im2col}$ and $W$, the total weight matrix of shape $\left(K^{D}C_{in}\times C_{out} \right)$. Although it also overlaps memory accesses with computations through pipelining and minimizes write-back traffic, it still incurs noticeable redundant computations.
TorchSparse++ \cite{tang2023torchsparse++} leverages the strengths of the different processing methods by proposing an adaptive computation dataflow selection.

%%%%%%%%%%%%%%%%%%%%%%%%%%%%%%%%%%%%%%%%%%%%%%%%%%%%%%%%%%%%%%%%%%%%%%%%%%%%%%
\section{Mixture of Experts} \label{app:moe}
%%%%%%%%%%%%%%%%%%%%%%%%%%%%%%%%%%%%%%%%%%%%%%%%%%%%%%%%%%%%%%%%%%%%%%%%%%%%%%

Model scalability—specifically, increasing the number of parameters—is widely recognized as a key factor in improving the performance of Large Language Models (LLMs). However, compute resources pose a significant constraint in scaling such models.  To balance quality with computational efficiency, it is often desirable to design models that contain a large number of parameters while minimizing the required compute. Mixture of Experts (MoE) \cite{jacobs1991adaptive,shazeer2017outrageously,fedus2022switch} is a technique that enables the training of models with substantially more parameters, with far less compute compared to dense LLMs. MoE models can achieve comparable accuracy to their dense counterparts, while offering significantly faster training and inference times. In practice, this means that MoE models can scale in both model and data size to a larger extent under the same compute budget as dense models.

The core concept behind MoE layers is that each layer comprises multiple independent dense sub-networks, referred to as experts. Each sub-network or expert is responsible for processing a different subset of the input data and learns to specialize in a particular region of the input space. To route inputs appropriately, MoE layers incorporate a gating network—also known as a router—that determines which experts should be activated for a given input. During training, both the experts and the gating network are jointly trained. The gating network learns to assign each input to the most suitable expert(s), thereby enabling specialization.

\begin{figure}[b!]
    \centering
    \includegraphics[width=.75\linewidth]{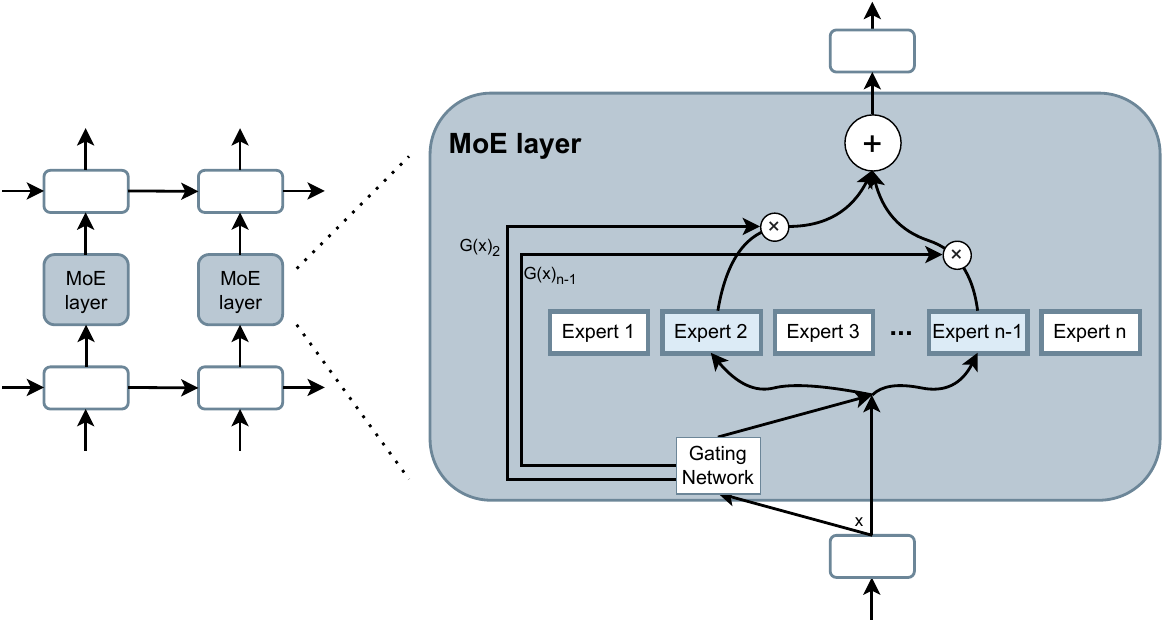}
    \caption{A sparse Mixture of Experts (MoE) layer.}
    \label{fig:moe}
\end{figure}

Figure~\ref{fig:moe} shows an example of a sparse MoE layer. A sparse MoE layer consists of multiple experts (e.g., 8 experts), where each expert is usually implemented as a standard dense Feed-Forward Network (FFN) using conventional \texttt{GEMM} kernels. More advanced configurations are also possible, such as hierarchical MoEs, in which each expert is itself an MoE layer. A gating network (also referred to as a router) determines which experts are activated for each input token. In most cases, only a small subset of experts (e.g., top-2 or top-k) is selected per input token, meaning that each token is routed to one or a few experts, rather than all of them.

As a result, during training, the MoE technique enables significantly improved computational efficiency, as the gating network learns to effectively route input tokens to the most relevant experts. This efficiency extends to inference as well: although MoE models may contain a comparable number of parameters to dense LLMs, only a subset of these parameters—corresponding to the activated experts—is utilized for each input. Consequently, inference is substantially faster than in dense models with an equivalent parameter count. However, all experts must still reside in main memory (e.g., DRAM), resulting in a memory footprint that remains comparable to that of dense models.

In summary, unlike dense LLMs that process all input tokens through the entire network, MoEs introduce sparsity via conditional computation. This approach ensures that only a subset of the network—specifically, a few selected experts—is activated for each token. Although the activated experts are themselves dense feed-forward networks (FFNs) implemented using standard \texttt{GEMM} kernels, the overall sparse (conditional) computation enables substantial compute savings. This allows MoE models to scale to significantly larger parameter counts—improving accuracy and model quality—without a corresponding increase in computational cost.

%%%%%%%%%%%%%%%%%%%%%%%%%%%%%%%%%%%%%%%%%%%%%%%%%%%%%%%%%%%%%%%%%%%%%%%%%%%%%%
\section{Quantization} \label{app:quantization}
%%%%%%%%%%%%%%%%%%%%%%%%%%%%%%%%%%%%%%%%%%%%%%%%%%%%%%%%%%%%%%%%%%%%%%%%%%%%%%
Deep learning models continue to grow in depth and parameter count to improve task performance, address increasingly complex problems, and capture intricate patterns in the vast and ever-growing datasets available for training. Deploying such models in real-world, real-time scenarios often requires running them on resource-constrained hardware platforms, such as edge devices (e.g., mobile phones, smartwatches, and IoT devices).
To tackle excessive memory footprint demands and provide computational efficiency in deep learning models, quantization \cite{gholami2022survey} has become an increasingly prominent technique, particularly for Large Language Models (LLMs) \cite{li2024evaluating,dettmers2022gpt3,frantar2022gptq,xiao2023smoothquant}.
Quantization reduces the numerical precision of the values used to represent model parameters (weights) and activations—typically mapping high-precision formats like 32-bit floating point (FP32) to lower-precision formats such as 16-bit floating point (FP16), 8-bit integers (INT8), or even more aggressive formats such as 4-bit floating point (FP4) or 4-bit integer representations (INT4).
This reduction in precision can enable the practical deployment of state-of-the-art models in constrained environments without significantly compromising accuracy.

Quantization can be applied to both model weights and activations, and by reducing the bit-width of these values, it offers several critical advantages for deep learning model deployment.
First, quantization significantly reduces memory footprint requirements. For example, reducing numerical precision from FP32 to INT8 formats inherently yields a 4$\times$ reduction in memory footprint. This reduction enables the deployment and execution of larger models on memory-constrained hardware (devices with limited memory capacity).
Second, quantization facilitates faster inference. Lower-precision values reduce data movement and computational overhead, alleviating memory bandwidth pressure in modern computing systems and accelerating processing—particularly when lower-precision formats are supported by hardware. For instance, 8-bit integer operations are typically more computationally efficient than 32-bit floating-point operations on modern CPUs and GPUs \cite{rodriguez2018lower,karumbunathan2022nvidia,andersch2022nvidia,salvator2022h100}. Moreover, recent hardware accelerators and GPU architectures are beginning to provide native support for even lower-precision formats, such as 4-bit floating point (NVIDIA Blackwell) \cite{nvidia2025blackwell}, further enhancing performance. 
Third, reduced compute and memory usage directly translates to improved energy efficiency and less power requirements. This is especially beneficial for real-time inference on battery-powered or energy-constrained platforms. 
Finally, many mobile and embedded systems may lack dedicated floating-point units (FPUs), thus integer quantization plays a pivotal role in enabling the deployment of highly accurate deep learning models including LLMs, on such constrained devices—facilitating practical use cases in mobile phones, wearables, and other edge computing scenarios.

However, reducing numerical precision—by lowering the bit-width of values—can negatively impact model accuracy. To address this challenge, researchers have developed a variety of quantization methods for deep learning models \cite{frantar2022gptq,dettmers2022gpt3,egiazarian2024extreme,kurtic2024give,lascorz2024atalanta}, aiming to minimize or eliminate accuracy degradation. 

With respect to how quantization is performed, there are two primary approaches:

\begin{enumerate}
    \item \textbf{Quantization-Aware Training (QAT) \cite{jacob2018quantization,krishnamoorthi2018quantizing,esser2019learned,nagel2021white}}: In QAT, the model is trained with the target low-precision data type (e.g., INT8). Quantization is simulated during training, allowing the model to learn and adapt to quantization-induced noise. While QAT is effective at preserving model accuracy—especially for models with sensitive activation distributions—it requires access to the original training dataset and significantly increases training time.

    \item \textbf{Post-Training Quantization (PTQ) \cite{krishnamoorthi2018quantizing,fang2020post,liu2021post,zhang2023post}}: PTQ applies quantization to an already-trained (pre-trained) model without modifying its original training process. While PTQ is simpler and less computationally demanding, it may result in larger accuracy drops compared to QAT. To mitigate this, some PTQ methods incorporate light fine-tuning [cite], using a small number of additional gradient update steps on a calibration dataset. Notably, this calibration dataset may differ from the original training dataset.

\end{enumerate}

In terms of what to quantize, quantization methods are also classified as follows:

\begin{enumerate}
    \item \textbf{Static Quantization \cite{jacob2018quantization,zhang2025selectq,wang2025mergequant}}: Both weights and activations are quantized prior to inference. A calibration dataset is used to estimate the dynamic range of activations, enabling the computation of appropriate scaling factors. This method tends to provide superior inference performance and efficiency, but requires access to a representative calibration dataset.

    \item \textbf{Dynamic Quantization \cite{santini2025probabilistic,xu2018dnq,song2020drq,sun2021dynamic}}: Weights are quantized offline, while activations remain in full precision (typically FP32) and are quantized dynamically at runtime (on-the-fly) during inference. Although dynamic quantization typically offers lower performance gains compared to static quantization, it is easier to apply and does not require a calibration dataset.
\end{enumerate}

In summary, the choice of quantization method involves a trade-off between ease of application, computational cost, and the impact on model accuracy. For deployment in resource-constrained environments, the combination of static quantization and QAT often yields the most efficient and accurate results, albeit at the cost of increased complexity during model preparation.

Recent research \cite{kuzmin2023pruning,harma2024effective,chen2024low} has shown that quantization to lower bit-widths can naturally induce sparsity in deep learining tensors. This effect arises because quantization compresses the range of values of tensors, causing small-magnitude elements—those falling below a certain threshold—to be rounded to zero. As a result, quantized tensors often contain a substantial proportion of zero-valued elements. For example, \cite{chen2024low} demonstrate that quantizing state-of-the-art diffusion models to 4-bit floating point (FP4) introduces up to 34\% sparsity in the model weights.
Similarly, \cite{kuzmin2023pruning} show that the sparsity ratio increases as the bit-width decreases: quantizing tensors to 8-bit, 4-bit, and 2-bit formats leads to approximately 13\%, 35\%, and 59\% sparsity, respectively.

As deep learning models—particularly LLMs—increasingly adopt aggressive quantization strategies (e.g., 8-bit, 4-bit, or even 2-bit precision) \cite{chee2023quip,li2024evaluating,panferov2025quest}, induced sparsity is expected to become more pronounced. This trend presents a growing opportunity for systems and architecture researchers to design hardware and software optimizations that leverage sparsity to improve inference efficiency.

However, it is important to note that sparsity induced by quantization typically ranges from 30\% to 65\% \cite{chen2024low,kuzmin2023pruning}, which is substantially lower than the inherent sparsity observed in certain data domains. For instance, input data in GNNs often exhibit sparsity levels ranging from 85\% to 99\% \cite{giannoula2024pygim}. As such, while quantization-induced sparsity offers meaningful performance benefits, it may require different optimization strategies compared to naturally sparse data such as graph data.

Recent work \cite{harma2024effective} investigates the interplay between quantization and pruning, demonstrating that these two techniques are not orthogonal; rather, their combination can amplify sparsity—producing a large number of zero-valued elements in tensors. This effect arises because quantization can distort the relative magnitudes and importance of tensor elements, which can influence the selection of elements during subsequent pruning.
Additionally, quantization can naturally introduce sparsity by forcing values that are close to zero to become exactly zero, thereby directly increasing the number of zero-valued elements in tensors. 
The study further shows that applying pruning followed by quantization, or vice versa, both lead to sparsity in the resulting quantized tensors. However, through both theoretical analysis and empirical evaluations, the authors find that applying pruning prior to quantization is the more optimal sequence. This ordering minimizes quantization-induced error and better preserves overall model accuracy. 
In contrast, applying quantization before pruning can degrade performance significantly, as quantization may obscure the true importance of certain weights, leading to the erroneous removal of critical parameters during the pruning step.

\newpage
\bibliography{main}
\bibliographystyle{tmlr}

\end{document}